\documentclass[12pt]{article}
\usepackage{setspace}
\usepackage[margin=1in]{geometry}

\usepackage{authblk}

\usepackage{mathptmx}
\usepackage{times}
\usepackage{amsmath}
\usepackage{amssymb}

\usepackage[labelfont=bf]{caption}
\usepackage[labelsep=period]{caption}

\usepackage{endnotes}
\let\footnote=\endnote

\usepackage{float}

\usepackage{longtable}

\usepackage{titlesec}
\titleformat*{\section}{\Large \bfseries}
\titleformat*{\subsection}{\large \itshape}

\usepackage[dvipsnames]{xcolor}

\newcommand{\hl}[1]{{\leavevmode\color{black}#1}}

\usepackage{comment} %

\usepackage{caption}
\DeclareCaptionLabelFormat{alphatable}{Table A#2}

\newcommand*\samethanks[1][\value{footnote}]{\footnotemark[#1]}

\title{\textbf{ \Large Rooted America: Immobility and Segregation of the Intercounty Migration Network\thanks{This research was supported by the National Science Foundation (SES-1826589).}}}

\author{Peng Huang\thanks{Departments of Sociology, and Statistics, University of California, Irvine} ~ and Carter T. Butts\samethanks{} ~\thanks{Departments of Computer Science, and EECS, University of California, Irvine. Email: buttsc@uci.edu}}
\date{\vspace{-0.7cm}\small{(\href{https://doi.org/10.1177/00031224231212679}{\emph{American Sociological Review} 2023)}}}

\usepackage{natbib}
\usepackage{graphicx}

\begin{document}
\maketitle

\vspace{-0.7cm}
\noindent \textbf{Abstract}

\noindent \hl{Despite the popular narrative that the United States is a``land of mobility,'' the country may have become a ``rooted America'' after a decades-long decline in migration rates. This article interrogates the lingering question about the social forces that limit migration, with an empirical focus on internal migration in the United States. We propose a systemic, network model of migration flows, combining demographic, economic, political, and geographic factors and network dependence structures that reflect the internal dynamics of migration systems. Using valued temporal exponential-family random graph models, we model the network of intercounty migration flows from 2011 to 2015. Our analysis reveals a pattern of \emph{segmented immobility}, where fewer people migrate between counties with dissimilar political contexts, levels of urbanization, and racial compositions. Probing our model using ``knockout experiments'' suggests one would have observed approximately 4.6 million (27 percent) more intercounty migrants each year were the segmented immobility mechanisms inoperative. This article offers a systemic view of internal migration and reveals the social and political cleavages that underlie geographic immobility in the United States.}

\vspace{0.2cm}
\noindent \textbf{Keywords}

\noindent migration, social networks, political polarization, immobility, segregation
\clearpage

\noindent \hl{
While the active drivers of migration have been extensively studied, there has been less attention to the factors that \emph{hinder} migration -- a research gap that has been called the ``mobility bias'' within the migration literature (\citealt{schewel_understanding_2020}). The relatively overlooked phenomenon of immobility is important in its own right, having substantial consequences for the social world. As migration influences the functioning of labor market (\citealt{hyatt_interstate_2018}), the landscape of stratification and social mobility (\citealt{jasso_migration_2011}), and the sociocultural meanings in everyday lives (\citealt{bauman_liquid_2000,mata-codesal_ways_2015}), mechanisms that impede migration can have outcomes that extend far beyond the migration system itself.

Understanding immobility is an especially apt challenge in the context of the modern United States. Long thought of as a ``rootless society'' (\citealt{fischer_ever-more_2002}) with high geographical mobility (\citealt{long_residential_1991, steinbeck_grapes_1939}), the U.S has arguably turned into a ``rooted America'' after a decades-long decline in migration rates (\citealt{dewaard_changing_2020,frey_great_2009}). While the reality of low migration rates is clear, explanations for current population immobility are less well-developed. Macroeconomic studies have so far found that demographic and socioeconomic structures are not sufficient to explain observed levels of immobility, and neither are the business composition of labor market nor properties of the housing market (\citealt{hyatt_recent_2013,hyatt_interstate_2018,molloy_internal_2011,molloy_job_2017}).  A broader sociological view suggests the potential for cultural, political, and other social forces as possible explanatory factors \citep{tiebout_pure_1956,massey_american_1993,gimpel_seeking_2015,stockdale_editorial_2018}. Moreover, the migration system has its own intrinsic feedback mechanisms that could endogenously sustain or undermine further migration \citep{bakewell_relaunching_2014,de_haas_internal_2010},  which may also play a role in the population immobility. Probing the combined influence of these myriad factors requires a \emph{systemic} treatment of the U.S. internal migration, allowing us to simultaneously examine the joint impact of social, economic, political, and demographic mechanisms on flows of migrants throughout the country.  This paper pursues such an analysis, with the objective of identifying the factors associated with both mobility and immobility in contemporary America.  %
}

Broadly, extant research on drivers of U.S. migration and immobility shares two characteristics. First, most research examines migration from an economic perspective, assuming that most, if not all, migration is \textit{labor migration}, driven by economic incentives.\footnote{As an example, \citeauthor{eeckhout_gibrats_2004} (2004:1431) contends that ``the central thesis in this paper: population mobility is driven by economic forces.''} Yet, decisions regarding residential settlement are not purely economic (\citealt{ryo_deciding_2013}): political climate, racial composition, and urbanization of local communities are potential contributors to the phenomenon (\citealt{brown_measurement_2021,cramer_politics_2016,massey_suburbanization_2018}). This paper incorporates the sociocultural and political perspectives into the analysis of U.S. immobility.

A second dominant characteristic of the extant literature on U.S. migration is an approach that treats migration as a feature of geographical areas, examining the correlates between net migration rates into or out of states or counties and their demographic or economic characteristics. Although convenient, this practice of reducing the interconnected migration system into local features of areal units introduces two limitations. First, by aggregating across origins and destinations for migrants emigrating from or immigrating into a given area, it obscures the \emph{interactive effects} from the sending and receiving areas, such as their political or cultural similarity and differences in employment rates. Second, it does not allow for treatment of the \emph{internal dynamics} of the migration system (\citealt{de_haas_internal_2010}), in particular the presence of mechanisms such as return or stepwise migration, where the flow of migrants from one place to another can in turn affect the flow of migrants from that destination to others. Since migration is a relational process between places of origin and destination, and migration flows can influence each other, this paper takes a systemic, network approach that shifts analysis from the migration rates of areal units to the migration flows \emph{between} areal units. By leveraging migration systems theory and social network methods, we show that dissimilarities between counties are important contributors to the immobility of American society. %

To advance our understanding of the social forces behind geographical immobility in modern America, we here adopt a comprehensive theoretical framework incorporating geographical, demographic, economic, political, and social influences on migration and perform a systemic analysis of internal migration as an evolving valued network of migration flows.\footnote{By valued network (or weighted network), we refer to networks whose ties are not binary (present or absent), but are associated with a quantitative value; specifically, tie values in this study indicate the volume of migration flows between directed pairs of U.S. counties.} Using valued temporal exponential-family random graph models (valued TERGMs), this paper analyzes the network of intercounty migration flows of the United States from 2011 to 2015. We identify a pattern of \emph{segmented immobility}, where, net of other factors, less migration happens between counties with dissimilar political contexts, levels of urbanization, and racial compositions. We probe this mechanism using an \emph{in silico} ``knockout experiment,'' which suggests that in a counterfactual world without segmented immobility (but holding all other factors constant), we would expect to have seen approximately 4.6 million (27\%) more intercounty migrants in the United States each year. This implies that social and political cleavages in America are substantial contributors to immobility, and potentially exacerbate growing trends towards geographical segregation. Further, we also examine the relationship between internal and international migration flows, showing that - contrary to the balkanization thesis (\citealt{frey_immigration_1995,frey_immigration_1995-1}) - international migration into a county is positively associated with its overall domestic mobility, and does not promote net outflows of residents. The model also identifies the internal dynamics of migration systems (\citealt{de_haas_internal_2010}), including a suppression of what we dub ``waypoint'' flows (i.e., balanced in- and out-flows of an areal unit) alongside strong patterns of reciprocity and perpetuation. 
\hl{While the data availability constraints us to focus on understanding population immobility in the 2010s, the empirical evidence together with our proposed theoretical and methodological frameworks opens the door to unpack the long-term phenomenon of population immobility. This paper thus joins the growing literature that grapples with the mobility bias in migration studies (\citealt{schewel_understanding_2020}), demonstrating how a comprehensive analytical framework and a systemic, network approach offers new insights about immobility, and more broadly, the dynamics of population movement among social and geographical spaces.}

\section*{THEORY}
\hl{
Existing literature defines immobility as ``continuity in an individual’s place of residence over a period of time'' (\citeauthor{schewel_understanding_2020} 2020:344). Since immobility is not only an individualistic phenomenon, but also a population and social one, here, we offer a macrosociological definition of immobility, which is a lack of population exchange between localities. Drivers of immobility, in terms of this framework, are defined as factors that \emph{reduce} migration rates relative to what would be expected in their absence. The scarcity of migration in an immobile society has substantial impacts. Since migration is a critical channel for people to respond to fluctuations of local economy, population immobility implies a rigid labor market with lower productivity, higher unemployment rate, and more prolonged recession when experiencing economic shocks (\citealt{hyatt_interstate_2018}). Moreover, migration also serves as a way of improving life chances (\citealt{jasso_migration_2011,weber_Wirtschaft_1922}) and coping with adverse events (\citealt{spring_migration_2021}). Population immobility thus has important ramifications for  social mobility, stratification, and poverty  (\citealt{briggs_moving_2010,clark_reexamining_2008,jasso_migration_2011}).

Immobility is not merely the flip side of mobility, but carries its own sociocultural meanings. As the aspiration-ability model argues, migration requires both aspiration to migrate and the ability to realize that aspiration (\citealt{carling_revisiting_2018}). This means that immobility is not necessarily a passive outcome of simply staying in place, but can be a conscious choice to remain. In line with this view, recent literature has begun augmenting the widely discussed notion of ``cultures of migration'' with the notion of ``cultures of staying'' that facilitate and maintain immobility (\citealt{stockdale_editorial_2018}).
The level of population (im)mobility can in turn impact the broader social norms of a society; a mobile society may have a prevailing nomadic culture, while the dominant culture of an immobile society may be sedentary (\citealt{bauman_liquid_2000,mata-codesal_ways_2015}).

Understanding immobility is especially relevant in the American case. From the earliest observations of \cite{tocqueville_democracy_1834} and \cite{ravenstein_laws_1885} to \cite{steinbeck_grapes_1939}, America has long been considered a ``restless'' or ``rootless'' society with high geographical mobility. Yet, after a decades-long decline in its migration rate, the contemporary America has arguably become a ``rooted'' society with considerable population immobility. However, as \citeauthor{herting_social_1997} (1997:267) have noted, sociological research on U.S. mobility has ``narrowed and now focused almost exclusively on mobility of a purely economic or occupational variety,'' with much less focus on mobility across geographic space. In migration studies, research has been historically focused on studying the social forces that lead to migration, but largely neglected the counter forces that \emph{inhibit} people from moving, a tendency that \cite{schewel_understanding_2020} described as the ``mobility bias.'' A lack of research on geographical mobility in American sociology, together with the scarcity of theoretical and empirical work on immobility in migration studies, has led to gap in our knowledge regarding the mechanisms behind population immobility in contemporary American society.
}

\subsection*{Culture and Politics of Immobility}
While the immobility of the U.S. population has received less sociological attention, economists and geographers have conducted empirical analyses on this matter (e.g., \citealt{cooke_internal_2013,jia_economics_2022,kaplan_understanding_2017,treyz_dynamics_1993}). These studies have identified important connections between the labor market and migration rates, but their findings largely rely on the assumption that most, if not all, migration is \emph{labor migration}, driven by economic incentives. The economic perspective has a fundamental role in explaining migration and immobility; the relative gains in moving, and costs associated with both transaction costs and losses of specialized local investments \emph{are} factors that shape migration. But there also exist other factors, such as regionally specific cultural values and locally conventional ways of understanding opportunity (\citealt{carling_migration_2002,carling_revisiting_2018}), as well as preferences for particular local policies or political regimes (\citealt{tiebout_pure_1956}). Indeed, recent research on American economy has shown that over the past several decades, migration has not been effective in responding to fluctuations and shocks in labor markets (\citealt{dao_regional_2017,jia_economics_2022}). Relatedly, macroeconomic factors have not been found to have a strong correlation with migration rates in the U.S. (\citealt{hyatt_interstate_2018,hyatt_recent_2013,molloy_job_2017}). Therefore, while economic forces are important ingredients in a viable model of the migration system,  a comprehensive analysis of immobility demands considerations of other social institutions.

Although thinking on internal migration in the large has been dominated by labor market considerations, sociologists have given considerable attention to other factors when studying migration at smaller scales (e.g., across neighborhoods).  For instance, research on residential segregation has long identified how people with different racial identities and political beliefs become segregated from each other (\citealt{bishop_big_2009,krysan_cycle_2017,massey_american_1993}), including the accumulated influence of even relatively weak preferences for same-group interaction (\citealt{schelling_models_1969,sakoda_checkerboard_1971}); the latter can act as a powerful macro-level sorting force, even in the presence of economic or other factors (e.g., \citealt{butts_models_2007}). While much of this work has focused on racial segregation, more recent work has also probed segregation along political or cultural axes.  For instance, \cite{brown_measurement_2021} found that a large proportion of American adults live in neighborhoods where most residents share the same partisanship. \cite{gimpel_seeking_2015} used a survey experiment to show that people evaluate more favorably towards properties in areas with predominantly co-partisan neighborhoods. As social cleavages might deter people from settling 
in places with distinct identities and beliefs, the social gaps between rural and urban areas and those among different parts of the continent such as the South and the coastal regions (\citealt{cramer_politics_2016,hochschild_strangers_2018}), may also contribute to the inhibition of geographical movement. At another scale, in the contexts of international migration, migration studies have long stressed the roles of cultures and politics in shaping population mobility (\citealt{castles_age_2013,cohen_cultures_2011,jennissen_causality_2007,massey_worlds_1999,vogtle_global_2022,waldinger_transnationalism_2004}). Following this thread, this paper incorporates the political, racial and rural-urban structures in investigating American immobility.

\subsection*{Systemic Theories of Migration}
The second characteristic of the extant literature on U.S. immobility is that studies usually view migration as a feature of geographical areas. This approach examines the characteristics of an areal unit that influence its net immigration and emigration rates, such as percentages of current residents who are immigrants and/or emigrants. It is in essence a marginal approach that sums up (i.e., marginalizes) the migration flows from/to each areal unit across all destinations/origins to describe the overall mobility of each place. The marginal approach is empirically straightforward, and has unquestionably contributed to our understanding regarding the driving forces of migration by identifying the associations between demographic and economic features of an areal unit and the scale of its population inflows or outflows (e.g., \citealt{partridge_dwindling_2012,treyz_dynamics_1993}). Yet, migration - by definition, population moving from one place to another - is \emph{inherently} relational, having properties that cannot be reduced to the features of individual areal units.  For instance, studies considering net in- or out-migration rates in isolation must choose either the sending or receiving area as focus of analysis (thereby obscuring the joint roles of areas as origins and destinations), or must merge in- and out- migration to obtain a net migration rate (which confounds inflows and outflows). Beyond the fact that every pairwise migration flow among sending and receiving areas depends on both the properties of the sender and the receiver, such studies are unable to account for relational factors, such as geographical proximity and political difference between areal units. Neither can this approach consider the interactions among migration flows, such as reciprocal population exchange ($A\rightarrow B$ \& $B\rightarrow A$) arising from return migration. Probing such mechanisms requires a different theorization of the migration process, a systems theory of migration.

\hl{Such systemic thinking has a long tradition in migration studies under the umbrella of migration systems theory (MST, \citealt{bakewell_relaunching_2014,fawcett_networks_1989,kritz_international_1992,mabogunje_systems_1970,massey_worlds_1999}). A comprehensive theory that concerns the complex interactions among various elements related to migration, such as flows of people, information, (formal and informal) institutions, and strategies (\citealt{bakewell_relaunching_2014}), MST identifies \textit{interconnectivity} as a key feature of migration.}
As \citeauthor{de_haas_internal_2010} (2010:1593) summarized, a migration system is ``a set of places linked by flows and counter-flows of people, goods, services and information, which tend to facilitate further exchange, including migration, between the places.'' 
The theoretical focus on flows \textit{between} origin and destination suggests a relational analysis of migration, integrating push and pull factors in one single analytical framework (\citealt{lee_theory_1966}). \cite{fawcett_networks_1989} demonstrates this with a theoretical framework of ``linkages'' in MST, focusing on how various linkages between origin and destination shape the migration in between. Among the linkages \citeauthor{fawcett_networks_1989} (1989:677) discusses, here we focus on the relational linkages, ``derived from comparison of two places.'' Instead of studying how a state’s or a county's political climate influences its net marginal migration rate (e.g., \citealt{charyyev_complex_2019,preuhs_state_1999}), an analysis of relational linkages examines how the \emph{difference} in political climates between counties influences the number of people migrating from one to the other.
Another critical implication from the interconnectivity feature of migration systems is the presence of internal dynamics of migration (\citealt{bakewell_introduction_2016,de_haas_internal_2010,mabogunje_systems_1970}). As \citeauthor{mabogunje_systems_1970} (1970:16) put it, the migration system is ``a circular, interdependent, progressively complex, and self-modifying system in which the effect of changes in one part can be traced through the whole of the system.''  Similarly, \citeauthor{fawcett_networks_1989} (1989:673) argued that the migration systems framework ``brings into focus the interconnectedness of the system, in which one part is sensitive to changes in other parts.'' \hl{This means that migration is not a pure product of exogenous social forces. It forms a system with endogenous processes, where one migration flow can promote or suppress another migration flow.} For example, since migrants transmit information and social connections when they move, the migration flow from Arizona to Texas brings job information and personal contacts along, potentially inspiring migration in the opposite direction. Internal dynamics like this can lead to an endogenous accumulation of migration net of exogenous social and economic influences.

\subsection*{Migration Systems Through a Network Lens}

\begin{figure}[t]
\centering
\includegraphics[scale=0.23]{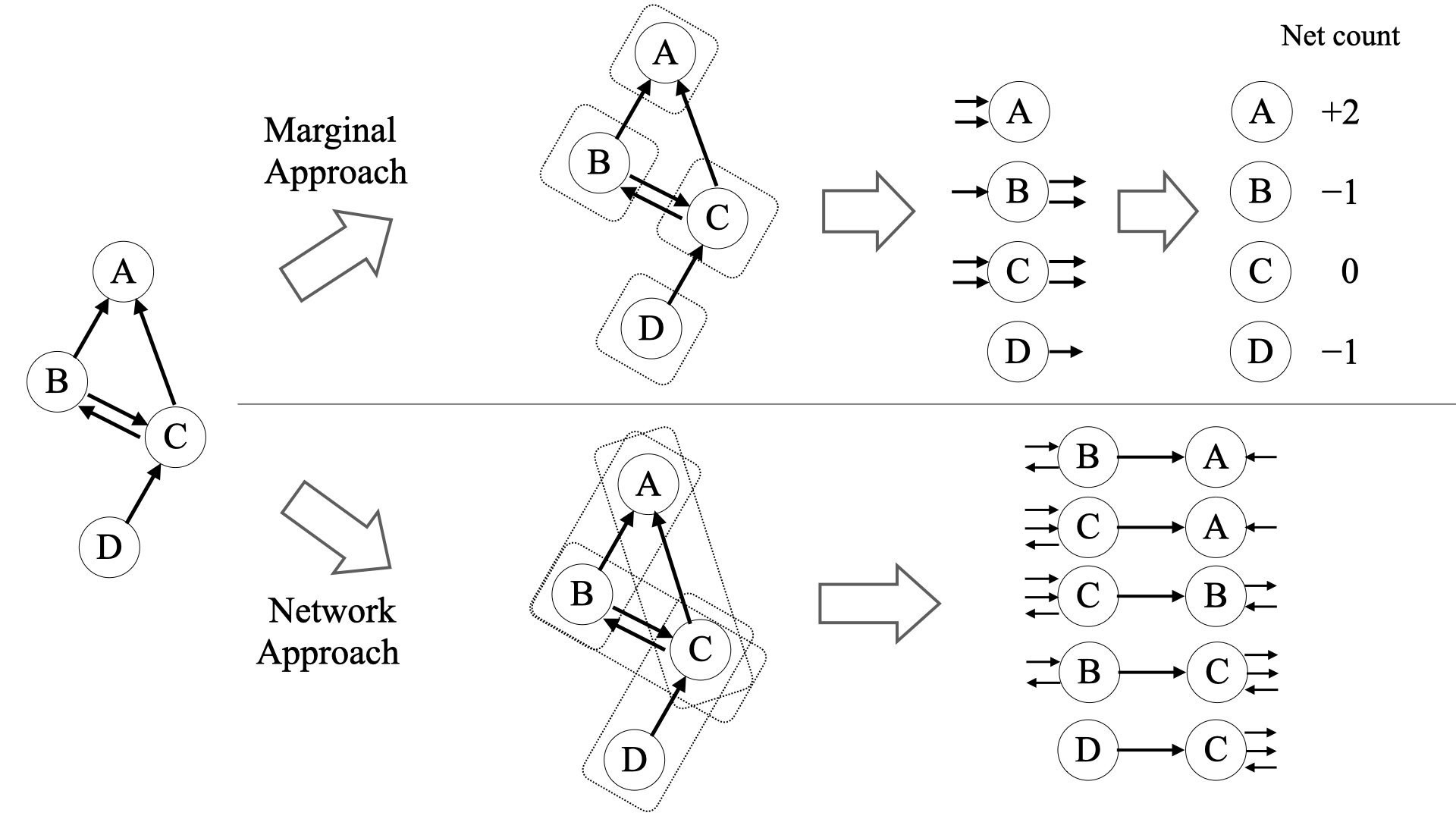}
\caption{Schematic illustration of the marginal approach versus the network approach \\ \emph{Note:} The marginal approach takes geographical areas as units of analysis, and tends to condense the in- and out- migration flows into a single number about net migration rate/count of a geographical area. The network approach takes each migration flow between a directed pair of geographical areas as an analytical unit. This approach incorporates origin and destination in understanding their joint influence on migration flows; it also preserves the local structural properties of migration flows, allowing systemic patterns to be examined.}
\label{fig:relational}
\end{figure}

\hl{The insight of interconnecitivity from MST resonates with that of social network analysis. Indeed, past research has employed social network analysis to study migration systems (\citealt{charyyev_complex_2019,desmarais_statistical_2012,dewaard_migration_2012,dewaard_resituating_2019,dewaard_changing_2020,hauer_migration_2017,leal_network_2021,liu_migration_2019,nogle_systems_1994,vogtle_global_2022,windzio_network_2018,windzio_network_2019}). This school of MST, called by \cite{bakewell_relaunching_2014} the ``abstract system,'' interrogates the macro-level migration patterns by analyzing migration networks consisting of localities (in network terms, \emph{nodes}) and migration flows between each directed pairs of localities (in network terms, \emph{edges}).\footnote{\cite{bakewell_reflections_2010,bakewell_relaunching_2014} and \cite{dewaard_resituating_2019} have debated about whether and how studies of migration networks contribute to MST. As this paper shows, echoing \cite{leal_network_2021}, we agree with \cite{dewaard_resituating_2019} that network analysis is an effective way of theorizing and testing the structures and dynamics of migration across geography; we also recognize Bakewell's critique that network analysis of migration flows is one of the many approaches to study migration systems, and that students of MST should beware the pitfall of abstract and static descriptions of migration systems. In this regard, this paper leverages theories and empirical findings in migration studies to motivate tests about structures and patterns of migration networks. We also call for more research with different levels of analysis to triangulate our findings for a comprehensive understanding of migration and immobility.}}
Network analysis effectively captures the two critical implications of MST, relational linkages and internal dynamics of migration systems, bringing new perspectives compared to the marginal approach of migration, commonly employed in studies of U.S. immobility. Rather than viewing localities/places as units of analysis, the network approach takes migration flows between places as analytical units. This perspective preserves information regarding emigration and immigration processes, enabling analysis of how characteristics of origin and destination \emph{interact} to influence migration flows, a relational account of linkages in migration systems. 
\hl{The network approach also examines the internal dynamics of migration systems, by studying the dependence structure among migration flows. The dependence structure identifies how migration flows are associated with each other, net of the exogenous contexts such as the economic and political environments. Taking the above-mentioned example of reciprocity, the network approach measures whether and to what extent, an increase of one migration flow (e.g., Los Angeles to Baltimore) is associated an increase in its opposite flow (Baltimore to Los Angeles), net of other factors. The dependence structure can further go beyond a pair of places and describes how the whole network system of migration flows are interconnected, such as how the migration inflows of Denver are associated with its outflows, which in turn serve as the inflows of another places, Dallas, Atlanta, etc.} 
Figure~\ref{fig:relational} illustrates the network approach in contrast to the marginal approach.

\hl{While the network approach introduces unique perspectives overlooked by the marginal approach, its insights has not yet been fully appreciated. One notable characteristic of prior research on migration networks is the focus on the ``diversity'' rather than the ``intensity'' of migration flows (\citealt{dewaard_resituating_2019,leal_network_2021,vogtle_global_2022,windzio_network_2018,windzio_network_2019}). In other words, extant research examines the \emph{number} of migration flows rather than their \emph{magnitudes}. This is associated with the practice of dichotomizing migration flows into two statuses, either no migrants versus at least one migrant, or few migrants versus many migrants (though \cite{windzio_network_2018} and \cite{windzio_network_2019} divide them into more (5) statuses in some parts of their research). This approach is compatible with the common practice in social network research of approximating social relations by a binary form, facilitating the use of existing network theories and methods to describe the migration system. While analyzing the ``diversity'' of migration flows offers useful knowledge about the migration system, it ignores the rich information about the variation in migration magnitudes. The intensity of migration flows becomes a critical question when it comes to understanding population immobility. In particular, \cite{dewaard_changing_2020} find that the decline of U.S. migration is not due to the decline in the diversity of migration flows (the number of county pairs with population exchange), but the decline in the intensity of migration flows (their average count of migrants). Studying the intensity of migration flows requires describing migration networks in a valued form, where the edges are not binary, but take quantitative values. Since the quantitative feature of migration intensity is critical in grappling with the question of population immobility, this paper bridges migration systems theory and recent advances in statistical and computational methods for valued network analysis (\citealt{huang_parameter_2024,krivitsky_exponential-family_2012}). We formally theorize the relational linkages and internal dynamics in the expressions of valued networks, developing a roadmap to quantitatively describe and test the interconnectivity of population flows.

On the side of migration systems, new theoretical insights are needed for studies of immobility. MST is not an exception from the mobility bias critique of migration theories (\citealt{schewel_understanding_2020}). As \cite{de_haas_internal_2010} argues, MST has historically focused extensively on migration-facilitating mechanisms that lead to the perpetuation of migration flows, but largely overlooked the migration-undermining mechanisms that lead to the decline of migration flows. Building on this critique, a line of theoretical and empirical research studies why some instances of pioneer migration drive the formation of migration systems while others do not, and the endogenous mechanisms that can undermine the migration system (\citealt{bakewell_migration_2012,bakewell_beyond_2016,de_haas_internal_2010}). \cite{bakewell_beyond_2016} further go beyond the MST framework, as they pursue the notion of  incorporating scenarios where the migration systems fail to form or perpetuate. Unquestionably, this is a promising direction to further the theorization of migration dynamics. Yet, for our focus of internal migration in the contemporary U.S., the migration system has been in existence for generations, and is unlikely to vanish in the near future. Therefore, the migration system is still a useful research subject and perspective, where we explore the social mechanisms that immobilize population from migrating.

The network approach inspires us to consider population immobility from a relational perspective. We conceptualize the pattern of \emph{segmented immobility}, that in a society where people cluster in geographical segments based on their cultural and political traits, immobility can occur due to people's tendency to avoid migrating towards places with divergent environments. By jointly incorporating origin and destination in an analytical framework, the relational perspective allows us to examine the influence of dissimilarity between counties on the magnitude of migrant populations moving between them, connecting population immobility with segregation and polarization.
Apart from examining the pattern via a hypothesis testing lens, we further utilize the idea of ``knockout experiments'' broadly employed in the experimental sciences to directly quantify its contribution to immobility. Originating in biomedical research, a knockout experiment probes the functional role of a system component by removing or inactivating it, comparing normal system behavior with behavior when the component is ``knocked out'' (\citealt{hall_overview_2009,vogel_knockout_2007}). In social sciences, knockout experiments are performed \emph{in silico}, where researchers simulate the potential social outcomes when certain social forces were removed. The knockout experiment can be considered as a model-based thought experiment (\textit{Gedankenexperiment}, \citealt{einstein_can_1935}), in which we predict the social outcomes of interest under a counterfactual scenario where certain social effects are inoperative. In our case, we compare the total number of migrants observed in the real world to that simulated when segmented immobility mechanisms are knocked out. This theoretical exercise allows us to leverage the power of modern, generative network models to gain insights into the functioning of the migration systems.}

\section*{HYPOTHESES}
\subsection*{Relational Linkages: Political Segregation and Segmented Immobility}
Decisions about migration, a behavior aiming at improving life chances (\citealt{jasso_migration_2011}), typically come out of a comparison between place of departure and destination. Moving from current place of residence, will the destination be adaptive? One critical dimension in drawing an answer is the political environment of the origin and the putative destination communities. Rising political polarization has divided Americans along the party lines (\citealt{levendusky_partisan_2009}), where social cleavage by political ideology extends to a growing array of public opinions (\citealt{baldassarri_partisans_2008,dellaposta_pluralistic_2020}) and choice of lifestyles (\citealt{dellaposta_why_2015}), and has lead to segregated social networks and tensions in relationships such as familial interactions (\citealt{chen_effect_2018,diprete_segregation_2011}). This political alignment also happens across space, with distinct political consciousness across geographical regions, rural and urban lands, and local neighborhoods (\citealt{bishop_big_2009,cramer_politics_2016,hochschild_strangers_2018}). Recent spatial analysis on partisan isolation reveals that a large fraction of American adults lives in places where almost no one in their neighborhood votes in a manner opposed to their own (\citealt{brown_measurement_2021}). They also found that this pattern is prevalent nationwide and is a distinct pattern from segregation in other dimensions such as across racial lines. This state of affairs is also overtly recognized within American political discourse, where media outlets routinely make distinctions between ``red'' (conservative) and ``blue'' (liberal) regions, and ascribe (correctly or not) a large body of cultural and political traits to both the regions and their inhabitants (\citealt{badger_suburbs_2018,wallace_true_2020}).  To the extent that individuals are likely to both affiliate with the political culture of their area, and regard their opposites on the political spectrum with suspicion and even hostility (\citealt{iyengar_affect_2012, iyengar_origins_2019}), people may be unwilling to migrate between regions with differing political cultures. Even setting aside motivations arising from political culture, according to the public choice theory and the consumer-voter model, people should still be more willing to migrate to regions whose governments most closely match their own policy preferences (\citealt{dye_american_1990,tiebout_pure_1956}), with those from solidly ``red'' areas preferring to move to other ``red'' areas, and likewise for those from ``blue'' areas. Empirical analyses using various data and methods generally confirm the existence of migration preference towards co-partisanship (\citealt{tam_cho_voter_2013,gimpel_seeking_2015,liu_migration_2019}), though with some counter evidence (\citealt{mummolo_why_2016}). Together they motivate the following hypothesis:

\emph{Hypothesis 1.1: \emph{Ceteris paribus,} the more dissimilar counties are in their average political orientation, the lower the migration flow between them.}

\hl{The limited population exchange between geographical segments with dissimilar social environments, or what we call \emph{segmented immobility,} may not be unique to the political dimension, but would rather be a pervasive pattern arising from people's evaluation of places along multiple dimensions. One of the underlying mechanisms that can lead to such a pattern is homophily. Homophily refers to people's tendency to be connected to and interact with those similar to themselves in various characters such as racial and ethnic identity, religious belief, political ideology, personality, and normative inclination like altruism (\citealt{diprete_segregation_2011,leszczensky_what_2019,mcpherson_birds_2001,moody_race_2001,smith_social_2014,wilson_human_2009}). Homophily occurs not only within personal networks, but is also a spatial phenomenon, where people tend to live close to others with similar racial identity, economic background, or political ideas (\citealt{bishop_big_2009,massey_american_1993,intrator_segregation_2016}). A social process that can give rise to this spatial pattern is that residents choose to migrate towards places where people similar to them concentrate, but avoid destinations with identities different from their own (\citealt{crowder_neighborhood_2012,massey_migration_1994,schelling_models_1969}). Although literature about this residential sorting process focuses primarily on mobility among neighborhoods in urban areas, we argue that a similar process may also work at a larger scale. When choosing a county to reside in, people may favor places with a significant presence of their co-ethnics and those that host like-minded residents. Likewise, opportunities to migrate may be turned down if they would lead to settings in which the mover would find themselves socially isolated or targets of discrimination.

Segmented immobility can also arise in more subtle ways: even if individuals do not avoid living with dissimilar others, they may exclude potential migration destinations that are not able to offer the lifestyle and cultural consumption they are used to. Moving from Manhattan to rural Texas, the New Yorker would miss the coffee shop at the street corner, while a Texan migrating in reverse might feel nostalgia for the country music scene back home. Hence, migration between rural and urban areas, and across culturally different states is likely to be disfavored. Racial demographics can also be a determinant of the cultural and economic conditions of a place, where a racially diversified area not only offers a diversity of cultural affordances (as reflected by cuisines and music genres, for example), but also provides vital economic opportunities and ethnic capital for ethnic minorities (\citealt{fernandez-kelly_back_2008,lee_why_2017,zhou_chinatown_1992}). Similarly, migrants from rural counties might find themselves excluded from jobs in urban areas because they demand skills hard to obtain in their rural hometown, potentially leading to circulation of poor rural migrants among non-metropolitan counties (\citealt{lichter_intercounty_2022}). These together suggest an economic dimension to segmented immobility, in which migration between dissimilar places is suppressed when these places have different economic structures, making it difficult for migrants to utilize human capital accumulated in their place of origin. As services, cultural activities, and modes of production become specialized to a local social ecology, those adapted to both producing and consuming within that ecology will find it increasingly difficult to utilize opportunities in ecologically distinct localities.  Together, these mechanisms lead to the following hypotheses:}

\emph{Hypothesis 1.2: \emph{Ceteris paribus,} the more dissimilar counties are in their levels of urbanization, the lower the migration flow between them.}

\emph{Hypothesis 1.3: \emph{Ceteris paribus,} the more dissimilar counties are in their racial compositions, the lower the migration flow between them.}

The hypothesis of segmented immobility is based on the assumption that most residents and migrants identify themselves with their current residence, which is also the place of departure. However, if we were to suppose that the majority of the migrating population moved to \emph{escape} their current residence in favor of one more to their liking, then migration flows would preferentially occur between dissimilar areas; this would lead to ``mobility across segments,'' in contrast to ``segmented immobility.'' This type of process was proposed by \cite{tiebout_pure_1956} as a mechanism of political sorting, and at the micro-level similar processes have been occur in personnel turnover (\citealt{krackhardt_snowball_1986}) and cascade-like relocation phenomena (\citealt{schelling_micromotives_1978}).  We contend that such sorting flows are unlikely to be the major force of the contemporary internal migration in the United States. This is because research has not documented substantive social changes that drove massive redistribution of American population since the fading of the Great Migration of Black Americans in 1970s (\citealt{sharkey_geographic_2015,tolnay_african_2003}), and the continuing decline of internal migration for the past decades seems to suggest a scenario of equilibrium, or ``an inflection point'' (\citealt{molloy_internal_2011}: 173). Analyses of voting behaviors also reveal that internal migrants tend to hold political orientations consistent with those of their origins (\citealt{preuhs_pack_2020}). Nevertheless, we consider it as a competing hypothesis to the segmented immobility hypotheses above, and will directly test them in our analysis.

\subsection*{Internal Dynamics: Reciprocity and Perpetuation}

The network approach also brings the opportunity to formally examine the interrelationships among migration flows themselves, thereby revealing the internal dynamics of the migration system. This is particularly true for the valued network models used here, which allow us to examine quantitative questions that go beyond the simple presence or absence of migration.  Here, we focus on several mechanisms motivated by prior theory on migration behavior at the micro-level, which lead to hypotheses regarding interdependence among macroscopic migration flows. 

We begin by considering the relationship between one migration flow (e.g., from Seattle to Austin) and its opposite flow (e.g., from Austin to Seattle). As has been argued by the transnationalism school in the context of international migration, migration is not a one-way process, but an enduring reciprocal exchange of people, goods, and cultures between sending and receiving countries (\citealt{schiller_immigrant_1995,waldinger_immigrant_2013}). These same mechanisms could also apply to movement within countries: in his classic work, \citeauthor{ravenstein_laws_1885} (1885:187) documented the ``universal existence''  of ``counter-currents of migration''  between counties in the United Kingdom, where population not only moved from agricultural areas to commercial and industrial areas, but each of these migration currents corresponded to a current running in the reverse direction. Considering that migration control policies suppress the circulation of international migrants between states (\citealt{czaika_effect_2017,massey_why_2016}), we expect even stronger reciprocity of migration flows in the context of internal migration in the U.S., where there is no state control over migration. The reciprocity can arise from the sharing exogenous properties of the bidirectional flow; for example, geographical proximity is a driver of reciprocal population exchange, as it facilitates migration in both directions. Nevertheless, we argue that reciprocity is also an internal dynamic of the migration flow system, such that net of exogenous factors, a larger migration flow in one direction is still associated with a larger migration flow in the opposite direction.

\hl{The endogenous, systemic pattern of reciprocity could result from at least two micro-mechanisms in the American migration system.} First, migration in one direction actively motivates the flow in the opposite direction. Migrants bring information and social connections from their origin to destination, inspiring and facilitating migration in the opposite direction. Second, return migrants participate in flows in both directions, contributing to the positive association between the pair of flows.
\hl{For example, \cite{spring_migration_2021} find family ties to be a decisive factor for people separated from their spouses or cohabiting partners to return to their hometowns. \cite{von_reichert_impacts_2014,von_reichert_reasons_2014} show that migrants returning from urban to rural areas are mainly driven by social connections rather than economic opportunities, and they usually bring people in their family network along when they return.} Given the plausibility of both mechanisms, we posit the following macro-level hypothesis:

\emph{Hypothesis 2.1: \emph{Ceteris paribus,} the flow of migration from county A to county B increases with the flow of migration from county B to county A.}

As is implied above, an important feature underlying the macro-level pattern of reciprocity is the presence of (interpersonal) migrant networks that link persons in the sending and receiving regions, so as literature on transnationalism points out (\citealt{lubbers_social_2020,mouw_binational_2014,verdery_communication_2018}). Migrant networks, according to the definition of \citeauthor{massey_theories_1993} (1993:448), ``are sets of interpersonal ties that connect migrants, former migrants, and nonmigrants in origin and destination areas through ties of kinship, friendship, and shared community origin.''  We have argued that, theoretically, migrant networks should contribute to the reciprocity of migration-flow networks, by migrants bringing resources to destination and triggering population moving in the opposite direction, and by motivating return migrants moving between regions in both directions. Yet, reciprocity is not the only pattern that emerges from migrant networks. As the cumulative causation theory argues, the formation and development of migrant networks are a key contributor to the perpetuation of migration flows, which suggests the presence of inertia (aka. a positive association) of the same migration flow over time (\citealt{massey_social_1990,massey_theories_1993}). Specifically, migrants not only bring information and social connections of origin to their destination, triggering migration in reverse, but also take those kinds of resources from destination back to their origin, by returning home or via communication with nonmigrants back home; this lowers the costs and potentially raises the aspiration of migrating to the same destination, making future migration more likely to happen (\citealt{garip_social_2008,garip_network_2016,liang_cumulative_2008,liang_migration_2013,lu_emigration_2013,massey_continuities_1994,palloni_social_2001}). Therefore, we hypothesize the perpetuation of migration flows in the system:

\emph{Hypothesis 2.2: \emph{Ceteris paribus,} the flow of migration from county A to B increases with the past flow of migration from county A to county B.}
\hl{\subsection*{Waypoint Flows}
We now turn to the internal dynamics at the level of triads, i.e., among three localities (\citealt{davis_structure_1972}). Specifically, we examine the \emph{waypoint structure} in the migration flow networks. Similar to a layover airport that mainly serves connecting flights, the ``waypoint'' is a place where its scales of migrant inflows and outflows are similar to each other. Demonstrated in Figure~\ref{fig:Waypoint}, County A, B, C have the same amount of associated migration events in total (six), but their distributions of immigration and emigration are different. County A is an example of waypoint, where inflows and outflows are evenly distributed, while County C is a counter-example that has few inflows but many outflows, and County B is in between. The difference can be represented by the measure of \textit{waypoint flow}, which is the total amount of migration flows moving in and out of a focal place. When we hold constant the total number of migration events, a high volume of waypoint flow represents a high level of equality between their inflows and outflows. In Figure~\ref{fig:Waypoint}, the volume of waypoint flows for County A, B, C are three, two, one, respectively, indicating that County A has the most balanced inflows and outflows, followed by B and C. 

\begin{figure}[t]
\centering
\includegraphics[scale=0.06]{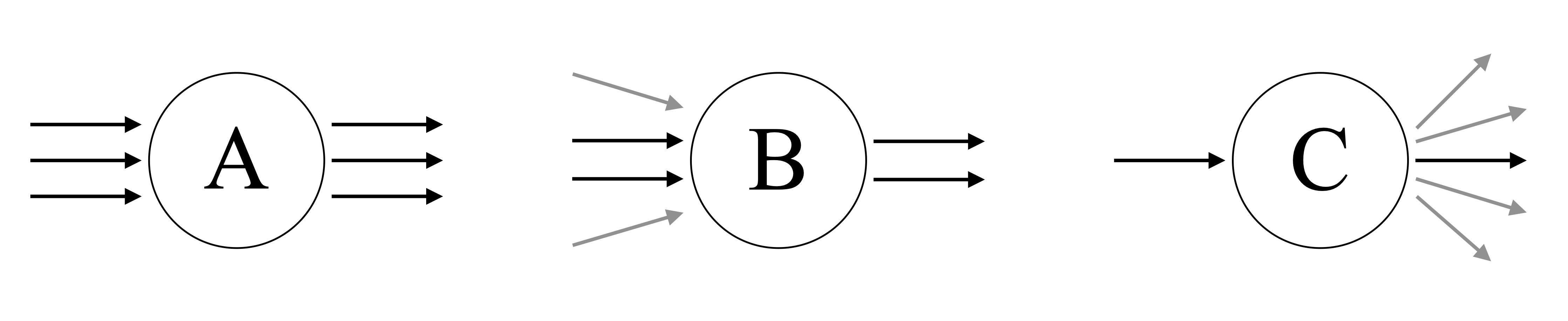}
\caption{Waypoint Flows \\ \emph{Note:} County A, B and C have the same number (six) of associated migration events, but their levels of equality in the in- and out-migration flows vary. This is reflected on their volumes of waypoint flow, three for the most equal County A, two for the medium equal County B, and one for the least equal County C.}
\label{fig:Waypoint}
\end{figure}

Waypoint flows can arise from chain-like migration processes (\citealt{leal_network_2021}), such as stepwise migration and relay migration. Stepwise migration refers to movements of migrants that pass through at least one waypoint before reaching the final destination (\citealt{conway_step-wise_1980}). Originally theorized in the classic piece of \cite{ravenstein_laws_1885}, stepwise migration has been widely documented to happen under various social contexts (\citealt{freier_impact_2019,paul_stepwise_2011,paul_multinational_2017,riddell_urban_1972}), including internal migration in the United States (\citealt{dewaard_population_2016}). Stepwise migration usually happens when the final destination is not directly reachable because of the high financial burden or the hardship in acquiring visas for international migration; migrants respond to this challenge by first migrating to waypoints that facilitate their accumulattion of capital of various kinds before moving to their ultimate stop (\citealt{paul_stepwise_2011}). Another migration process that gives rise to waypoints is relay migration, where exodus of local residents leave vacancies in the labor market that attract inflows of migrants (\citealt{durand_new_2010}). Relay migration can also happen in the reverse order, where the influx of migrants triggers outflows of local residents (\citealt{leal_network_2021}). The key difference between stepwise migration and relay migration is that the former is about the same migrant taking a multiple-step move, but the latter involves different populations participating in the inflows and outflows of waypoints.\footnote{We thank an anonymous reviewer for pointing out this distinction.} The two processes are not distinguishable in aggregate migration flows, but both reflect the interconnectedness of the migration system, where the change of one migration flow could alter another via their shared connection at the waypoint.

While existing literature has studied the migration processes that can generate waypoint flows, less is known about their prevalence in the migration systems. This knowledge gap drives us to further theorize chain-like migration processes by considering them against other migration processes. Since migration is an arduous undertaking with substantial risks, costs and barriers (\citealt{carling_revisiting_2018,liang_cumulative_2008,schewel_understanding_2020}), prolonging one-step migration into stepwise is not a desirable choice unless necessary. Compared to international migration, internal migration in the U.S. is usually more affordable and less constrained by state regulations; an American internal migrant is thus less likely to opt for stepwise migration than a Filipino who wishes to settle in Spain. Relay migration is not a universal pattern, either. It requires substantial inflows or outflows that can alter the local labor and housing market or socio-political contexts to trigger further migration flows. This means that waypoint flows arising from relay migration is conditioned on uncommon incidents such as major economic shocks or environmental disasters that bring mass population movements.

Moreover, a \emph{deficit} in waypoint flows can also be a structural signature of inequality in migration flow networks,  where the majority of counties either receive many migrants but send few, or send many migrants but receive few. This imbalance between in- and out-migration flows can arise when the difference in the level of attractiveness across places remain unaccounted for; in this case, a county is either popular so to attract and retain migrants, or the reverse. A lack of waypoint flows can also occur endogenously.  For instance, potential migrants may take current levels of migration rate as social or economic signals about the long-term desirability of an area, and adjust their own decisions accordingly.  This tendency creates a feedback loop in which influx of migrants to an area leads potential out-migrants from the area to instead remain, which in turn feeds an imbalance between in- and out-migration ($in>out$) that motivates yet more potential migrants to move in. By turns, this same mechanism may lead to a Schelling-like exit cascade (\citealt{schelling_micromotives_1978}), in which an initial out-migration shock both encourages further exit from those now in the location and makes the location appear less-desirable to potential in-migrants, thus leading to poorer in/out balance ($in<out$), and further net out-migration.

Clearly, then, there are interesting and plausible hypotheses in both directions. For simplicity, we hypothesize a high-waypoint scenario, reflected by a balanced distribution of inflows and outflows:

\emph{Hypothesis 3: \emph{Ceteris paribus}, the inflows of migration to a county increase with its outflows.}

It should be noted that the waypoint flow is a network structure related to but distinct from the transitive hierarchy studied in some international migration network research (\citealt{leal_network_2021,windzio_network_2018}). Both are triadic structures concerning migration flow among three places ($i,j,k$). The waypoint flow is the backbone of the transitive hierarchy, as the former considers migration flows of $i\rightarrow j$  and $j\rightarrow k$, while the latter involves the co-presence of $i\rightarrow k$ flow. This means that networks with a lack of waypoint flow will have few closed transitive triads ($i\rightarrow j, j\rightarrow k,i \rightarrow k$).\footnote{This is because transitive hierarchy is a network structure built on waypoint flow, and an underrepresentation of the former necessarily implies an underrepresentation of the latter. It is possible that in this circumstance there can be net tendency for waypoint flows to be transitively rather than cyclically closed \textit{where they occur}. But one will still see fewer transitive closures (as there are fewer paths to close in the first place) than one would expect by chance. Put another way, standard transitivity effects measure the overrepresentation of both waypoint flow and transitive closure, not merely the latter.} We thus focus on the more fundamental waypoint flow structure to explore the more basic form of endogenous mechanism in the migration network.

}

\subsection*{Internal Migratory Response to Immigration}

\begin{figure}[t]
\centering
\includegraphics[scale=0.06]{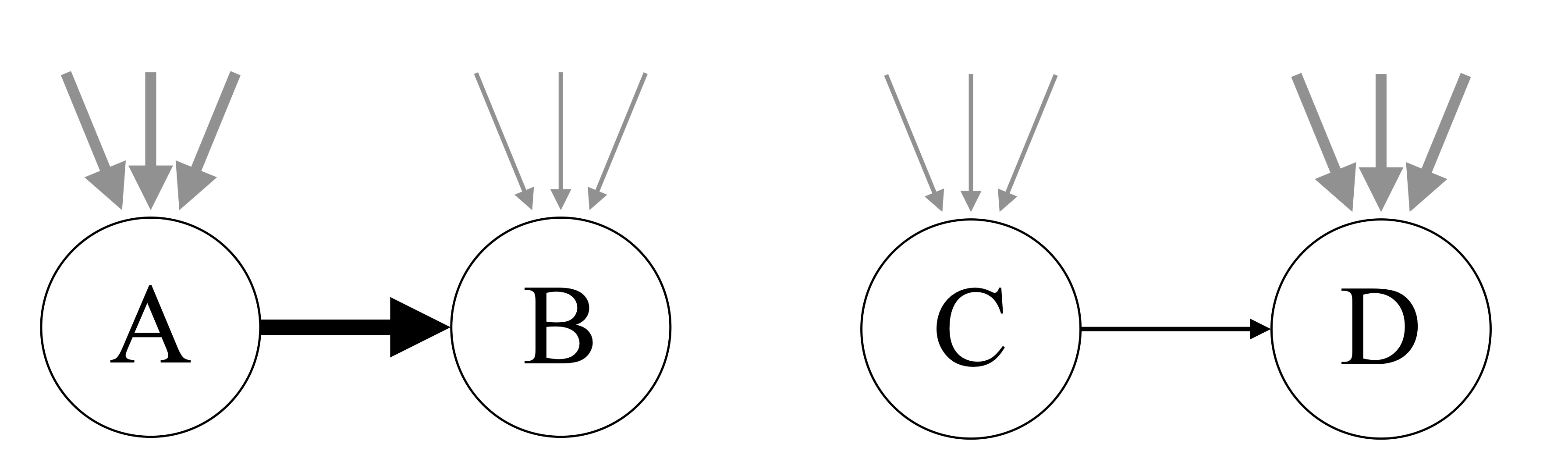}
\caption{Hypothesized relation between internal and international migration \\ \emph{Note:} vertical grey arrows denote international immigration flows and horizontal dark arrows denote internal migration flows. Arrow width denotes the magnitude of migration flows. According to the hypothesis of \cite{frey_immigration_1995}, larger population are expected to migrate from County A that has high immigrant inflows towards County B that has low immigrant inflows, while less population would leave County C that has low immigrant inflows towards County D that has high immigrant inflows, net of other factors.}
\label{fig:balkanization}
\end{figure}

Lastly, this paper considers the relationship between international migrant (i.e., immigrant) inflows and internal migrant flows in the United States. Debates about the impact of immigration on internal migration provoked much research in 1990s, which provided insights about the demographic and economic influence of immigration, the structure of labor markets, and the social cohesion of American society. \cite{frey_immigration_1995} hypothesized that immigration to the U.S. would lead to demographic balkanization, in which immigrant inflows trigger outflows of internal migrants and deter their inflows. Figure 3 visualizes this hypothesis from the perspective of internal migration flows, where larger population are expected to migrate from County A that has high immigrant inflows towards County B that has low immigrant inflows, while less population would leave County C that has low immigrant inflows towards County D that has high immigrant inflows, net of other factors. This mechanism was proposed to lead to a ``balkanized'' regionalization of the U.S., with immigrants and natives increasingly segregated in different regions.\footnote{Since the phrasing of ``balkanization'' can be construed to carry certain normative connotations regarding immigration, we follow the practice in \cite{kritz_impact_2001}, and phrase the phenomenon as the internal migratory response to immigration.} Empirical findings were inconclusive about the relationship between internal and international migration flows, with some supporting evidence for \citeauthor{frey_immigration_1995}'s (1995a) hypothesis (\citealt{borjas_native_2006,frey_immigration_1995,frey_immigration_1995-1,white_effect_1998}), and other opposing evidence (\citealt{card_immigrant_2001,kritz_impact_2001,wright_linkage_1997}). This paper revisits this debate with new data about migration in 2010s of all U.S. counties. Following \citeauthor{frey_immigration_1995}'s (1995a) proposal, we hypothesize, from the perspective of internal migration flows, that:

\emph{Hypothesis 4: \emph{Ceteris paribus,} an internal migration flow increases with international immigration inflow in the sending county, but decreases with international immigration inflow in the receiving county.}

\section*{DATA AND METHODS}
\subsection*{Valued TERGMs}

\hl{We use the valued temporal exponential-family random graph models (valued TERGMs) to study the intercounty migration-flow network within the United States. Exponential-family random graph models (ERGM) offer a flexible framework that describes the probability of observing certain network structure as a function of their nodes' covariates, edges' covariates, and the dependence structure among edges (\citealt{hunter_ergm_2008,wasserman_logit_1996}). This empowers us to simultaneously model the characteristics of areal units (nodes' covariates), the relational linkages (edges' covariates), and the internal dynamics (dependence structure) hypothesized to characterize migration-flow networks. Previous research has employed ERGMs in a wide range of social network settings, including friendship networks in schools (\citealt{goodreau_birds_2009,mcfarland_network_2014,mcmillan_tied_2019}), inmate power relationships in prison (\citealt{kreager_where_2017}), collaboration networks in firms (\citealt{srivastava_culture_2011}), online social networks (\citealt{lewis_limits_2013,lewis_preferences_2016,wimmer_beyond_2010}), and various types of gang networks (\citealt{lewis_rules_2019,papachristos_corner_2013,smith_trust_2016}). While most studies model social relations as binary networks (i.e., encoding only whether or not relationships exist), it is more accurate and informative to model migration-flow systems as valued networks, where edges represent the size of population migrating between county pairs. Although valued ERGMs (VERGMs) are to date less well-studied than binary ERGMs, we employ the count-data ERGM framework of \cite{krivitsky_exponential-family_2012} to capture migration rates in a quantitative fashion. Our model also incorporates temporal effects (the perpetuation pattern), making it a valued temporal ERGM, or valued TERGM).}

\hl{We detail the model setup, computation methods and procedures in Part B of the supplement. We also develop and report a model adequacy check for VTERGMs, detailed in Part D of the supplement.}

\subsection*{Knockout Experiments}
Exploiting our ability to quantitatively model the magnitude of migration flows using VTERGMs, we perform \emph{in silico} ``knockout experiments'' to show the impact of modelled social mechanisms in influencing the size of the migrant population, tackling the question of how particular social forces give rise to immobility. Originating and widely used in the experimental sciences (\citealt{hall_overview_2009,vogel_knockout_2007}), this way of thinking has also been applied in the social sciences (e.g., \citealt{han_quantifying_2021,lakon_simulating_2015,xie_long-term_2019}), especially in the context of agent-based modelling (\citealt{miller_complex_2009}). For social science research, the knockout experiment can be considered as a model-based thought experiment (\textit{Gedankenexperiment}, \citealt{einstein_can_1935}), where we use models to predict social outcomes of interest (e.g.,~total number of migrants) under a counterfactual scenario where certain social mechanisms are removed (e.g.,~the political segmentation effect) while other factors are held constant.  This approach is particularly powerful for nonlinear, systemic models like those used here, where seemingly small, local effects can have global consequences.

\hl{Our knockout experiments are performed as follows.  Starting with a VTERGM calibrated using empirical migration data, we compute the total expected number of intercounty migrants when either the political segmentation mechanism (\emph{per se}) or all of the three segmentation mechanisms (jointly) are knocked out, and compare this number with the observed migrant population size. The differences in total migrant population between these scenarios thus offer an insight about the scale of mobility suppression from these segmentation mechanisms - i.e., if we could ``turn them off,'' what would we hypothetically expect to see? The counterfactual scenario was simulated by the Markov chain Monte Carlo (MCMC) algorithm based on the network model with zero coefficient values for the specified knockout social effects.\footnote{We also simulated networks using the full model (without knockouts), and calculated the difference in the total migrant size between the full-model simulation and the observed, as a measurement of bias introduced in the procedure. We then corrected the total population sizes in knockout scenarios by extracting that difference. As the difference is 0.7\% of the observed migration volume, corrected and uncorrected estimates are nearly identical.} Since the network model specifies the dependence structure between migration flows, it accounts for both direct impacts of the segmentation between each county pair on their own migration flows, and the indirect impact arising from the internal dynamics of migration systems that spillover this exogenous impact. It thus offers a systemic depiction of the segmented immobility pattern.}

\subsection*{Data}
We analyze the inter-county migration flow data for from the American Community Survey (ACS). 
\hl{As a political unit with reliable demographic and economic data, counties serve as a level of geographical area that effectively describes the social contexts of residents such as political environments and rurality (\citealt{lobao_local_2019,mueller_ethnically_2023,schroeder_across_2021}). Movement across a county boundary is a frequently-used definition of internal migration in the literature (\citealt{brown_conceptualizing_2016,dewaard_changing_2020,hauer_migration_2017,partridge_dwindling_2012}).}
Administered by U.S. Census Bureau, ACS surveys respondents’ location of residence one year ago and estimates the population size that migrated between each pair of counties each year.\footnote{Another dataset that reports counts of county-to-county migration flows is offered by the Internal Revenue Service (IRS) (\citealt{hauer_irs_2019}). While ACS is a nationally representative demographic survey, the representativeness is a potential concern of the IRS data, as it only contains people filing tax returns, and therefore are not representative of the elder, the low-income, and the immigrant populations. Further, the IRS data of the post 2011-2012 period currently suffers from systemic problems that are not yet resolved (\citealt{dewaard_user_2022}). Nonetheless, the IRS reports migration data annually, and can be useful for fine-grained dynamic analysis of migration before 2011.} Their released data reports the averaged annual migrant counts in a five-year time window in order to have enough monthly samples for reliable estimation at the inter-country level. The outcome of interest is the count of migrant population flowing between 3,142 counties in the United States during 2011-2015.%

The explanatory variables are from 2010 United States Census and ACS 2006-2010. Specifically, the intercounty distance was calculated based on the 2010 Census by \cite{national_bureau_of_economic_research_county_2016}. We use presidential election turnout in 2008 to indicate the political climate of each county (\citealt{mit_election_data_and_science_lab_county_2018}). Data sources for each covariate are listed in Part A of the supplement.

\subsection*{Variables}
\textit{Dependent edge variable.} The model predicts the count of migrants moving between each directed pair of counties during 2011-2015 from the American Community Survey. Because the count-valued ERGM effectively operates through a logarithmic link (see \citealt{krivitsky_exponential-family_2012}), \hl{we are able to directly predict untransformed migrant counts in the model.} %

\vspace{0.1in}
\textit{Dissimilarity score for segmented immobility.}
The segmented immobility thesis contends that less migration happens between places with different political climates, levels of urbanization, and racial compositions. 
\hl{To test the hypotheses, we measure the dissimilarity within each pair of counties along these dimensions as edge covariates for migration flows.\footnote{We use the L1 Euclidean distance measure, or what was called the dissimilarity score in social segregation literature (\citealt{massey_dimensions_1988}).}}
For difference in political climates, we follow \cite{liu_migration_2019} and calculate the absolute difference in percentage of votes for the Democratic candidate in the 2008 presidential election, a behavioral measure of partisanship.\footnote{Given how Hawaii and Alaska calculate their election results, we conduct the following operations to map their local election data to counties. Since Kalawao County, HI is regarded as a part of Maui County, HI for election purposes, we input the election results of both counties with their pooled results. Election results in Alaska were reported by election districts rather than counties. We used the map to match election results of the 40 districts with the 28 counties. The result of a county was input with that of its district if the county was affiliated with one single district. We take the mean of the results of the districts that a county spans if the county is affiliated with multiple districts. The approximation would underestimate the political difference between counties, but the bias should be minor as the affected county takes less than 1\%  of the sample. We thank the election offices of Hawaii and Alaska for clarification and maps of the election districts during 2002-2013 in Alaska.} For levels of urbanization, we calculate the absolute difference in percentage of population residing in rural areas, a standard urbanization measurement reported in 2010 Census. For racial/ethnic composition, we use a function of the sum of absolute differences in population share for each racial category. Formally, we describe relationship between counties $A$ and $B$ by
$$
R_{AB}=\frac{1}{2} \sum_{i=1}^n{\left| \frac{P(A)_i}{P(A)}-\frac{P(B)_i}{P(B)} \right|}
$$
\noindent where $R_{AB}$ is the dissimilarity score of racial composition between county A and county B, $P(A)$ is the total population size of county A and $P(A)_i$ is the population size of the i-th racial group in county A. We follow the Census to consider the following five racial/ethnic categories, Hispanic or Latino, Non-Hispanic Black or African American, Non-Hispanic Asian, Non-Hispanic White, and population with the other racial identifications. The difference is divided by two to make the theoretical value of the score range from 0 to 1. The higher the dissimilarity score, the more different the two counties are in the measured dimension, and the less migration is expected according to the hypotheses.

\vspace{0.1in}
\textit{Network covariates.} 
\hl{We utilize the mutuality statistic in the ergm.count R package to measure reciprocity in migration flows (\citealt{krivitsky_modeling_2013}).\footnote{\hl{The reciprocity statistic calculates the summation of minimum value of each pair of edges by dyad. Formally,
$
g_m(y)=\sum _{(i,j) \in \mathbb{Y}} min(y_{ij}, y_{ji} )
$, where $\mathbb{Y}$ denotes the set of all ${i,j}$ pairs.
}} 
A positive coefficient for indicates reciprocity within the network, such that a large migration flow is more likely to have a larger counter current rather than a smaller one, \emph{ceteris paribus}.}

The model also includes the number of migrants in the past 5-year window during 2006-2010 in log scale from ACS as an edgewise covariate, to account for the association of migration flows over time, utilizing the temporal feature of TERGMs. A positive coefficient for this term suggests the perpetuation of migration flows over time, while a negative coefficient suggests negative dependence between past and present flows.

Waypoint flow is captured by the summation of the volumetric flow for each county in the network.  Intuitively similar to the notion of the flow volume ``through'' or ``across'' an areal unit in the field of fluid mechanics, the flow associated with a given unit is the minimum of its total inflows and its total outflows.\footnote{\hl{Formally:
$g_f=\sum_{i \in \mathbb{V}} min \{  \sum_{j \in \mathbb{V}, j \neq i} y_{ij} , \sum_{k \in \mathbb{V}, k \neq i} y_{ki}      \}$,
where $\mathbb{V}$ is the set of all vertices/nodes (counties), and  $y_{ij} , y_{ki} $  are values of the edge from county $i$ to $j$ and $k$ to $i$, respectively. The term is similar to the 2-paths or mixed-2-stars in binary ERGMs, which is the number of times a node receives an edge and sends another (\citealt{morris_specification_2008}). }}
A positive coefficient for the flow term indicates that the observed network has larger volumes of waypoint flows than would be expected given all other mechanisms and covariates specified in the model, suggesting a relatively equal distribution of in- and out-migration flows across counties, and a negative coefficient would indicate otherwise.

To examine the relationship between internal and international migration flows, for each intercounty migration flow, the model measures its associations with the total immigrant inflows of its sending and receiving counties in the same time window (2011-2015). The international immigrant population is transformed by taking the natural logarithm.

\vspace{0.1in}
\textit{Demographic covariates.} The model also accounts for areal characteristics that might influence intercounty migration. These include demographic characteristics of the sending and receiving counties, from basic geo-demographic statistics to demographic compositions.

Classic models from spatial econometrics (a.k.a.~the gravity model) suggest that migration rates are positively associated with the population sizes of the sending and receiving regions, but negatively associated with their distance, with a general power law form (\citealt{boyle_exploring_2014,poot_gravity_2016,zipf_p1_1946,zipf_human_1949}). Such models can be expressed by a linear combination of population and distance in the log space. Formally,
$$
log(M_{AB})=\beta_0+\beta_1 log(P_A)+\beta_2 log(P_B)+\beta_3 log(D_{AB})+\varepsilon
$$
\noindent where $M_{AB}$ is the migration volume from $A$ to $B$, $P$ is the regional population, $D$ is the inter-regional distance, $\beta$ is a covariate vector, and $\varepsilon$ is the residual.  \cite{almquist_predicting_2015} suggest that this may arise from the volume of interpersonal contacts between regions, which also frequently scales in power law form.  Although we do not use a regression model of this type here, we emulate this class of effects within our own model by incorporating (1) the log populations for the sending and receiving counties and (2) the log distance between counties (in kilometers) as predictors of inter-county migration rates; this means that our models can be considered as an extension of the gravity model. We also include population densities of sending and receiving counties (in thousand people per squared-kilometer), since \cite{cohen_international_2008} has shown that population density is a critical factor in predicting international migration flows. We use data from the 2010 Census for the covariates listed above.

For demographic composition, the model first considers the age structure of sending and receiving counties, as \cite{kim_determinants_2010} found that migrants are more likely to leave younger countries towards older countries in the context of international migration. Using the 2010 Census, the potential support ratio (PSR) equals to the ratio of population aging 15-64 over population aging 65+, which is the inverse of dependency ratio in demography literature; the higher PSR, the younger the population.

Racial composition could influence the mobility of population as well, as extant literature found different patterns of internal migration between racial groups (\citealt{crowder_neighborhood_2012, sharkey_geographic_2015}). Hence, besides the dissimilarity of racial composition between counties, we also consider the racial composition of the sending county to account for the varying mobility of different groups, as measured by the proportion of each racial category in the population.  

\vspace{0.1in}
\textit{Economic covariates.} Economic structures of origins and destinations could potentially influence their migration flows. Since renters on average are more mobile than house owners (\citealt{frey_great_2009,molloy_internal_2011}) even after controlling for demographic and socioeconomic factors (\citealt{jia_economics_2022}), the model includes the percentages of housing units occupied by renters for both origin and destination, using 2010 Census data. The model also controls the percentage of population with a college degree using the 2006-2010 ACS. This is because human capital may offer greater ability and opportunities for migration, and previous analysis found that population with higher education attainments have higher migration rates in the U.S. (\citealt{frey_great_2009}).

Neoclassical economic theory predicts that people migrate towards economic opportunities (\citealt{massey_theories_1993,todaro_internal_1976}). The theory also predicts that regions with more economic opportunities will send more migrants, since their population have more capital to finance their migration (\citealt{massey_whats_1997}). We thus include the unemployment rate of the origin, and the difference in the unemployment rate between the destination and the origin. In combination of neoclassical economic theory and the aspiration-ability model (\citealt{carling_migration_2002}), we hypothesize that more migration will come from counties with lower unemployment rate given their greater ability to move, and more migration will happen when the destination has lower unemployment rate than the origin, offering more economic opportunities and higher aspiration for migration. Similarly, the models incorporate the logarithm of median monthly housing costs of the origin and the difference in log housing costs between destination and origin. %

\vspace{0.1in}
\textit{Geographical covariates.} Besides distance between counties, the model also controls for regional differences in mobility. Previous research found that migration rates and their trends in different parts of America vary significantly (\citealt{frey_great_2009}). We believe that the regional difference may not be fully explained by difference in social contexts indicated by the covariates above. Dummy variables are created to indicate whether the origin and destination is in the West, the Midwest, the South, with the Northeast as the reference group, based on the definition of \cite{us_census_bureau_census_2013}.

Administrative boundaries are likely to influence migration flows as well. \cite{charyyev_complex_2019} found that, marginally speaking, the majority of inter-county migration in the U.S. happens within a state, and in this paper we further examine whether state boundary influences migration flows after controlling for distance and dissimilarity between counties. Intrastate intercounty migration could be more prominent than cross-state migration because compared to intrastate migration, the cross-state migration creates extra burdens ranging from adaptation to unfamiliar legal and cultural environments, to navigation of administrative procedures such as change in occupational licensing for workers in certain occupations (\citealt{johnson_is_2020}). Yet, the opposite hypothesis is plausible under the consumer-voter model, which contends that people vote by their feet (\citealt{dye_american_1990,tiebout_pure_1956}); as means of pursuing favorable policies, cross-state migration is more effective if people migrate to seek lower tax rates or more welcoming policies and climates for immigrants (\citealt{preuhs_state_1999,schildkraut_tale_2019}). The model creates a dummy variable indicating whether the two counties are affiliated with the same state. A positive coefficient would suggests that intrastate intercounty migration is more prominent, and a negative coefficient suggests that inter-state migration is more prominent.

\vspace{0.1in}
\textit{Variable Setup.}  We report two models in the following results section. The first model contains every covariate except the rural dissimilarity score, which is later included in the second model, the full model. Since the level of urbanization is strongly associated with their political environment, comparison between the two models could reveal how much of the total effect of political dissimilarity might be explained by their difference in the level of urbanization. Besides the sum term serving as an intercept, we add to models a term that counts the number of nonzero dyads of the network to account for the zero-inflation of migration flow data (\citealt{krivitsky_modeling_2013}). Its negative coefficients in Table~\ref{tab1} indicate the sparsity of migration flow network, that a county pair is more likely to have no migrants moving between than otherwise, even after controlling for all the covariates in the model. Summaries of descriptive statistics and data sources are attached in Part A of the supplement.

\section*{RESULTS}

\begin{figure}[t]
\centering
\includegraphics[width=\textwidth]{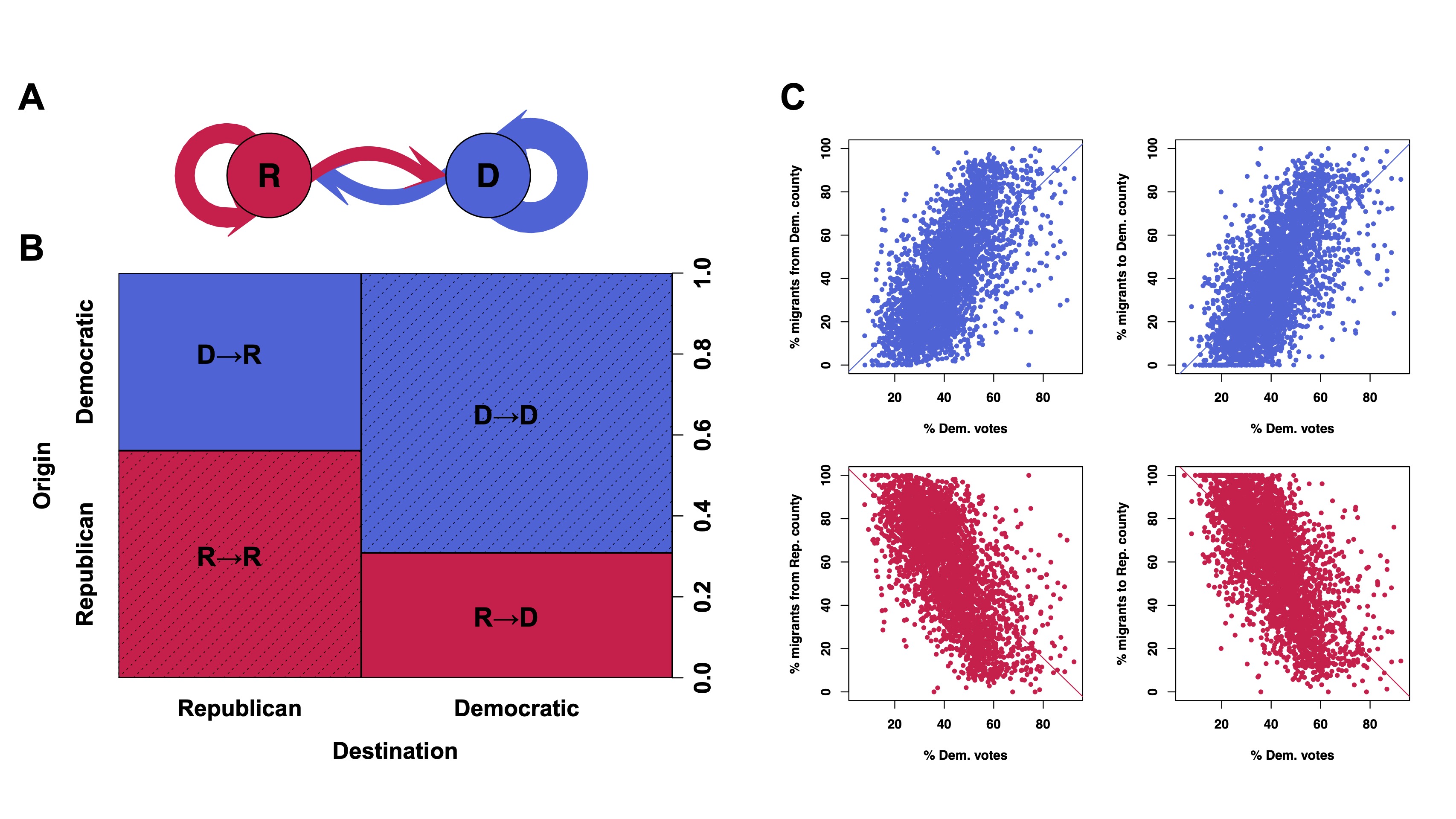}
\caption{Immobility from political division \\ \emph{Note:} The sociogram (A) represents the magnitude of migration flow within and between Democratic counties (node D in blue) and Republican counties (node R in red), which is proportional to the width of the edge. The spineplot (B) represents the magnitude of migration flow within and between the two groups by the area of each block. The shaded blocks represent migration within each group. Scatterplots (C) show the relationship between percentage of Democratic votes in 2008 of a county and the composition of its in-migrants and out-migrants. The lines are fitted bivariate linear regression lines.}
\label{fig:poli}
\end{figure}

\subsection*{Bivariate Analyses of Migration and Political Division}
To explore the pattern of segmented immobility by political orientation, we first perform bivariate analyses between intercounty migration and political division, as visualized in Figure 4. We divide counties into two broad groups, Democratic counties and Republican counties. Democratic counties are counties where the Democratic candidate (Obama) received more votes than the Republican candidate (McCain) in the 2008 presidential election, and vice versa for the Republican counties. The sociogram in Panel A of Figure 4 visualizes the magnitude of migration within and between Democratic and Republican counties, which is proportional to the width of edges. The sociogram shows that migration flows within each group has thicker edges than flows between, suggesting that more migration happens from one Democratic county to another, or from one Republican county to another, than between a Democratic county and a Republican county. The spineplot in Panel B represents the magnitude of migration flow within and between groups by the area of each block. The shaded blocks are migration happening within Democratic or Republican county groups, suggesting again that more migration happens on either side of the party line than across it. The color of each block indicates whether the origin of the migration flow is from a Democratic (blue) county or a Republican (red) county. The spineplot indicates that only 31\% of the migrants moving into a Democratic county come from a Republican county, and just 44\% of the migrants moving into a Republican county come from a Democratic county.

Panel C of Figure 4 visualizes the relationship between the percentage of the Democratic votes in the 2008 election and the composition of the in-migrants and out-migrants for each county. The upper left panel shows that the higher the Democratic vote in 2008, the larger the proportion of migrants coming from a Democratic county, and the smaller the proportion of migrants coming from a Republican county, as shown in the lower-left panel. Similarly, the right-hand column suggests that a larger share of 2008 Democratic votes within a county is associated with a larger proportion of out-migrants moving to a Democratic county, and a smaller proportion to a Republican county. Overall, the figures reveal a clear and strong pattern of political sorting, where less population migrate between counties with distinct political environments than those with similar political environments.

\subsection*{Segmented Immobility}
The bivariate analysis is suggestive that intercounty migration is immobilized by political divisions in the United States. We further examine this using VTERGMs that incorporate the demographic, economic, geographical and political factors at the county and inter-county levels, together with explicit specifications of internal dynamics of migation systems.  Table 1 displays the results. Model 1 suggests that, holding all other factors constant, a larger difference in political environments between counties predicts less migration between them. Since the political environment is associated with the level of urbanization of a county (\citealt{cramer_politics_2016}), Model 2 further includes the dissimilarity of urbanization between counties. From Model 1 to Model 2, the effect size of political dissimilarity becomes modestly smaller, suggesting that the effect of political difference can be partly (but not completely) explained by their difference in the level of urbanization. The smaller BIC of Model 2 further indicates that difference in the level of urbanization is effectively explaining the variation in the magnitude of migration flows. Nonetheless, in Model 2, larger political dissimilarity is still a statistically significant predictor of less migration between counties, offering empirical evidence for Hypothesis 1.1. Holding other factors constant, a pair of counties with 10\% larger difference in 2008 voting outcome is expected to have 2.5\% (i.e., $[1-\exp(-0.256 \times 10\%)]$) fewer migrants than another county pair. Similar to political segmentation, Model 2 also reveals that larger differences in levels of urbanization and racial compositions of two counties predict fewer migrants moving between, holding other factors constant, lending support for Hypotheses 1.2 and 1.3. The VTERGM results do suggest that migration is inhibited between places with dissimilar political contexts, levels of urbanization, and racial compositions.

\singlespacing

\begin{longtable}[t]{lr@{\hskip -0.8pt}rrr@{\hskip -0.8pt}rr}
\caption{Valued TERGMs for Inter-county Migration Flows, 2011-2015} 
\label{tab1}
\\ \hline
& \multicolumn{3}{c}{Model 1} & \multicolumn{3}{c}{Model 2} \\
\cline{2-7}
& \multicolumn{2}{c}{Estimate}  & \multicolumn{1}{c}{SE} & \multicolumn{2}{c}{Estimate} & \multicolumn{1}{c}{SE} \\ 
\hline
\endfirsthead
\caption{\textit{(continued)} Valued TERGMs for Inter-county Migration Flows, 2011-2015}
  \\ \hline
& \multicolumn{3}{c}{Model 1} & \multicolumn{3}{c}{Model 2} \\
\cline{2-7}
& \multicolumn{2}{c}{Estimate}  & \multicolumn{1}{c}{SE} & \multicolumn{2}{c}{Estimate} & \multicolumn{1}{c}{SE} \\ 
\hline
\endhead
  \hline
  \textit{Segmented Immobility} & & & & & & \\ 
Political dissimilarity & -.368 & *** & .007 & -.256 & *** & .007 \\ 
  Rural dissimilarity &  &  &  & -.399 & *** & .004 \\ 
  Racial  dissimilarity & -.361 & *** & .006 & -.217 & *** & .006 \\ 
  \textit{Network Patterns} & & & & & & \\
  Mutuality & .054 & *** & .002 & .045 & *** & .002 \\ 
  Log(past migrant flow) & .303 & *** & $<$.001 & .300 & *** & $<$.001 \\
  Waypoint flow & -.014 & *** & .001 & -.015 & *** & .001 \\ 
  Destin.log(immigrant inflow) & .062 & *** & .001 & .056 & *** & .001 \\ 
  Origin.log(immigrant inflow) & .040 & *** & .001 & .035 & *** & .001 \\
  \textit{Demographics} & & & & & & \\
  Destin.log(population size) & .351 & *** & .002 & .351 & *** & .002 \\ 
  Origin.log(population size) & .370 & *** & .002 & .373 & *** & .002 \\ 
  Destin.log(population density) & -.077 & *** & .001 & -.083 & *** & .001 \\ 
  Origin.log(population density) & -.062 & *** & .001 & -.069 & *** & .001 \\ 
  Destin.PSR & .018 & *** & .001 & .017 & *** & .001 \\ 
  Origin.PSR & .013 & *** & .001 & .013 & *** & .001 \\ 
  Origin.P(White) & \multicolumn{6}{c}{(reference group)} \\
  Origin.P(Hispanic) & -.012 &  & .007 & -.064 & *** & .007 \\ 
  Origin.P(Black) & .147 & *** & .008 & .117 & *** & .008 \\ 
  Origin.P(Asian) & .408 & *** & .020 & .467 & *** & .020 \\ 
  Origin.P(other race) & 1.031 & *** & .015 & .993 & *** & .015 \\ 
  \textit{Economics} & & & & & & \\
  Destin.P(renter) & .405 & *** & .011 & .348 & *** & .011 \\ 
  Origin.P(renter) & .507 & *** & .012 & .476 & *** & .012 \\ 
  Destin.P(higher education) & .327 & *** & .011 & .359 & *** & .011 \\ 
  Origin.P(higher education) & .157 & *** & .012 & .153 & *** & .012 \\
  Difference.log(housing costs) & -.135 & *** & .004 & -.153 & *** & .004 \\ 
  Origin.log(housing costs) & -.248 & *** & .005 & -.277 & *** & .005 \\
  Difference.P(unemployment) & -1.305 & *** & .040 & -1.300 & *** & .040 \\ 
  Origin.P(unemployment) & -3.039 & *** & .052 & -3.012 & *** & .052 \\  
  \textit{Geographics} & & & & & & \\
  Log(distance) & -.563 & *** & .001 & -.568 & *** & .001 \\ 
  Same state & .501 & *** & .002 & .510 & *** & .002 \\ 
  \hl{Northeast} & \multicolumn{6}{c}{\hl{(reference group)}} \\
  Destin.South & .258 & *** & .003 & .253 & *** & .003 \\ 
  Origin.South & .047 & *** & .003 & .046 & *** & .003 \\ 
  Destin.West & .384 & *** & .004 & .374 & *** & .004 \\ 
  Origin.West & .193 & *** & .004 & .184 & *** & .004 \\ 
  Destin.Midwest & .203 & *** & .003 & .197 & *** & .003 \\ 
  Origin.Midwest & .085 & *** & .003 & .080 & *** & .003 \\ 
  \textit{Baseline} & & & & & & \\
  Sum & -1.609 & *** & .040 & -1.193 & *** & .040 \\ 
  Nonzero & -13.966 & *** & .028 & -13.917 & *** & .028 \\ 
   \emph{BIC} & \multicolumn{3}{c}{2,221,363} & \multicolumn{3}{c}{2,210,125} \\
   \hline
   \multicolumn{7}{l}{\emph{Note:} *$p<0.05$; **$p<0.01$; ***$p<0.001$ (two-tailed tests).}
\end{longtable}

\begin{table}[h]
\renewcommand{\arraystretch}{1.25} %
    \centering
    \caption{Migrant Population Sizes under Observed and Knockout Scenarios}
    \begin{tabular}{lrr}
    \hline
         &  Total Migrants & Increment in Count and Rate   \\ \hline
      Observed & 17,176,675 & \\
      Remove political segregation~~~~ & 17,965,336  & 788,661 ~~~~~ 4.6\% \\
      Remove all segmentation~~~~ & 21,741,021 & 4,564,346 ~~~ 26.6\% \\ \hline
    \end{tabular}
    \label{tab2}
\end{table}

\doublespacing

\hl{
To quantify the contribution of segmented effects to immobility, we perform knockout experiments to compute the total migrant population under counterfactual scenarios where these effects are inoperative, and compare that with the observed scenario. Table~\ref{tab2} shows that when the political segregation effects on migration flows were knocked out, the expected intercounty migrant population each year would increase by 789 thousand, 4.6\% higher than the observed. At the absence of all three segmentation patterns, we would expect to observe 26.6\% more internal migrants in the United States, that is 4.56 million more people moving from one county to another each year.\footnote{\hl{We note that this conclusion depends on the assumption that the context dissimilarity influences people's decision of whether to migrate or not, and not merely influencing their choice of destination. We would thus not expect this model to accurately predict involuntary migration in response to events like political turmoil or natural disasters, which dominate people's decision of migrating or not under those circumstances. However, these seem unlikely to have been significant drivers of internal migration in the U.S. during the study period. We thank the anonymous reviewer for pointing out this assumption.}}

Results of the VTERGMs and knockout experiments together suggest that segmented immobility serves as a critical and substantial social mechanism behind the immobility of the contemporary American society. These social mechanisms may be partly driven by economic forces (although supplementary analysis shows that dual labor and housing markets make little impact on the described segmentation pattern, see Part C of the supplement); %
it may also reflect people's preference for residing in an environment that is culturally and politically familiar to them. This tendency not only implies social cleavages along party lines, between urban and rural lands, and across communities with varying racial demographics; it could also contribute to a growing geographical segmentation along those lines. As has been known since the classic works of Sakoda and Schelling, even a small preference for homophily can lead to substantial segregation in residential settlement patterns (\citealt{fossett_ethnic_2006,sakoda_checkerboard_1971, schelling_models_1969}).
}

\subsection*{Network Dynamics Influencing Migration Flows}
The VTERGMs also consider the network patterns of the migration flow system. That all coefficients are significant in the \emph{Network Patterns} section in Model 2 of Table~\ref{tab1} confirms that they play a significant role in determining the directions and magnitudes of intercounty migration flow. 
In Model 2, the positively significant mutuality term confirms Hypothesis 2.1, that reciprocity is present in the migration-flow networks: a larger flow from county A to B is positively associated with a larger flow from county B to A, holding other effects constant. Joining research on global migration and intercounty migration in U.K. (\citealt{ravenstein_laws_1885,windzio_network_2018}), we show that reciprocity is also a network pattern found within U.S. migration. It is interesting to note that some prior studies not observe reciprocity effects in their analyses (\citealt{desmarais_statistical_2012,windzio_network_2019}); this might come from omission of some regional characteristics that influence the attractiveness of regions to migrants, or their operation of data transformation for the migrant count variable. Future research may replicate the analysis of reciprocity using count-data network models under various social contexts to understand whether reciprocity is a prevalent phenomenon, or can be suppressed by some social forces.

Model 2 also reveals that a larger migration flow during 2006-2010 is significantly associated with a larger migration flow during 2011-2015, even after holding all exogenous and endogenous factors constant. This confirms Hypothesis 2.2 regarding the perpetuation of the migration flow system, showing that migration-facilitating mechanisms offer the system its own momentum, promoting future migration net of exogenous factors such as demographic structures of a region (\citealt{de_haas_internal_2010}). 

\hl{The significantly negative coefficient of the flow term indicates a lack of waypoint structures of inter-county migration, refuting Hypothesis 3. The negative waypoint flow effect implies that relatively little migration is proceeding in the chain-like manner such as stepwise and relay migration. After holding other factors constant, counties generally have an imbalance or inequality in the scales of their migration inflows and outflows, either sending many migrants but receiving few, or receiving many migrants but sending few. This may represent emergent attractiveness effects, in which in-migration makes a county seem more attractive to other possible migrants, and out-migration makes a county seem correspondingly less attractive. It may also reflects unobserved heterogeneity in attractiveness arising from other factors; the specification of waypoint flows in the model thus controls for this possible source of autocorrelation, beyond its substantive interest.
}

Note that the inequality identified by a lack of waypoint flows in this inter-county migration network is different from the inequality captured by an abundance of transitive hierarchy in other cross-national migration networks (e.g., \citealt{leal_network_2021}). The transitive hierarchy requires many waypoints serving as the ``mildly structurally attractive position,'' between the highly and the minimally ``structurally attractive positions'' (\citealt{leal_network_2021}: 1086). In analogy, that implies a multi-layer hierarchy of the global system with countries positioned in the core, the semi-periphery and the periphery (\citealt{wallerstein_modern_2011}). On the contrary, in this network with a lack of waypoint flows, there is an \emph{absence} of semi-periphery areas serving as waypoints between the core and the periphery; in comparison with the international migration system, the U.S. migration system is relatively bipolar, with counties tending to be, \textit{ceteris paribus}, either structurally attractive or unattractive, with few in the middle ground.

The model also examines the relationship between internal and international migration. It shows that larger immigrant inflows from 2011 to 2015 are positively associated with larger inter-county inflows and outflows in the same period. This finding does not correspond to either side in the debate about internal migratory response to immigration, which contends that large immigrant inflows are either associated with small internal migrant inflow and large outflows, or not associated with  internal migrant flows. Rather, the results suggest that counties with large immigrant inflows are active in \emph{both} sending and receiving intercounty migrants. Further, the larger coefficient of destination effect than the origin effect suggests that, increasing immigrant inflows to a county is associated with larger increase of internal inflow than internal outflow. In other words, immigration is actually associated with net population increase from internal migration. Overall, the finding shows a common mobility pattern for internal and international migration, wherein counties popular among international immigrants are also popular in receiving and active in sending internal migrants.\footnote{Since this is an aggregate-level analysis of population flows, the finding does not distinguish the characteristics of internal migrants, such as their race and ethnicity or socioeconomic status. Hence, we do not directly engage with more fine-grained debates about whether immigration deters in-migration and promotes out-migration of certain population categories as predicted by some literature (\citealt{frey_immigration_1995}), which requires more detailed data.}

\subsection*{Demographic, Economic, and Geographic Determinants of Migration}

Alongside segmented immobility and network patterns, the models also consider other factors that could influence intercounty migration. For demographic characteristics, Model 2 confirms findings from spatial econometrics (gravity) models that population sizes in both sending and receiving regions are positively associated with migrant flow (\citealt{boyle_exploring_2014,zipf_p1_1946,zipf_human_1949}). A 10\% increase in destination's population size is associated with a 3.4\% (i.e., $[1.1^{0.351}-1]$) increase in the number of migrants, and a 10\% increase in origin's population size is associated with a 3.6\% (i.e., $[1.1^{0.373}-1]$) increase in the number of migrants, holding other factors constant. Population density has a significantly negative effect for both the number of in-migrants and out-migrants, holding population size and other factors constant. One possible mechanism is that higher population density leads to larger shares of local connections for their residents (\citealt{butts_geographical_2012,hipp_extrapolative_2013,thomas_geographical_2022}), where more job transitions and housing transactions can happen locally thanks to these connections, reducing migration across county borders. %

With respect to demographic composition, larger migration flows are significantly more likely to be observed between counties with younger populations, in line with the migration schedule literature finding that younger adults are more mobile than older adults (\citealt{raymer_applying_2007,rogers_model_1981}). The model also shows that counties with larger shares of Hispanic population tend to send fewer migrants, but counties with larger shares of Non-Hispanic Black, Non-Hispanic Asian and Other races populations tend to send more intercounty migrants. Note that these effects do not directly describe the mobility of each racial/ethnic population, since they are predicting the magnitude of migration flow for all racial and ethnic populations. Decomposing migration flows into migrants of each racial/ethnic population is necessary to further reveal the variation of mobility between people with different racial/ethnic identities. 

Economic covariates in Model 2 show that larger migration flows exist between counties with higher shares of renters and people with college degrees, consistent with previous literature observing that renters and people with higher education credentials are more mobile than their counterparts (\citealt{frey_great_2009}). We also see that larger migration flows happen when the route offers greater declines in housing costs, indicating a tendency of mobility towards cheaper housing (\citealt{plantinga_housing_2013}). Holding other factors constant, counties with lower housing costs have higher out-migration. This might be due to the better financial conditions renters have in low housing cost areas, enabling them to move and relocate. It is also compatible with previous findings that lower housing equity is associated with higher mobility rates (\citealt{coulson_mobility_2013}). For unemployment rates, the model suggests that the lower the unemployment rate at the origin, and the larger the decline in unemployment rate from origin to destination, the more intercounty migration. These results are compatible with the cost-benefit model of the neoclassical economic theory of migration that population move towards economic opportunities (\citealt{todaro_internal_1976}), and that more economic opportunities financing migration makes migration more likely to happen (\citealt{massey_whats_1997}). The relational approach employed here enables empirical analysis of the aspiration-ability model (\citealt{carling_migration_2002,carling_revisiting_2018}), revealing that both the aspiration, as influenced by the relative economic conditions of origin and destination, and the ability, as influenced by the economic conditions of the origin, matter to migration behaviors.

In terms of geographical factors, the model suggests a negative association between distance and number of migrants flowing between two counties, as the gravity model predicts (Zipf 1946, 1949). A $10\%$  increase in distance between two counties is associated with a $5.3\%$ (i.e., $[1-1.1^{-0.568}]$) decrease in intercounty migration.  Administrative boundaries also influence migration flows; migration flows within the same state are expected to be larger than those across states, holding other factors constant. Additionally, different U.S. regions have varying mobilities. The model indicates that compared to the Northeast, every other region receives and sends more intercounty migrants, \emph{ceteris paribus}. This suggests the existence of some latent characteristics inhibiting the mobility of the Northeast, which deserves more examination in future work.

Lastly, to check the model adequacy, we simulate networks based on Model 2 (the full model) in Table 1 using MCMC algorithms. We then calculate the total in-migrant and out-migrant count for each county, and compare the observed distribution with the simulated distribution. We find that the fitted model recapitulates the county-level migration data (see Part D of the supplement). We also calculate the Pearson's correlation between observed and simulated distributions, which are all above 0.95. We conclude that the model effectively reproduces the quantitative features of observed migration flow networks.

\section*{DISCUSSION AND CONCLUSION}
This paper offers a comprehensive analysis of the inter-county migration structure encompassing not only economic, demographic and geographical factors, but also political, cultural factors and internal dynamics of the migration system. Network models reveal a pattern of segmented immobility in America, in which less migration happens between counties with dissimilar political environments, levels of urbanization, and ethnic/racial compositions. Yet, we do not observe segmentation between internal migrants and international immigrants; rather, the model shows that counties active in receiving many international immigrants are active in both sending and receiving many internal migrants as well. Our analysis also suggests the significance of internal dynamics of the migration flow system; we observe strong patterns of reciprocity and perpetuation, along with a suppression of waypoint structure. 
These results lend empirical evidence to the systemic theory of migration (\citealt{bakewell_relaunching_2014,de_haas_internal_2010,mabogunje_systems_1970,fawcett_networks_1989}), showing that the population flows assemble an interdependent network system that carries its own momentum.

This paper identifies segmentation as a critical mechanism behind population immobility in the contemporary American society, which could potentially have deterred millions of people from migrating each year, as suggested by the knockout experiments. This finding implies people's tendency of choosing residency in localities that match with their political affiliations and sociocultural attributes, potentially leading to geographical segmentation between people with different political identities (\citealt{brown_measurement_2021}) and increasing the homogeneity of their social relations (\citealt{diprete_segregation_2011}).  Such sorting could possibly reinforce political polarization (\citealt{dellaposta_center_2015}), and can also serve as a mechanism that maintains and even exacerbates residential segregation along other dimensions (\citealt{fossett_ethnic_2006,sakoda_checkerboard_1971,schelling_models_1969}). While classic analyses of segregation have focused on local communities within urban areas (\citealt{bishop_big_2009}), the effects seen here could potentially contribute to macro-level segmentation across the whole country (\citealt{liu_migration_2019}). From a migration perspective, although internal migration in the U.S. does not involve border-crossing in international migration or other forms of governmental restrictions (such as the household registration system in China, \textit{hukou}), population movement is never free of constraints. Rather, as our analysis shows, Americans today are separated by the invisible borders and walls standing along the party lines, at the midway between rural and urban landscapes, and over the gap across communities with varying racial demographics.

The analytical framework in this paper provides an example of structural and systemic analysis of mobility and immobility, broadly defined. The relational approach connects the perspectives of emigration and immigration to examine how characteristics of origin and destination \textit{jointly} influence migration, which enables revelation of the segmented immobility in the U.S. migration system. The formal specification of the interdependence between migration flows under the ERGM framework identifies the structural signature of networks, reflecting the internal dynamics of migration systems. The knockout experiment offers model-based insights into how the system might react to social change. Lastly, leveraging advances in scalable VERGM estimation and simulation allows quantitative analysis of the magnitude of population flows and their determinants in large social systems. The applicability of this framework extends beyond the population movement between geographical areas, encompassing mobility in the occupational system for the study of social stratification and mobility (\citealt{cheng_flows_2020}), the exchange of personnel between organizations (\citealt{sparrowe_process_1997}), and the migration of scholars between institutions and research domains in the sociology of knowledge (\citealt{burris_academic_2004,gondal_duality_2018,mcmahan_creative_2021}). 

\hl{
While our study enables a much richer examination of the mechanisms driving or inhibiting internal migration at a larger scale than what has been possible in extant literature, it is not without its own limitations.  First, as a macrosociological study about the ``functioning of a social system'' (\citealt{coleman_social_1986}:~1312), this paper informs an aggregate-level social phenomenon, i.e.,~population immobility. While analysis of the migration flow network facilitates a systemic understanding of migration and its relation to segmentation from a holistic viewpoint, it does not directly describe the patterns of individual migration behavior. Although we can test for the structural signatures of such micro-level processes, unpacking those fine details requires information on decision making and behavior patterns at the individual level. For example, distinguishing stepwise migration and relay migration requires data about the migration trajectories of individual migrants. Studies like this are hence complementary to micro-level analyses (both quantitative and qualitative) that could shed further light on processes at the individual and household levels (e.g., \citealt{deluca_why_2019,fitchen_residential_1994,lichter_intercounty_2022,quillian_comparison_2015}). Research that aims to bridge individual behaviors and aggregate social outcomes are deemed to be fruitful, which is still an open problem in sociology, but a promising program to pursue (\citealt{cetina_advances_2014,coleman_social_1986}).
}

Second, since the American Community Survey did not start collecting data until 2005, our analysis only includes migration-flow networks for two time points (2006-2010, 2011-2015). This data limitation prevents us from conducting dynamic analysis about changes in intercounty migration patterns throughout the past decades, and therefore, our findings do not speak directly to the reasons behind the long-term decline of migration. Yet, our identification of drivers and especially inhibitors behind migration flows could serve as a starting point for this inquiry. For example, since political division across geographical areas deters migration, it may be worthwhile for future research to examine how the geography of politics and preference about political homophily have changed over time, and how the evolution of political landscapes and polarization relates to the long-term decline of migration. Studies of the changing patterns of immigrant inflows and the relationship between internal and international migration flows can illuminate the change of population dynamics over time. Integration of knockout experiments via network simulation and historical data about political climate and migration/immigration flows might be one approach to advance the inquiry into the social forces behind the growing immobility in the United States. In addition, future research might also benefit from exploring the changing balance of forces of the competing internal dynamics of the migration system over the past decades. Given that the VTERGM framework we employ here is capable of handling networks with multiple time steps, our analytical framework could be employed for dynamic analysis once migration-flow data for more time points becomes available.

In like vein, the time period we analyzed covers the Great Recession (\citealt{grusky_great_2011}). Despite our controls of various economic factors, it is possible that some aspects of our findings may be particular to this period, as economic shocks can influence migration patterns (\citealt{monras_economic_2018}; cf. \citealt{molloy_internal_2011}). Specifically, since economic recession can suppress migration, it is possible that fewer waypoint flows are consequence of the period effect that temporarily suppresses stepwise migration. Nevertheless, the formal expressions of relational linkages and network patterns, and the modeling of migration-flow networks using ERGMs are generally applicable to study migration flows of different periods and regions at different scales. Future research may consider replicate and compare analysis of relational and network patterns of migration flows in different time and space using similar frameworks; they will reveal what patterns are context-specific in certain spatial-temporal settings, and which are generalizable to migration in other societies.

Furthermore, another fruitful direction for future work is to complicate the analysis of internal dynamics of migration system by examining higher-order dependence structure of (valued) networks. One example is network transitivity, a structural feature associated with hierarchy within the migration system (\citealt{leal_network_2021}). We do not observe a strong transitive hierarchical system in the U.S. internal migration system, as indicated by the lack of waypoint flows, \textit{ceteris paribus}.\footnote{As discussed in Hypotheses, both waypoint flow and transitivity are triadic features that concern edge structure in an ($i,j,k$) triple; waypoint flow captures the``backbone'' of flow within the triple ($i\rightarrow j\rightarrow k$), while transitive triads involve the co-presence of waypoint flow and a direct $i\rightarrow k$ flow. The negative effect for waypoint flow in our models means that triples with strong $i\rightarrow j\rightarrow k$ paths are suppressed, which also necessarily suppresses transitive triples net of other effects in the model. Interestingly, while the waypoint flow (and its binary-network version, two-paths) is a more basic lower-level dependence structure, which carries motivations from social behavior patterns such as those detailed in this paper, it receives relatively less examination in the network literature. We hope this paper helps draw more attention to waypoint flow and other triadic network structures of potential substantive importance for flow networks.} Nevertheless, transitivity is in general a theoretically-interesting dependence structure for study of mobility networks, and should ideally be examined in valued networks so to consider the quantitative feature of migration flows. This requires theoretical and methodological developments in formal specification of dependence terms in the valued network setting, e.g., clarifying the properties of different definitions of transitivity and their relationship to network degeneracy (\citealt{krivitsky_exponential-family_2012}). It also demands further advancements in computational methods for valued network models to allow for evaluation of more complicated dependence structures in large networks.

\hl{Last but not least, as population immobility has become a long-term phenomenon in the U.S., it poses important questions about its broader social implications. Future research could explore the relationship between geographical mobility and social mobility, and how the divergent geographical mobility patterns across various social groups may influence their life chances and well being. A lack of population exchange, especially between localities with different cultural and political climates, could have ramifications on the social divisions of the country. Two decades ago, Putnam's (\citeyear{putnam_bowling_2000}) \textit{Bowling Alone} embarked the great debates about the ``collapse of American communities,'' marked by the detachment and disengagement of individuals from their communities. Observing the population segmentation and immobility, it raises the question whether we are witnessing the ``tribalization of American communities,'' where local communities diverge in their demographics, culture, and policy, with limited interaction, communication, and cooperation among people and organizations from dissimilar local communities.}

In conclusion, grappling with the mobility bias in migration studies, this paper utilizes migration systems theory and network methods to study the mechanisms behind population immobility in the United States. We identify segmentation as a significant feature of the American migration landscape, which has potentially immobilized millions of intercounty migration each year in the 2010s. The paper demonstrates how network and simulation methods can contribute to a systemic understanding of mobility and population dynamics. We also call for more theoretical and empirical research about the interrelationships between migration, segregation, and polarization, and how they shape the foundation of social lives in America and beyond.

\section*{SUPPLEMENT}
\subsection*{Part A. Descriptive Statistics}
\begin{table}[h]
\renewcommand{\thetable}{S\arabic{table}}  %
\setcounter{table}{0}  %
\centering
\caption{Descriptive Statistics of Inter-county Migration Flow Networks} 
\label{tab:a1}
\begin{tabular}{lrr}
  \hline
 & 2011-2015 & 2006-2010 \\ 
  \hline
  Vertices & 3,142 & 3,142 \\ 
  Edges & 274,197 & 241,526 \\ 
  Density & 0.028 & 0.024 \\ 
  Mean degree $^1$ & 175 & 154 \\ 
  Total migrants & 17,176,675 & 17,248,855 \\ 
  Mean migrants per county $^2$ & 10,934 & 10,980 \\ 
  Mean migrants per flow & 63 & 71 \\ 
   \hline
\end{tabular}
\end{table}
{\footnotesize\emph{Note:} 1. The reported degree is the total degree (Freeman degree), which equals the summation of in and out degree (Freeman 1978). For a closed network system, the mean in-degree equals to the mean out-degree. 2. Similarly, the mean migrant per county is the summation of mean in-migrants and mean out-migrants per county (and mean in-migrants equals to mean out-migrants). } \\

\singlespacing
\begin{table}[H]
\centering
\renewcommand{\thetable}{S\arabic{table}}  %
    \caption{Descriptive Statistics of Vertex/County Covariates}
    \label{tab:a2}
\begin{tabular}{lrrl}
  \hline
 & Mean & Str Dev & Source \\ 
  \hline
  \emph{Networks} &   &   &  \\ 
  Immigrant inflow & 593.9 & 2744.3 & ACS2011-2015 \\ 
  \emph{Politics} &   &   &  \\ 
  Democrat poll (\%) & 41.6 & 13.9  & Presidential Election Returns 2008 \\ 
  \emph{Demographics} &   &   &  \\ 
  Population size & 98,262.0 & 312,946.7 & Census2010 \\ 
  Population density (/km$^2$) & 100.0 & 665.8 & Census2010 \\
  Potential Support Ratio (PSR) & 4.4 & 1.5 & Census2010 \\ 
  White (\%) &  78.3 & 19.9 &  Census2010 \\
  Black (\%) & 8.7 & 14.4 &  Census2010\\ 
  Hispanic (\%) & 8.3 & 13.2 &  Census2010\\ 
  Asian (\%) & 1.1 & 2.5 & Census2010 \\ 
  Other race  (\%) & 3.5 & 8.3 & Census2010 \\ 
  \emph{Economics} &   &   &  \\ 
  Renter (\%) & 27.8 & 7.7 & Census2010 \\
  Higher education (\%) & 19.0 & 8.7 & ACS2006-2010 \\
  Unemployment (\%) & 7.5 & 3.4 & ACS2006-2010 \\
  Housing costs (\$/month) & 707.5 & 272.3 & ACS2006-2010 \\
  \emph{Geographics} &   &   &  \\ 
  Rural population (\%) & 58.7 & 31.5 & Census 2010 \\
  Northeast  (\%) & 6.8 &   &   U.S. Census Bureau \\
  South  (\%) & 45.3 &  & U.S. Census Bureau \\ 
  West  (\%) & 14.3 &  & U.S. Census Bureau \\ 
  Midwest  (\%) & 33.6 &  & U.S. Census Bureau \\ 
   \hline
\end{tabular}
\end{table}

\begin{table}[h]
\centering
\renewcommand{\thetable}{S3}  %
    \caption{Descriptive Statistics of Dyadic/County-Pair Covariates}
    \label{tab:a3}
\begin{tabular}{lrrl}
  \hline
 & Mean & Str Dev & Source \\ 
  \hline
  \emph{Dissimilarity} &   &   &  \\ 
  Political dissimilarity (\%) & 15.6 & 11.8 & Presidential Election Returns 2008 \\ 
  Rural dissimilarity (\%) & 36.1 & 26.1 & Census2010 \\ 
  Racial dissimilarity (\%) & 24.8 & 20.0 & Census2010 \\ 
  \emph{Geographics} &   &   &  \\ 
  Distance (km) & 1,439.5 & 961.3 & Census2010 \\
   \hline
\end{tabular}
\end{table}

\doublespacing
\hl{\subsection*{Part B. Model Setup and Computation}
\textit{Model Setup.}
The exponential-family random graph models, formally speaking, specify the probability of observing a specific network configuration, out of all possible configurations, given the vertex/node set and covariates (\citealt{hunter_ergm_2008}). In our context, this constitutes the probability of obtaining the observed migration flow structure among the 3,142 U.S. counties, versus the other structures that could have been observed. Formally, our model family is defined by
\begin{equation}
    \Pr(Y^t=y^t)=\frac{h(y^t)\exp\left(\theta^Tg\left(x,y^t,y^{t-k}_{t-1}\right)\right)}{\kappa(\theta)} \label{e_ergm}
\end{equation}
\noindent where $Y^t$ is the valued network structure at time $t$ (with observed value $y^t$),  $g(x,y^t,y^{t-k}_{t-1})$  is a vector of sufficient statistics, and  $\theta$  is a vector of coefficients. The model statistics are functions of (valued) graph structure, together with covariates ($x$) and past history (represented by the lagged variable $y^{t-k}_{t-1}=(y^{t-1},\ldots,y^{t-k})$).   The exogenous covariates are characteristics not determined by network configurations, including demographic, economic, political, and geographical attributes of individual counties or county pairs.  $g$ may also include statistics relating to endogenous structure,  such as reciprocity and the waypoint structure. The use of lagged network predictors allows us to incorporate predictors related to network dynamics (\citealt{hanneke_discrete_2010}) in analogy to a lagged multivariate regression model. 

In equation~\ref{e_ergm},  $\kappa(\theta)$  is the normalizing factor for the exponential family model, the summation of the quantity of the numerator for all possible network configurations:
$$
\kappa(\theta)=\sum _{y' \in \mathfrak{Y} } h(y')\exp(\theta^Tg(x,y',y^{t-k}_{t-1})     
$$
\noindent where $\mathfrak{Y}$ is the set of all possible network configurations.   The \emph{reference measure}, $h(y)$, is a function that determines the baseline distribution to which the model converges as $\theta \to 0$.  This is particularly important for valued ERGMs (\citealt{krivitsky_exponential-family_2012}), as different choices of $h$ can lead to very different distributions of edge values (just as different choices of reference measures for one-dimensional exponential families differentiate e.g. the binomial, geometric, and Poisson random variables in generalized linear models).  Here, we use the form
$$
h(y)=1/ \prod_{(i,j) \in \mathbb{Y}} y_{i,j}!
$$
\noindent where $\mathbb{Y}$ is the set of all dyads, and $y_{i,j}$  is the value of the edge from county $i$ to $j$.  This leads to a model where edge values are Poisson-like in the absence of dependence effects; just as a conventional ERGM can be viewed as a network logistic regression extended to incorporate dependence, this family can be viewed as a network Poisson regression with dependence among the edge variables.  (Note, however, that the addition of dependence terms renders the model non-Poissonian, and a model of this form can accommodate e.g., over-dispersion and/or zero-inflation by appropriate choice of terms, as is done here.)

\textit{Computation.}
Estimation for ERGMs can be computationally demanding, since the normalizing factor $\kappa$ required for likelihood computation is too expensive to compute directly.  This is especially the case for large and high-variance valued networks such as intercounty migration-flow networks, where the number of edge variables is extremely large (nearly ten million), and the edge value ranges from zero to tens of thousands. Researchers typically use Markov Chain Maximum Likelihood Estimation (MCMLE) for ERGM parameter estimation when modeling binary networks, which uses a Markov Chain Monte Carlo (MCMC) algorithm to produce importance samples of network structures that are subsequently used for approximate maximum likelihood inference (\citealt{geyer_constrained_1992,snijders_markov_2002}).  Unfortunately, current MCMLE implementations for valued networks (including both Geyer-Thompson-Hummel and stochastic approximation methods) do not scale well enough to be feasible for networks studied here, with such large network size and edge variance, as the space of potential random networks is too large for the MCMC sampler to converge in feasible time.  Instead, we employ an efficient Subsampled Regularized Maximum Pseudo-Likelihood Estimation (\citealt{huang_parameter_2024}). Maximum Pseudo-Likelihood Estimation (MPLE) is a classical computational method for ERGMs (\citealt{besag_spatial_1974,strauss_pseudolikelihood_1990,wasserman_logit_1996}), which continues to be employed as an efficient strategy for computationally intensive ERGMs (\citealt{an_fitting_2016,schmid_exponential_2017}). MPLE approximates the network likelihood function as the product of the conditional probability functions of each edge; this approximation becomes exact in the limit of weak edgewise dependence. This allows parallel computing and subsampling of edges to reduce computation time. While MPLE performance is not always equal to MCMLE performance for binary ERGMs, simulation studies find that for count-valued ERGMs with large edge variance, MPLE can match or even outperform MCMLE with reliable and nearly unbiased estimates, while its computational cost is substantially smaller (\citealt{huang_parameter_2024}). We further incorporate an L2 (ridge) regularizer to our likelihood function, making it Regularized MPLE, following the suggestion of using regularization to improve MPLE performance in the literature (\citealt{van_duijn_framework_2009}). 

Since calculating the likelihood function of valued TERGMs based on all 9.87 million dyads is computationally infeasible, we use the subsampling implementation of MPLE to approximate the likelihood function using one million dyads. The sampled dyads were selected using Tie-No-Tie (TNT) sampling, which puts equal weight on selecting zero-valued dyads and nonzero-valued dyads. TNT is an efficient sampling method for ERGMs when the network density is low, as in the present case where the majority of county pairs have no migrants. This results in including all the 274 thousand directed county pairs with at least one migrant, and randomly sampling the other 726 thousand directed county pairs without any migrant; this stratification is then corrected for using inverse probability weighting in the likelihood calculation.\footnote{The standard errors calculated from the sampling method are expected to be slightly larger than using the full network data (\citealt{huang_parameter_2024}), as the former uses fewer observations, making our inference more conservative. Since our main findings are statistically significant at the level of 0.001 under a conservative protocol, we expect the effects to be significant if the computational constraints had allowed us to use the full network for calculation.} L2 regularization (parameter=0.01) was incorporated in the likelihood function to avoid the convex hull problem and improve estimation performance (\citealt{van_duijn_framework_2009}).

\subsection*{Part C. Supplementary Analysis}
This section considers a potential confounding social process that could be correlated with our main findings.\footnote{We thank the anonymous reviewer for suggesting this possible social process.} Since people with and without college degrees may respond to different labor market segments under the dual labor market theory (\citealt{piore_2018_dual}), it is possible that college graduates are more likely to migrate to another place with concentrated college graduates, where there tends to be more job opportunities that they fit into; the same logic holds for people without college degrees. This means that larger migration can occur between counties with similar proportion of college graduates, because of the match in labor market supply and demand. Similarly, we may observe larger migration flow between counties with similar proportion of renters. The proportions of college graduates and renters of a county correlate with its political environment, urbanization level, and racial compositions, so it is desirable to examine how the hypothesized processes may be associated with our major findings. We add covariates of the absolute differences in college graduate share and renter share in Model 3 of Table~\ref{tab:a4}, displayed alongside with the model in the main text (Model 2). The results show that in Model 3, the coefficients representing segmented immobility patterns all remain statistically significant, with little change in its scale for the political covariate, a modest decrease for the urbanization covariate, and a modest increase for the racial composition covariate. The analysis shows that this social process does not confound our main findings.

\singlespacing
\captionsetup[table]{labelformat=alphatable} %

\begin{longtable}[t]{lr@{\hskip -0.8pt}rrr@{\hskip -0.8pt}rr}
\caption{Supplementary Valued TERGMs for Inter-county Migration Flows, 2011-5} 
\label{tab:a4}
\\ \hline
& \multicolumn{3}{c}{Model 2} & \multicolumn{3}{c}{Model 3} \\
\cline{2-7}
& \multicolumn{2}{c}{Estimate}  & \multicolumn{1}{c}{SE} & \multicolumn{2}{c}{Estimate} & \multicolumn{1}{c}{SE} \\ 
\hline
\endfirsthead
\caption{\textit{(continued)} Supplementary Valued TERGMs for Inter-county Migration Flows, 2011-5}
  \\ \hline
& \multicolumn{3}{c}{Model 1} & \multicolumn{3}{c}{Model 2} \\
\cline{2-7}
& \multicolumn{2}{c}{Estimate}  & \multicolumn{1}{c}{SE} & \multicolumn{2}{c}{Estimate} & \multicolumn{1}{c}{SE} \\ 
\hline
\endhead
  \hline
  \textit{Segmented Immobility} & & & & & & \\ 
Political dissimilarity & -.256 & *** & .007 & -.257 & *** & .008 \\ 
  Rural dissimilarity & -.399 & *** & .004 & -.298 & *** & .004 \\ 
  Racial  dissimilarity & -.217 & *** & .006 & -.246 & *** & .006 \\ 
  \textit{Network Patterns} & & & & & & \\ 
  Reciprocity & .045 & *** & .002 & .036 & *** & .002 \\ 
  Waypoint flow & -.015 & *** & .001 & -.016 & *** & .001 \\ 
  Log(past migrant flow) & .300 & *** & $<$.001 & .296 & *** & $<$.001 \\ 
  Destin.log(immigrant inflow) & .056 & *** & .001 & .056 & *** & .001 \\ 
  Origin.log(immigrant inflow) & .035 & *** & .001 & .034 & *** & .001 \\ 
  \textit{Demographics} & & & & & & \\ 
  Destin.log(population size) & .351 & *** & .002 & .355 & *** & .002 \\ 
  Origin.log(population size) & .373 & *** & .002 & .376 & *** & .002 \\ 
  Destin.log(population density) & -.083 & *** & .001 & -.086 & *** & .001 \\ 
  Origin.log(population density) & -.069 & *** & .001 & -.073 & *** & .001 \\ 
  Destin.PSR & .017 & *** & .001 & .020 & *** & .001 \\ 
  Origin.PSR & .013 & *** & .001 & .016 & *** & .001 \\  
  Origin.P(White) & \multicolumn{6}{c}{(reference group)} \\
  Origin.P(Hispanic) & -.064 & *** & .007 & -.029 & *** & .007 \\ 
  Origin.P(Black) & .117 & *** & .008 & .106 & *** & .008 \\ 
  Origin.P(Asian) & .467 & *** & .02 & .592 & *** & .02 \\ 
  Origin.P(other race) & .993 & *** & .015 & .972 & *** & .016 \\ 
  \textit{Economics} & & & & & & \\ 
  Destin.P(renter) & .348 & *** & .011 & .318 & *** & .012 \\ 
  Origin.P(renter) & .476 & *** & .012 & .409 & *** & .012 \\
  Dissimilarity.P(renter) &  & & & .309 & *** & .010 \\
  Destin.P(higher education) & .359 & *** & .011 & .624 & *** & .012 \\ 
  Origin.P(higher education) & .153 & *** & .012 & .440 & *** & .013 \\
  Dissimilarity.P(higher education) & & & & -1.051 & *** & .009 \\
  Difference.P(unemployment) & -1.3 & *** & .040 & -.969 & *** & .040 \\ 
  Origin.P(unemployment) & -3.012 & *** & .052 & -2.297 & *** & .052 \\ 
  \textit{Geographics} & & & & & & \\ 
  Log(distance) & -.153 & *** & .004 & -.171 & *** & .004 \\ 
  Same state & -.277 & *** & .005 & -.327 & *** & .005 \\ 
  Northeast & \multicolumn{6}{c}{(reference group)} \\
  Destin.South & -.568 & *** & .001 & -.572 & *** & .001 \\ 
  Origin.South & .510 & *** & .002 & .518 & *** & .002 \\ 
  Destin.West & .253 & *** & .003 & .255 & *** & .003 \\ 
  Origin.West & .046 & *** & .003 & .053 & *** & .003 \\ 
  Destin.Midwest & .374 & *** & .004 & .379 & *** & .004 \\ 
  Origin.Midwest & .184 & *** & .004 & .191 & *** & .004 \\ 
  \textit{Baseline} & & & & & & \\ 
  Sum & .197 & *** & .003 & .197 & *** & .003 \\ 
  Nonzero & .080 & *** & .003 & .086 & *** & .003 \\ 
   \emph{BIC} & \multicolumn{3}{c}{2,210,125} & \multicolumn{3}{c}{2,196,016} \\
   \hline
   \multicolumn{7}{l}{\emph{Note:} *$p<0.05$; **$p<0.01$; ***$p<0.001$ (two-tailed tests).}
\end{longtable}
 \doublespacing
\subsection*{Part D. Model-Adequacy Checks}
We develop a procedure to evaluate the model adequacy for VERGMs emulating that for binary ERGMs (\citealt{hunter_goodness_2008}). Based on Model 2 in Table 1 (the full model), we simulate 100 networks using MCMC algorithms using \texttt{ergm.count} package in \textsf{R}, and calculate the distributions of in- and out-volumes of edge values, i.e. the total in-migrants and out-migrants, for each node (county). We plot the distributions of in- and out-volumes by county against their observed statistics, and calculate the correlations between them. An adequate model specification is expected to reproduce the observed statistics with distribution centered on and close to observed values, and have high correlations between observed and simulated statistics.

Figure S1 and S2 visualize the simulated distributions against the observed statistics of in- and out-migrants, respectively. To make the comparison more viewable than stacking 3,142 counties in one plot, we sample four states from the West, the Midwest, the Northeast, and the South; the y-axis is re-scaled by taking logarithm considering their high skewness. Noticeably, the simulated interval of each county looks very narrow compared to the full range of in- and out-migrant count in each state, especially compared to the common goodness-of-fit plots for binary ERGMs. This indicates that the support of the valued ERGMs is tremendously huge, suggesting that reproducing the migrant distribution is a difficult task. The task is also more difficult because we compare simulated distribution with observed statistics for each county, rather than comparing network-level statistics for an unlabelled graph, which does not consider whether the simulated statistics match the observed at the node level. Nevertheless, the figures show that the simulated distribution is centered around and close to the observed statistics, and the Pearson's Correlations are all above 95\%. We conclude that the model effectively reproduces the migrant distribution across counties, and accounts for its variation.

}

\section*{Data Note}
Replication data and code can be found at https://doi.org/10.7910/DVN/I7HT9T.

\begin{figure}[h!]
\renewcommand{\thefigure}{S\arabic{figure}}  %
\setcounter{figure}{0}  %
\centering
\includegraphics[scale=0.13]{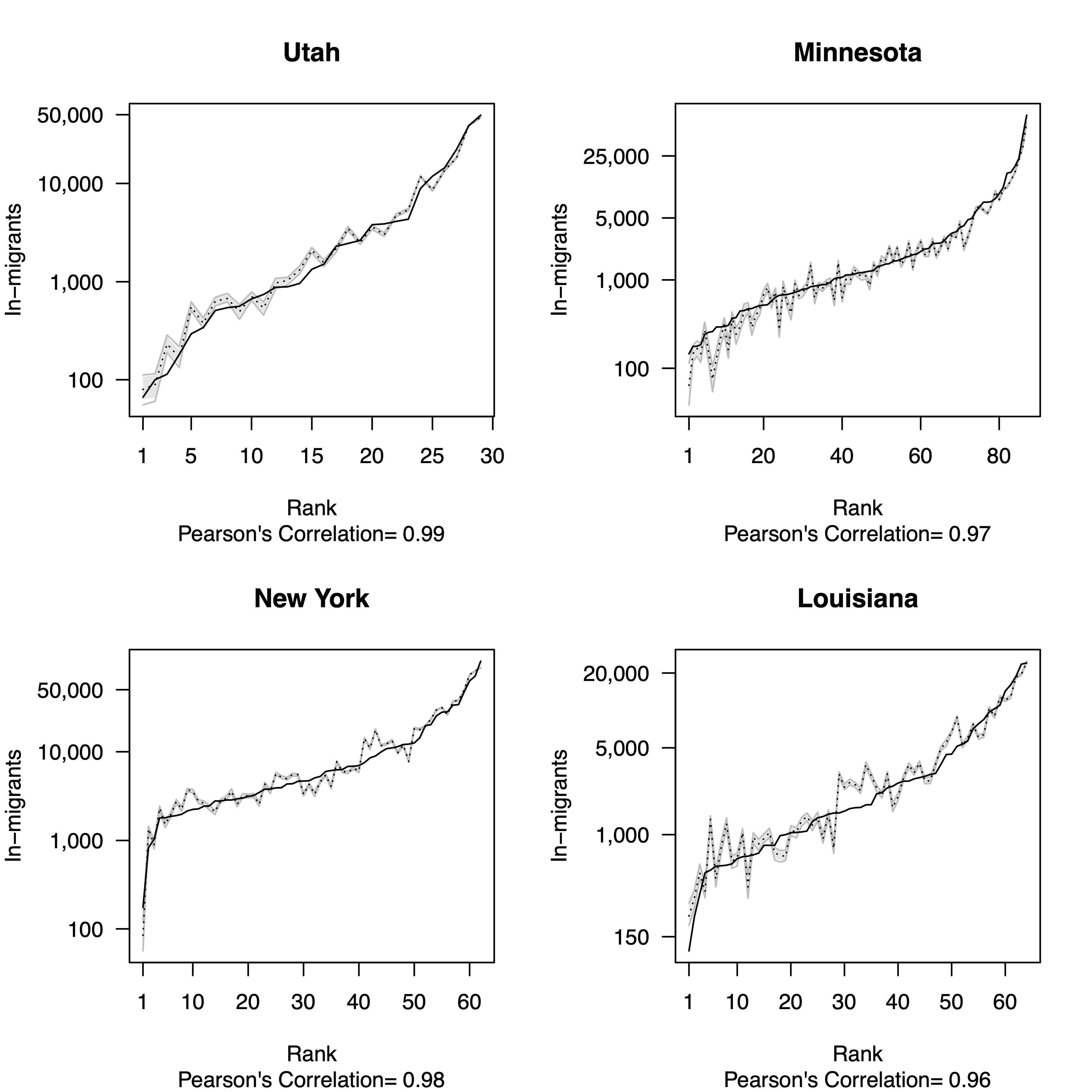}
\caption{Model-adequacy plots (involume) \\ \emph{Note:} The solid dark lines indicates the observed statistics, the dotted dark lines is the median of simulated distributions, grey lines are the min and max values of simulated distributions and the grey polygon is the 95$\%$  simulation interval of the distribution. }
\label{fig:adequacy1}
\end{figure}

\clearpage  %

\begin{figure}[H]
\renewcommand{\thefigure}{S\arabic{figure}}  %
\centering
\includegraphics[scale=0.13]{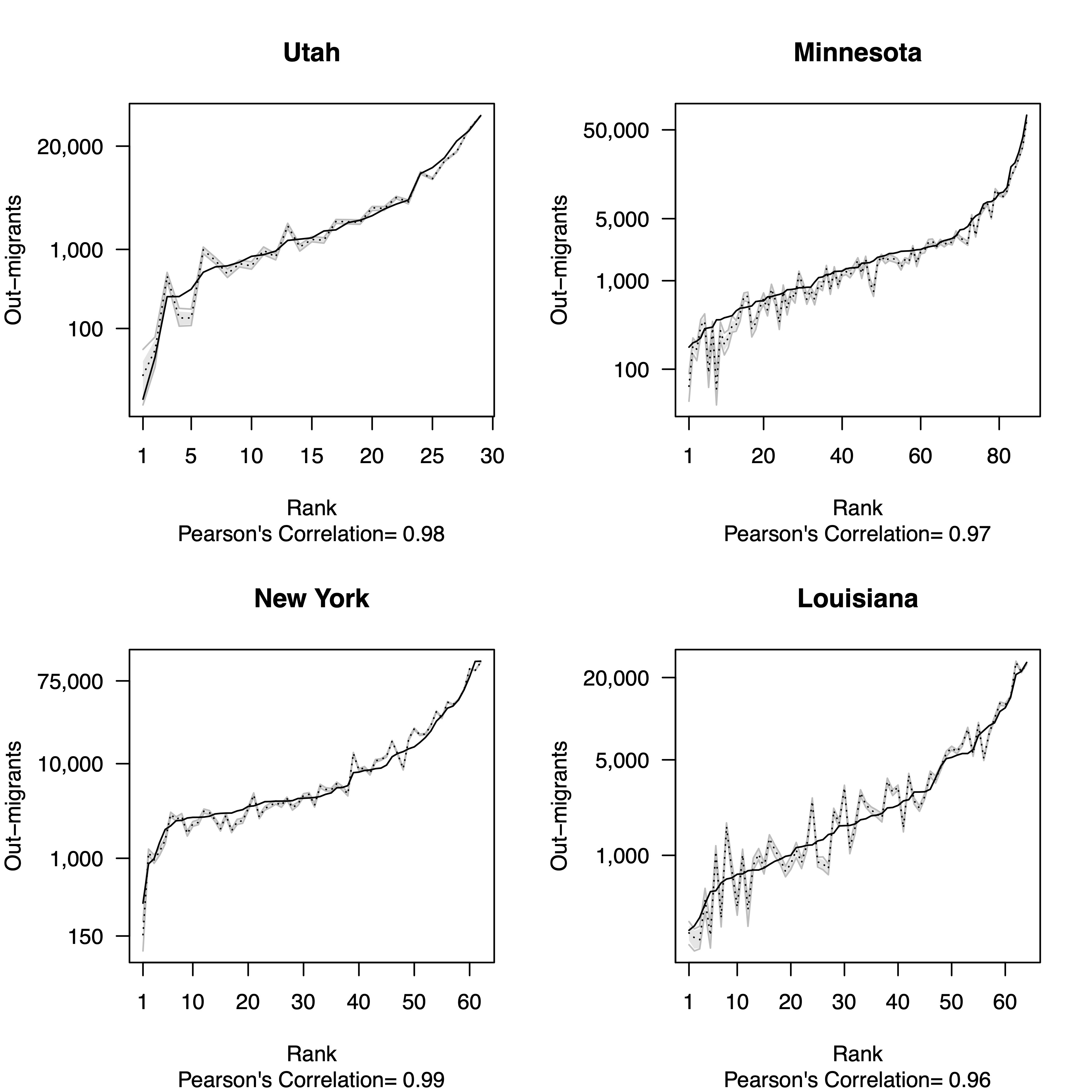}
\caption{\leftline{ Model-adequacy plots (outvolume)}}
\label{fig:adequacy2}
\end{figure}

\theendnotes

\singlespacing


\begin{thebibliography}{}
\newcommand{\enquote}[1]{``#1''}

\bibitem[\protect\citeauthoryear{Almquist and Butts}{Almquist and Butts}{2015}]{almquist_predicting_2015}
Almquist, Zack~W. and Carter~T. Butts. 2015.
\newblock \enquote{Predicting {Regional} {Self}-{Identification} from {Spatial} {Network} {Models}.}
\newblock {\em Geographical Analysis\/} 47:50--72.

\bibitem[\protect\citeauthoryear{An}{An}{2016}]{an_fitting_2016}
An, Weihua. 2016.
\newblock \enquote{Fitting {ERGMs} on big networks.}
\newblock {\em Social Science Research\/} 59:107--119.

\bibitem[\protect\citeauthoryear{Badger, Bui, and Katz}{Badger et~al.}{2018}]{badger_suburbs_2018}
Badger, Emily, Quoctrung Bui, and Josh Katz. 2018.
\newblock \enquote{The {Suburbs} {Are} {Changing}. {But} {Not} in {All} the {Ways} {Liberals} {Hope}.}
\newblock {\em The New York Times\/} .

\bibitem[\protect\citeauthoryear{Bakewell}{Bakewell}{2010}]{bakewell_reflections_2010}
Bakewell, Oliver. 2010.
\newblock \enquote{Some {Reflections} on {Structure} and {Agency} in {Migration} {Theory}.}
\newblock {\em Journal of Ethnic and Migration Studies\/} 36:1689--1708.

\bibitem[\protect\citeauthoryear{Bakewell}{Bakewell}{2014}]{bakewell_relaunching_2014}
Bakewell, Oliver. 2014.
\newblock \enquote{Relaunching migration systems.}
\newblock {\em Migration Studies\/} 2:300--318.

\bibitem[\protect\citeauthoryear{Bakewell, De~Haas, and Kubal}{Bakewell et~al.}{2012}]{bakewell_migration_2012}
Bakewell, Oliver, Hein De~Haas, and Agnieszka Kubal. 2012.
\newblock \enquote{Migration {Systems}, {Pioneer} {Migrants} and the {Role} of {Agency}.}
\newblock {\em Journal of Critical Realism\/} 11:413--437.

\bibitem[\protect\citeauthoryear{Bakewell, Engbersen, Fonseca, and Horst}{Bakewell et~al.}{2016a}]{bakewell_beyond_2016}
Bakewell, Oliver, Godfried Engbersen, Maria~Lucinda Fonseca, and Cindy Horst. 2016a.
\newblock {\em Beyond {Networks}: {Feedback} in {International} {Migration}\/}.
\newblock Springer.

\bibitem[\protect\citeauthoryear{Bakewell, Kubal, and Pereira}{Bakewell et~al.}{2016b}]{bakewell_introduction_2016}
Bakewell, Oliver, Agnieszka Kubal, and Sónia Pereira. 2016b.
\newblock \enquote{Introduction: {Feedback} in {Migration} {Processes}.}
\newblock In {\em Beyond {Networks}: {Feedback} in {International} {Migration}\/}, edited by  Oliver Bakewell, Godfried Engbersen, Maria~Lucinda Fonseca, and Cindy Horst, Migration, {Diasporas} and {Citizenship}, pp. 1--17. London: Palgrave Macmillan UK.

\bibitem[\protect\citeauthoryear{Baldassarri and Gelman}{Baldassarri and Gelman}{2008}]{baldassarri_partisans_2008}
Baldassarri, Delia and Andrew Gelman. 2008.
\newblock \enquote{Partisans without {Constraint}: {Political} {Polarization} and {Trends} in {American} {Public} {Opinion}.}
\newblock {\em American Journal of Sociology\/} 114:408--446.

\bibitem[\protect\citeauthoryear{Bauman}{Bauman}{2000}]{bauman_liquid_2000}
Bauman, Zygmunt. 2000.
\newblock {\em Liquid {Modernity}\/}.
\newblock John Wiley \& Sons.

\bibitem[\protect\citeauthoryear{Besag}{Besag}{1974}]{besag_spatial_1974}
Besag, Julian. 1974.
\newblock \enquote{Spatial {Interaction} and the {Statistical} {Analysis} of {Lattice} {Systems}.}
\newblock {\em Journal of the Royal Statistical Society: Series B (Methodological)\/} 36:192--225.

\bibitem[\protect\citeauthoryear{Bishop and Cushing}{Bishop and Cushing}{2009}]{bishop_big_2009}
Bishop, Bill and Robert~G. Cushing. 2009.
\newblock {\em The {Big} {Sort}: {Why} the {Clustering} of {Like}-minded {America} is {Tearing} {Us} {Apart}\/}.
\newblock Houghton Mifflin Harcourt.

\bibitem[\protect\citeauthoryear{Borjas}{Borjas}{2006}]{borjas_native_2006}
Borjas, George~J. 2006.
\newblock \enquote{Native {Internal} {Migration} and the {Labor} {Market} {Impact} of {Immigration}.}
\newblock {\em Journal of Human Resources\/} 41:221--258.

\bibitem[\protect\citeauthoryear{Boyle, Keith~H., Vaughan, and Vaughan}{Boyle et~al.}{2014}]{boyle_exploring_2014}
Boyle, Paul, Halfacree Keith~H., Robinson Vaughan, and Robinson Vaughan. 2014.
\newblock {\em Exploring {Contemporary} {Migration}\/}.
\newblock Abingdon, United Kingdom: Routledge.

\bibitem[\protect\citeauthoryear{Briggs, Popkin, and Goering}{Briggs et~al.}{2010}]{briggs_moving_2010}
Briggs, Xavier de~Souza, Susan~J. Popkin, and John Goering. 2010.
\newblock {\em Moving to {Opportunity}: {The} {Story} of an {American} {Experiment} to {Fight} {Ghetto} {Poverty}\/}.
\newblock Oxford University Press.

\bibitem[\protect\citeauthoryear{Brown and Enos}{Brown and Enos}{2021}]{brown_measurement_2021}
Brown, Jacob~R. and Ryan~D. Enos. 2021.
\newblock \enquote{The measurement of partisan sorting for 180 million voters.}
\newblock {\em Nature Human Behaviour\/} 5:998--1008.

\bibitem[\protect\citeauthoryear{Brown and Bean}{Brown and Bean}{2016}]{brown_conceptualizing_2016}
Brown, Susan~K. and Frank~D. Bean. 2016.
\newblock \enquote{Conceptualizing {Migration}: {From} {Internal}/{International} to {Kinds} of {Membership}.}
\newblock In {\em International {Handbook} of {Migration} and {Population} {Distribution}\/}, edited by  Michael~J. White, International {Handbooks} of {Population}, pp. 91--106. Dordrecht: Springer Netherlands.

\bibitem[\protect\citeauthoryear{Burris}{Burris}{2004}]{burris_academic_2004}
Burris, Val. 2004.
\newblock \enquote{The {Academic} {Caste} {System}: {Prestige} {Hierarchies} in {PhD} {Exchange} {Networks}.}
\newblock {\em American Sociological Review\/} 69:239--264.

\bibitem[\protect\citeauthoryear{Butts}{Butts}{2007}]{butts_models_2007}
Butts, Carter~T. 2007.
\newblock \enquote{Models for {Generalized} {Location} {Systems}.}
\newblock {\em Sociological Methodology\/} 37:283--348.

\bibitem[\protect\citeauthoryear{Butts, Acton, Hipp, and Nagle}{Butts et~al.}{2012}]{butts_geographical_2012}
Butts, Carter~T., Ryan~M. Acton, John~R. Hipp, and Nicholas~N. Nagle. 2012.
\newblock \enquote{Geographical variability and network structure.}
\newblock {\em Social Networks\/} 34:82--100.

\bibitem[\protect\citeauthoryear{Card}{Card}{2001}]{card_immigrant_2001}
Card, David. 2001.
\newblock \enquote{Immigrant {Inflows}, {Native} {Outflows}, and the {Local} {Labor} {Market} {Impacts} of {Higher} {Immigration}.}
\newblock {\em Journal of Labor Economics\/} 19:22--64.

\bibitem[\protect\citeauthoryear{Carling}{Carling}{2002}]{carling_migration_2002}
Carling, Jørgen. 2002.
\newblock \enquote{Migration in the age of involuntary immobility: {Theoretical} reflections and {Cape} {Verdean} experiences.}
\newblock {\em Journal of Ethnic and Migration Studies\/} 28:5--42.

\bibitem[\protect\citeauthoryear{Carling and Schewel}{Carling and Schewel}{2018}]{carling_revisiting_2018}
Carling, Jørgen and Kerilyn Schewel. 2018.
\newblock \enquote{Revisiting aspiration and ability in international migration.}
\newblock {\em Journal of Ethnic and Migration Studies\/} 44:945--963.

\bibitem[\protect\citeauthoryear{Castles, Haas, and Miller}{Castles et~al.}{2013}]{castles_age_2013}
Castles, Stephen, Hein~de Haas, and Mark~J. Miller. 2013.
\newblock {\em The {Age} of {Migration}: {International} {Population} {Movements} in the {Modern} {World}\/}.
\newblock Palgrave Macmillan.

\bibitem[\protect\citeauthoryear{Cetina and Cicourel}{Cetina and Cicourel}{2014}]{cetina_advances_2014}
Cetina, Karin~Knorr and A.~V. Cicourel (eds.). 2014.
\newblock {\em Advances in {Social} {Theory} and {Methodology} ({RLE} {Social} {Theory}): {Toward} an {Integration} of {Micro}- and {Macro}-{Sociologies}\/}.
\newblock London: Routledge.

\bibitem[\protect\citeauthoryear{Charyyev and Gunes}{Charyyev and Gunes}{2019}]{charyyev_complex_2019}
Charyyev, Batyr and Mehmet~Hadi Gunes. 2019.
\newblock \enquote{Complex network of {United} {States} migration.}
\newblock {\em Computational Social Networks\/} 6:1--28.

\bibitem[\protect\citeauthoryear{Chen and Rohla}{Chen and Rohla}{2018}]{chen_effect_2018}
Chen, M.~Keith and Ryne Rohla. 2018.
\newblock \enquote{The effect of partisanship and political advertising on close family ties.}
\newblock {\em Science\/} 360:1020--1024.

\bibitem[\protect\citeauthoryear{Cheng and Park}{Cheng and Park}{2020}]{cheng_flows_2020}
Cheng, Siwei and Barum Park. 2020.
\newblock \enquote{Flows and {Boundaries}: {A} {Network} {Approach} to {Studying} {Occupational} {Mobility} in the {Labor} {Market}.}
\newblock {\em American Journal of Sociology\/} 126:577--631.

\bibitem[\protect\citeauthoryear{Clark}{Clark}{2008}]{clark_reexamining_2008}
Clark, William A.~V. 2008.
\newblock \enquote{Reexamining the moving to opportunity study and its contribution to changing the distribution of poverty and ethnic concentration.}
\newblock {\em Demography\/} 45:515--535.

\bibitem[\protect\citeauthoryear{Cohen, Roig, Reuman, and GoGwilt}{Cohen et~al.}{2008}]{cohen_international_2008}
Cohen, Joel~E., Marta Roig, Daniel~C. Reuman, and Cai GoGwilt. 2008.
\newblock \enquote{International migration beyond gravity: {A} statistical model for use in population projections.}
\newblock {\em Proceedings of the National Academy of Sciences\/} 105:15269--15274.

\bibitem[\protect\citeauthoryear{Cohen and Sirkeci}{Cohen and Sirkeci}{2011}]{cohen_cultures_2011}
Cohen, Jeffrey~H. and Ibrahim Sirkeci. 2011.
\newblock {\em Cultures of {Migration}: {The} {Global} {Nature} of {Contemporary} {Mobility}\/}.
\newblock University of Texas Press.

\bibitem[\protect\citeauthoryear{Coleman}{Coleman}{1986}]{coleman_social_1986}
Coleman, James~S. 1986.
\newblock \enquote{Social {Theory}, {Social} {Research}, and a {Theory} of {Action}.}
\newblock {\em American Journal of Sociology\/} 91:1309--1335.

\bibitem[\protect\citeauthoryear{Conway}{Conway}{1980}]{conway_step-wise_1980}
Conway, Dennis. 1980.
\newblock \enquote{Step-{Wise} {Migration}: {Toward} a {Clarification} of the {Mechanism}.}
\newblock {\em International Migration Review\/} 14:3--14.

\bibitem[\protect\citeauthoryear{Cooke}{Cooke}{2013}]{cooke_internal_2013}
Cooke, Thomas~J. 2013.
\newblock \enquote{Internal {Migration} in {Decline}.}
\newblock {\em The Professional Geographer\/} 65:664--675.

\bibitem[\protect\citeauthoryear{Coulson and Grieco}{Coulson and Grieco}{2013}]{coulson_mobility_2013}
Coulson, N.~Edward and Paul L.~E. Grieco. 2013.
\newblock \enquote{Mobility and mortgages: {Evidence} from the {PSID}.}
\newblock {\em Regional Science and Urban Economics\/} 43:1--7.

\bibitem[\protect\citeauthoryear{Cramer}{Cramer}{2016}]{cramer_politics_2016}
Cramer, Katherine~J. 2016.
\newblock {\em The {Politics} of {Resentment}: {Rural} {Consciousness} in {Wisconsin} and the {Rise} of {Scott} {Walker}\/}.
\newblock Chicago: University of Chicago Press.

\bibitem[\protect\citeauthoryear{Crowder, Pais, and South}{Crowder et~al.}{2012}]{crowder_neighborhood_2012}
Crowder, Kyle, Jeremy Pais, and Scott~J. South. 2012.
\newblock \enquote{Neighborhood {Diversity}, {Metropolitan} {Constraints}, and {Household} {Migration}.}
\newblock {\em American Sociological Review\/} 77:325--353.

\bibitem[\protect\citeauthoryear{Czaika and de~Haas}{Czaika and de~Haas}{2017}]{czaika_effect_2017}
Czaika, Mathias and Hein de~Haas. 2017.
\newblock \enquote{The {Effect} of {Visas} on {Migration} {Processes}.}
\newblock {\em International Migration Review\/} 51:893--926.

\bibitem[\protect\citeauthoryear{Dao, Furceri, and Loungani}{Dao et~al.}{2017}]{dao_regional_2017}
Dao, Mai, Davide Furceri, and Prakash Loungani. 2017.
\newblock \enquote{Regional {Labor} {Market} {Adjustment} in the {United} {States}: {Trend} and {Cycle}.}
\newblock {\em The Review of Economics and Statistics\/} 99:243--257.

\bibitem[\protect\citeauthoryear{Davis and Leinhardt}{Davis and Leinhardt}{1972}]{davis_structure_1972}
Davis, James~A and Samuel Leinhardt. 1972.
\newblock \enquote{The {Structure} of {Positive} {Interpersonal} {Relations} in {Small} {Groups}.}
\newblock In {\em Sociological {Theories} in {Progress}\/}, pp. 218--251. Boston: Houghton Mifflin.

\bibitem[\protect\citeauthoryear{de~Haas}{de~Haas}{2010}]{de_haas_internal_2010}
de~Haas, Hein. 2010.
\newblock \enquote{The {Internal} {Dynamics} of {Migration} {Processes}: {A} {Theoretical} {Inquiry}.}
\newblock {\em Journal of Ethnic and Migration Studies\/} 36:1587--1617.

\bibitem[\protect\citeauthoryear{DellaPosta}{DellaPosta}{2020}]{dellaposta_pluralistic_2020}
DellaPosta, Daniel. 2020.
\newblock \enquote{Pluralistic {Collapse}: {The} “{Oil} {Spill}” {Model} of {Mass} {Opinion} {Polarization}.}
\newblock {\em American Sociological Review\/} 85:507--536.

\bibitem[\protect\citeauthoryear{DellaPosta and Macy}{DellaPosta and Macy}{2015}]{dellaposta_center_2015}
DellaPosta, Daniel and Michael Macy. 2015.
\newblock \enquote{The center cannot hold. {Networks}, echo chambers, and polarization.}
\newblock In {\em Order on the {Edge} of {Chaos}, {Social} {Psychology} and the {Problem} of {Social} {Order}\/}, pp. 86--104. New York: Cambridge University Press.

\bibitem[\protect\citeauthoryear{DellaPosta, Shi, and Macy}{DellaPosta et~al.}{2015}]{dellaposta_why_2015}
DellaPosta, Daniel, Yongren Shi, and Michael Macy. 2015.
\newblock \enquote{Why {Do} {Liberals} {Drink} {Lattes}?}
\newblock {\em American Journal of Sociology\/} 120:1473--1511.

\bibitem[\protect\citeauthoryear{DeLuca, Wood, and Rosenblatt}{DeLuca et~al.}{2019}]{deluca_why_2019}
DeLuca, Stefanie, Holly Wood, and Peter Rosenblatt. 2019.
\newblock \enquote{Why {Poor} {Families} {Move} ({And} {Where} {They} {Go}): {Reactive} {Mobility} and {Residential} {Decisions}.}
\newblock {\em City \& Community\/} 18:556--593.

\bibitem[\protect\citeauthoryear{Desmarais and Cranmer}{Desmarais and Cranmer}{2012}]{desmarais_statistical_2012}
Desmarais, Bruce~A. and Skyler~J. Cranmer. 2012.
\newblock \enquote{Statistical {Inference} for {Valued}-{Edge} {Networks}: {The} {Generalized} {Exponential} {Random} {Graph} {Model}.}
\newblock {\em PLoS ONE\/} 7:e30136.

\bibitem[\protect\citeauthoryear{DeWaard, Curtis, and Fussell}{DeWaard et~al.}{2016}]{dewaard_population_2016}
DeWaard, Jack, Katherine~J. Curtis, and Elizabeth Fussell. 2016.
\newblock \enquote{Population recovery in {New} {Orleans} after {Hurricane} {Katrina}: exploring the potential role of stage migration in migration systems.}
\newblock {\em Population and Environment\/} 37:449--463.

\bibitem[\protect\citeauthoryear{DeWaard, Fussell, Curtis, and Ha}{DeWaard et~al.}{2020}]{dewaard_changing_2020}
DeWaard, Jack, Elizabeth Fussell, Katherine~J. Curtis, and Jasmine~Trang Ha. 2020.
\newblock \enquote{Changing spatial interconnectivity during the “{Great} {American} {Migration} {Slowdown}”: {A} decomposition of intercounty migration rates, 1990–2010.}
\newblock {\em Population, Space and Place\/} 26:e2274.

\bibitem[\protect\citeauthoryear{DeWaard and Ha}{DeWaard and Ha}{2019}]{dewaard_resituating_2019}
DeWaard, Jack and Jasmine~Trang Ha. 2019.
\newblock \enquote{Resituating relaunched migration systems as emergent entities manifested in geographic structures.}
\newblock {\em Migration Studies\/} 7:39--58.

\bibitem[\protect\citeauthoryear{DeWaard, Hauer, Fussell, Curtis, Whitaker, McConnell, Price, Egan-Robertson, Soto, and Castro}{DeWaard et~al.}{2022}]{dewaard_user_2022}
DeWaard, Jack, Mathew Hauer, Elizabeth Fussell, Katherine~J. Curtis, Stephan~D. Whitaker, Kathryn McConnell, Kobie Price, David Egan-Robertson, Michael Soto, and Catalina~Anampa Castro. 2022.
\newblock \enquote{User {Beware}: {Concerning} {Findings} from the {Post} 2011–2012 {U}.{S}. {Internal} {Revenue} {Service} {Migration} {Data}.}
\newblock {\em Population Research and Policy Review\/} 41:437--448.

\bibitem[\protect\citeauthoryear{DeWaard, Kim, and Raymer}{DeWaard et~al.}{2012}]{dewaard_migration_2012}
DeWaard, Jack, Keuntae Kim, and James Raymer. 2012.
\newblock \enquote{Migration {Systems} in {Europe}: {Evidence} {From} {Harmonized} {Flow} {Data}.}
\newblock {\em Demography\/} 49:1307--1333.

\bibitem[\protect\citeauthoryear{DiPrete, Gelman, McCormick, Teitler, and Zheng}{DiPrete et~al.}{2011}]{diprete_segregation_2011}
DiPrete, Thomas~A., Andrew Gelman, Tyler McCormick, Julien Teitler, and Tian Zheng. 2011.
\newblock \enquote{Segregation in {Social} {Networks} {Based} on {Acquaintanceship} and {Trust}.}
\newblock {\em American Journal of Sociology\/} 116:1234--83.

\bibitem[\protect\citeauthoryear{Durand and Massey}{Durand and Massey}{2010}]{durand_new_2010}
Durand, Jorge and Douglas~S. Massey. 2010.
\newblock \enquote{New world orders: {Continuities} and changes in {Latin} {American} migration.}
\newblock {\em The Annals of the American Academy of Political and Social Science\/} 630:20--52.
\newblock Publisher: Sage Publications Sage CA: Los Angeles, CA.

\bibitem[\protect\citeauthoryear{Dye}{Dye}{1990}]{dye_american_1990}
Dye, Thomas~R. 1990.
\newblock {\em American {Federalism}: {Competition} {Among} {Governments}\/}.
\newblock Lexington Books.

\bibitem[\protect\citeauthoryear{Eeckhout}{Eeckhout}{2004}]{eeckhout_gibrats_2004}
Eeckhout, Jan. 2004.
\newblock \enquote{Gibrat's {Law} for ({All}) {Cities}.}
\newblock {\em American Economic Review\/} 94:1429--1451.

\bibitem[\protect\citeauthoryear{Einstein, Podolsky, and Rosen}{Einstein et~al.}{1935}]{einstein_can_1935}
Einstein, A., B.~Podolsky, and N.~Rosen. 1935.
\newblock \enquote{Can {Quantum}-{Mechanical} {Description} of {Physical} {Reality} {Be} {Considered} {Complete}?}
\newblock {\em Physical Review\/} 47:777--780.
\newblock Cambridge, UK: Cambridge University Press.

\bibitem[\protect\citeauthoryear{Fawcett}{Fawcett}{1989}]{fawcett_networks_1989}
Fawcett, James~T. 1989.
\newblock \enquote{Networks, {Linkages}, and {Migration} {Systems}.}
\newblock {\em The International Migration Review\/} 23:671--680.

\bibitem[\protect\citeauthoryear{Fernández-Kelly}{Fernández-Kelly}{2008}]{fernandez-kelly_back_2008}
Fernández-Kelly, Patricia. 2008.
\newblock \enquote{The {Back} {Pocket} {Map}: {Social} {Class} and {Cultural} {Capital} as {Transferable} {Assets} in the {Advancement} of {Second}-{Generation} {Immigrants}.}
\newblock {\em The ANNALS of the American Academy of Political and Social Science\/} 620:116--137.

\bibitem[\protect\citeauthoryear{Fischer}{Fischer}{2002}]{fischer_ever-more_2002}
Fischer, Claude~S. 2002.
\newblock \enquote{Ever-{More} {Rooted} {Americans}.}
\newblock {\em City \& Community\/} 1:177--198.

\bibitem[\protect\citeauthoryear{Fitchen}{Fitchen}{1994}]{fitchen_residential_1994}
Fitchen, Janet~M. 1994.
\newblock \enquote{Residential {Mobility} {Among} the {Rural} {Poor}.}
\newblock {\em Rural Sociology\/} 59:416--436.
\newblock \_eprint: https://onlinelibrary.wiley.com/doi/pdf/10.1111/j.1549-0831.1994.tb00540.x.

\bibitem[\protect\citeauthoryear{Fossett}{Fossett}{2006}]{fossett_ethnic_2006}
Fossett, Mark. 2006.
\newblock \enquote{Ethnic {Preferences}, {Social} {Distance} {Dynamics}, and {Residential} {Segregation}: {Theoretical} {Explorations} {Using} {Simulation} {Analysis}.}
\newblock {\em The Journal of Mathematical Sociology\/} 30:185--273.

\bibitem[\protect\citeauthoryear{Freier and Holloway}{Freier and Holloway}{2019}]{freier_impact_2019}
Freier, Luisa~Feline and Kyle Holloway. 2019.
\newblock \enquote{The {Impact} of {Tourist} {Visas} on {Intercontinental} {South}-{South} {Migration}: {Ecuador}’s {Policy} of “{Open} {Doors}” as a {Quasi}-{Experiment}.}
\newblock {\em International Migration Review\/} 53:1171--1208.

\bibitem[\protect\citeauthoryear{Frey}{Frey}{1995a}]{frey_immigration_1995}
Frey, William~H. 1995a.
\newblock \enquote{Immigration and {Internal} {Migration} '{Flight}' from {US} {Metropolitan} {Areas}: {Toward} a {New} {Demographic} {Balkanisation}.}
\newblock {\em Urban Studies\/} 32:733--757.

\bibitem[\protect\citeauthoryear{Frey}{Frey}{1995b}]{frey_immigration_1995-1}
Frey, William~H. 1995b.
\newblock \enquote{Immigration and internal migration “flight”: {A} {California} case study.}
\newblock {\em Population and Environment\/} 16:353--375.

\bibitem[\protect\citeauthoryear{Frey}{Frey}{2009}]{frey_great_2009}
Frey, William~H. 2009.
\newblock \enquote{The great {American} migration slowdown: {Regional} and metropolitan dimensions.}
\newblock Technical report, Brookings Institution, Washington, DC.

\bibitem[\protect\citeauthoryear{Garip}{Garip}{2008}]{garip_social_2008}
Garip, Filiz. 2008.
\newblock \enquote{Social capital and migration: {How} do similar resources lead to divergent outcomes?}
\newblock {\em Demography\/} 45:591--617.

\bibitem[\protect\citeauthoryear{Garip and Asad}{Garip and Asad}{2016}]{garip_network_2016}
Garip, Filiz and Asad~L. Asad. 2016.
\newblock \enquote{Network {Effects} in {Mexico}–{U}.{S}. {Migration}: {Disentangling} the {Underlying} {Social} {Mechanisms}.}
\newblock {\em American Behavioral Scientist\/} 60:1168--1193.

\bibitem[\protect\citeauthoryear{Geyer and Thompson}{Geyer and Thompson}{1992}]{geyer_constrained_1992}
Geyer, Charles~J. and Elizabeth~A. Thompson. 1992.
\newblock \enquote{Constrained {Monte} {Carlo} {Maximum} {Likelihood} for {Dependent} {Data}.}
\newblock {\em Journal of the Royal Statistical Society: Series B (Methodological)\/} 54:657--683.

\bibitem[\protect\citeauthoryear{Gimpel and Hui}{Gimpel and Hui}{2015}]{gimpel_seeking_2015}
Gimpel, James~G. and Iris~S. Hui. 2015.
\newblock \enquote{Seeking politically compatible neighbors? {The} role of neighborhood partisan composition in residential sorting.}
\newblock {\em Political Geography\/} 48:130--142.

\bibitem[\protect\citeauthoryear{Gondal}{Gondal}{2018}]{gondal_duality_2018}
Gondal, Neha. 2018.
\newblock \enquote{Duality of departmental specializations and {PhD} exchange: {A} {Weberian} analysis of status in interaction using multilevel exponential random graph models ({mERGM}).}
\newblock {\em Social Networks\/} 55:202--212.

\bibitem[\protect\citeauthoryear{Goodreau, Kitts, and Morris}{Goodreau et~al.}{2009}]{goodreau_birds_2009}
Goodreau, Steven~M., James~A. Kitts, and Martina Morris. 2009.
\newblock \enquote{Birds of a feather, or friend of a friend? using exponential random graph models to investigate adolescent social networks.}
\newblock {\em Demography\/} 46:103--125.

\bibitem[\protect\citeauthoryear{Grusky, Western, and Wimer}{Grusky et~al.}{2011}]{grusky_great_2011}
Grusky, David~B., Bruce Western, and Christopher Wimer. 2011.
\newblock {\em The {Great} {Recession}\/}.
\newblock Russell Sage Foundation.

\bibitem[\protect\citeauthoryear{Hall, Limaye, and Kulkarni}{Hall et~al.}{2009}]{hall_overview_2009}
Hall, Bradford, Advait Limaye, and Ashok~B. Kulkarni. 2009.
\newblock \enquote{Overview: {Generation} of {Gene} {Knockout} {Mice}.}
\newblock {\em Current Protocols in Cell Biology\/} 44:19.12.1--19.12.17.

\bibitem[\protect\citeauthoryear{Han, Xu, Fan, Huang, Xu, and Gao}{Han et~al.}{2021}]{han_quantifying_2021}
Han, Xiaoyi, Yilan Xu, Linlin Fan, Yi~Huang, Minhong Xu, and Song Gao. 2021.
\newblock \enquote{Quantifying {COVID}-19 importation risk in a dynamic network of domestic cities and international countries.}
\newblock {\em Proceedings of the National Academy of Sciences\/} 118:e2100201118.

\bibitem[\protect\citeauthoryear{Hanneke, Fu, and Xing}{Hanneke et~al.}{2010}]{hanneke_discrete_2010}
Hanneke, Steve, Wenjie Fu, and Eric~P. Xing. 2010.
\newblock \enquote{Discrete temporal models of social networks.}
\newblock {\em Electronic Journal of Statistics\/} 4:585--605.

\bibitem[\protect\citeauthoryear{Hauer and Byars}{Hauer and Byars}{2019}]{hauer_irs_2019}
Hauer, Mathew and James Byars. 2019.
\newblock \enquote{{IRS} county-to-county migration data, 1990‒2010.}
\newblock {\em Demographic Research\/} 40:1153--1166.

\bibitem[\protect\citeauthoryear{Hauer}{Hauer}{2017}]{hauer_migration_2017}
Hauer, Mathew~E. 2017.
\newblock \enquote{Migration induced by sea-level rise could reshape the {US} population landscape.}
\newblock {\em Nature Climate Change\/} 7:321--325.

\bibitem[\protect\citeauthoryear{Herting, Grusky, and Rompaey}{Herting et~al.}{1997}]{herting_social_1997}
Herting, Jerald~R., David~B. Grusky, and Stephen E.~Van Rompaey. 1997.
\newblock \enquote{The {Social} {Geography} of {Interstate} {Mobility} and {Persistence}.}
\newblock {\em American Sociological Review\/} 62:267--287.

\bibitem[\protect\citeauthoryear{Hipp, Butts, Acton, Nagle, and Boessen}{Hipp et~al.}{2013}]{hipp_extrapolative_2013}
Hipp, John~R., Carter~T. Butts, Ryan Acton, Nicholas~N. Nagle, and Adam Boessen. 2013.
\newblock \enquote{Extrapolative simulation of neighborhood networks based on population spatial distribution: {Do} they predict crime?}
\newblock {\em Social Networks\/} 35:614--625.

\bibitem[\protect\citeauthoryear{Hochschild}{Hochschild}{2018}]{hochschild_strangers_2018}
Hochschild, Arlie~Russell. 2018.
\newblock {\em Strangers in {Their} {Own} {Land}: {Anger} and {Mourning} on the {American} {Right}\/}.
\newblock The New Press.

\bibitem[\protect\citeauthoryear{Huang and Butts}{Huang and Butts}{2024}]{huang_parameter_2024}
Huang, Peng and Carter~T. Butts. 2024.
\newblock \enquote{Parameter estimation procedures for exponential-family random graph models on count-valued networks: {A} comparative simulation study.}
\newblock {\em Social Networks\/} 76:51--67.

\bibitem[\protect\citeauthoryear{Hunter, Goodreau, and Handcock}{Hunter et~al.}{2008a}]{hunter_goodness_2008}
Hunter, David~R., Steven~M. Goodreau, and Mark~S. Handcock. 2008a.
\newblock \enquote{Goodness of {Fit} of {Social} {Network} {Models}.}
\newblock {\em Journal of the American Statistical Association\/} 103:248--258.

\bibitem[\protect\citeauthoryear{Hunter, Handcock, Butts, Goodreau, and Morris}{Hunter et~al.}{2008b}]{hunter_ergm_2008}
Hunter, David~R., Mark~S. Handcock, Carter~T. Butts, Steven~M. Goodreau, and Martina Morris. 2008b.
\newblock \enquote{ergm: {A} {Package} to {Fit}, {Simulate} and {Diagnose} {Exponential}-{Family} {Models} for {Networks}.}
\newblock {\em Journal of statistical software\/} 24:nihpa54860.

\bibitem[\protect\citeauthoryear{Hyatt, McEntarfer, Ueda, and Zhang}{Hyatt et~al.}{2018}]{hyatt_interstate_2018}
Hyatt, Henry, Erika McEntarfer, Ken Ueda, and Alexandria Zhang. 2018.
\newblock \enquote{Interstate {Migration} and {Employer}-to-{Employer} {Transitions} in the {United} {States}: {New} {Evidence} {From} {Administrative} {Records} {Data}.}
\newblock {\em Demography\/} 55:2161--2180.

\bibitem[\protect\citeauthoryear{Hyatt and Spletzer}{Hyatt and Spletzer}{2013}]{hyatt_recent_2013}
Hyatt, Henry~R. and James~R. Spletzer. 2013.
\newblock \enquote{The recent decline in employment dynamics.}
\newblock {\em IZA Journal of Labor Economics\/} 2:1--21.

\bibitem[\protect\citeauthoryear{Intrator, Tannen, and Massey}{Intrator et~al.}{2016}]{intrator_segregation_2016}
Intrator, Jake, Jonathan Tannen, and Douglas~S. Massey. 2016.
\newblock \enquote{Segregation by race and income in the {United} {States} 1970–2010.}
\newblock {\em Social Science Research\/} 60:45--60.

\bibitem[\protect\citeauthoryear{Iyengar, Lelkes, Levendusky, Malhotra, and Westwood}{Iyengar et~al.}{2019}]{iyengar_origins_2019}
Iyengar, Shanto, Yphtach Lelkes, Matthew Levendusky, Neil Malhotra, and Sean~J. Westwood. 2019.
\newblock \enquote{The {Origins} and {Consequences} of {Affective} {Polarization} in the {United} {States}.}
\newblock {\em Annual Review of Political Science\/} 22:129--146.

\bibitem[\protect\citeauthoryear{Iyengar, Sood, and Lelkes}{Iyengar et~al.}{2012}]{iyengar_affect_2012}
Iyengar, Shanto, Gaurav Sood, and Yphtach Lelkes. 2012.
\newblock \enquote{Affect, {Not} {IdeologyA} {Social} {Identity} {Perspective} on {Polarization}.}
\newblock {\em Public Opinion Quarterly\/} 76:405--431.

\bibitem[\protect\citeauthoryear{Jasso}{Jasso}{2011}]{jasso_migration_2011}
Jasso, Guillermina. 2011.
\newblock \enquote{Migration and stratification.}
\newblock {\em Social Science Research\/} 40:1292--1336.

\bibitem[\protect\citeauthoryear{Jennissen}{Jennissen}{2007}]{jennissen_causality_2007}
Jennissen, Roel. 2007.
\newblock \enquote{Causality {Chains} in the {International} {Migration} {Systems} {Approach}.}
\newblock {\em Population Research and Policy Review\/} 26:411--436.

\bibitem[\protect\citeauthoryear{Jia, Molloy, Smith, and Wozniak}{Jia et~al.}{2022}]{jia_economics_2022}
Jia, Ning, Raven Molloy, Christopher~L. Smith, and Abigail Wozniak. 2022.
\newblock \enquote{The {Economics} of {Internal} {Migration}: {Advances} and {Policy} {Questions}.}
\newblock Finance and {Economics} {Discussion} {Series} 2022-003.
\newblock Washington: Board of Governors of the Federal Reserve System.

\bibitem[\protect\citeauthoryear{Johnson and Kleiner}{Johnson and Kleiner}{2020}]{johnson_is_2020}
Johnson, Janna~E. and Morris~M. Kleiner. 2020.
\newblock \enquote{Is {Occupational} {Licensing} a {Barrier} to {Interstate} {Migration}?}
\newblock {\em American Economic Journal: Economic Policy\/} 12:347--373.

\bibitem[\protect\citeauthoryear{Kaplan and Schulhofer-Wohl}{Kaplan and Schulhofer-Wohl}{2017}]{kaplan_understanding_2017}
Kaplan, Greg and Sam Schulhofer-Wohl. 2017.
\newblock \enquote{Understanding the {Long}-{Run} {Decline} in {Interstate} {Migration}.}
\newblock {\em International Economic Review\/} 58:57--94.

\bibitem[\protect\citeauthoryear{Kim and Cohen}{Kim and Cohen}{2010}]{kim_determinants_2010}
Kim, Keuntae and Joel~E. Cohen. 2010.
\newblock \enquote{Determinants of {International} {Migration} {Flows} to and from {Industrialized} {Countries}: {A} {Panel} {Data} {Approach} beyond {Gravity}.}
\newblock {\em International Migration Review\/} 44:899--932.

\bibitem[\protect\citeauthoryear{Krackhardt and Porter}{Krackhardt and Porter}{1986}]{krackhardt_snowball_1986}
Krackhardt, David and Lyman~W. Porter. 1986.
\newblock \enquote{The snowball effect: {Turnover} embedded in communication networks.}
\newblock {\em Journal of Applied Psychology\/} 71:50--55.

\bibitem[\protect\citeauthoryear{Kreager, Young, Haynie, Bouchard, Schaefer, and Zajac}{Kreager et~al.}{2017}]{kreager_where_2017}
Kreager, Derek~A., Jacob~T.N. Young, Dana~L. Haynie, Martin Bouchard, David~R. Schaefer, and Gary Zajac. 2017.
\newblock \enquote{Where “{Old} {Heads}” {Prevail}: {Inmate} {Hierarchy} in a {Men}’s {Prison} {Unit}.}
\newblock {\em American Sociological Review\/} 82:685--718.

\bibitem[\protect\citeauthoryear{Kritz and Gurak}{Kritz and Gurak}{2001}]{kritz_impact_2001}
Kritz, Mary~M and Douglas~T Gurak. 2001.
\newblock \enquote{The impact of immigration on the internal migration of natives and immigrants.}
\newblock {\em Demography\/} 38:133--145.

\bibitem[\protect\citeauthoryear{Kritz, Lim, Zlotnik, and Lim}{Kritz et~al.}{1992}]{kritz_international_1992}
Kritz, Mary~M., Lin~Lean Lim, Hania Zlotnik, and Lin Lean Lin~Lean Lim. 1992.
\newblock {\em International migration systems: a global approach\/}.
\newblock Oxford University Press, USA.

\bibitem[\protect\citeauthoryear{Krivitsky}{Krivitsky}{2012}]{krivitsky_exponential-family_2012}
Krivitsky, Pavel~N. 2012.
\newblock \enquote{Exponential-family random graph models for valued networks.}
\newblock {\em Electronic journal of statistics\/} 6:1100--1128.

\bibitem[\protect\citeauthoryear{Krivitsky and Butts}{Krivitsky and Butts}{2013}]{krivitsky_modeling_2013}
Krivitsky, Pavel~N and Carter~T Butts. 2013.
\newblock \enquote{Modeling valued networks with statnet.}
\newblock {\em The Statnet Development Team\/} .

\bibitem[\protect\citeauthoryear{Krysan and Crowder}{Krysan and Crowder}{2017}]{krysan_cycle_2017}
Krysan, Maria and Kyle Crowder. 2017.
\newblock {\em Cycle of {Segregation}: {Social} {Processes} and {Residential} {Stratification}\/}.
\newblock Russell Sage Foundation.

\bibitem[\protect\citeauthoryear{Lakon, Hipp, Wang, Butts, and Jose}{Lakon et~al.}{2015}]{lakon_simulating_2015}
Lakon, Cynthia~M., John~R. Hipp, Cheng Wang, Carter~T. Butts, and Rupa Jose. 2015.
\newblock \enquote{Simulating {Dynamic} {Network} {Models} and {Adolescent} {Smoking}: {The} {Impact} of {Varying} {Peer} {Influence} and {Peer} {Selection}.}
\newblock {\em American Journal of Public Health\/} 105:2438--2448.

\bibitem[\protect\citeauthoryear{Leal}{Leal}{2021}]{leal_network_2021}
Leal, Diego~F. 2021.
\newblock \enquote{Network {Inequalities} and {International} {Migration} in the {Americas}.}
\newblock {\em American Journal of Sociology\/} 126:1067--1126.

\bibitem[\protect\citeauthoryear{Lee}{Lee}{1966}]{lee_theory_1966}
Lee, Everett~S. 1966.
\newblock \enquote{A theory of migration.}
\newblock {\em Demography\/} 3:47--57.

\bibitem[\protect\citeauthoryear{Lee and Zhou}{Lee and Zhou}{2017}]{lee_why_2017}
Lee, Jennifer and Min Zhou. 2017.
\newblock \enquote{Why class matters less for {Asian}-{American} academic achievement.}
\newblock {\em Journal of Ethnic and Migration Studies\/} 43:2316--2330.

\bibitem[\protect\citeauthoryear{Leszczensky and Pink}{Leszczensky and Pink}{2019}]{leszczensky_what_2019}
Leszczensky, Lars and Sebastian Pink. 2019.
\newblock \enquote{What {Drives} {Ethnic} {Homophily}? {A} {Relational} {Approach} on {How} {Ethnic} {Identification} {Moderates} {Preferences} for {Same}-{Ethnic} {Friends}.}
\newblock {\em American Sociological Review\/} 84:394--419.

\bibitem[\protect\citeauthoryear{Levendusky}{Levendusky}{2009}]{levendusky_partisan_2009}
Levendusky, Matthew. 2009.
\newblock {\em The {Partisan} {Sort}: {How} {Liberals} {Became} {Democrats} and {Conservatives} {Became} {Republicans}\/}.
\newblock University of Chicago Press.

\bibitem[\protect\citeauthoryear{Lewis}{Lewis}{2013}]{lewis_limits_2013}
Lewis, Kevin. 2013.
\newblock \enquote{The limits of racial prejudice.}
\newblock {\em Proceedings of the National Academy of Sciences\/} 110:18814--18819.

\bibitem[\protect\citeauthoryear{Lewis}{Lewis}{2016}]{lewis_preferences_2016}
Lewis, Kevin. 2016.
\newblock \enquote{Preferences in the {Early} {Stages} of {Mate} {Choice}.}
\newblock {\em Social Forces\/} 95:283--320.

\bibitem[\protect\citeauthoryear{Lewis and Papachristos}{Lewis and Papachristos}{2019}]{lewis_rules_2019}
Lewis, Kevin and Andrew~V. Papachristos. 2019.
\newblock \enquote{Rules of the {Game}: {Exponential} {Random} {Graph} {Models} of a {Gang} {Homicide} {Network}.}
\newblock {\em Social Forces\/} 98:1829--1858.

\bibitem[\protect\citeauthoryear{Liang, Chunyu, Zhuang, and Ye}{Liang et~al.}{2008}]{liang_cumulative_2008}
Liang, Zai, Miao David Chunyu, Guotu Zhuang, and Wenzhen Ye. 2008.
\newblock \enquote{Cumulative {Causation}, {Market} {Transition}, and {Emigration} from {China}.}
\newblock {\em American Journal of Sociology\/} 114:706--737.

\bibitem[\protect\citeauthoryear{Liang and Chunyu}{Liang and Chunyu}{2013}]{liang_migration_2013}
Liang, Zai and Miao~David Chunyu. 2013.
\newblock \enquote{Migration within {China} and from {China} to the {USA}: {The} effects of migration networks, selectivity, and the rural political economy in {Fujian} {Province}.}
\newblock {\em Population Studies\/} 67:209--223.

\bibitem[\protect\citeauthoryear{Lichter, Parisi, and Taquino}{Lichter et~al.}{2022}]{lichter_intercounty_2022}
Lichter, Daniel~T., Domenico Parisi, and Michael~C. Taquino. 2022.
\newblock \enquote{Inter‐{County} {Migration} and the {Spatial} {Concentration} of {Poverty}: {Comparing} {Metro} and {Nonmetro} {Patterns}.}
\newblock {\em Rural Sociology\/} 87:119--143.

\bibitem[\protect\citeauthoryear{Liu, Andris, and Desmarais}{Liu et~al.}{2019}]{liu_migration_2019}
Liu, Xi, Clio Andris, and Bruce~A. Desmarais. 2019.
\newblock \enquote{Migration and political polarization in the {U}.{S}.: {An} analysis of the county-level migration network.}
\newblock {\em PLOS ONE\/} 14:e0225405.

\bibitem[\protect\citeauthoryear{Lobao and Kelly}{Lobao and Kelly}{2019}]{lobao_local_2019}
Lobao, Linda and Paige Kelly. 2019.
\newblock \enquote{Local {Governments} across the {Rural}–{Urban} {Continuum}: {Findings} from a {Recent} {National} {County} {Government} {Study}.}
\newblock {\em State and Local Government Review\/} 51:223--232.

\bibitem[\protect\citeauthoryear{Long}{Long}{1991}]{long_residential_1991}
Long, Larry. 1991.
\newblock \enquote{Residential {Mobility} {Differences} among {Developed} {Countries}.}
\newblock {\em International Regional Science Review\/} 14:133--147.

\bibitem[\protect\citeauthoryear{Lu, Liang, and Chunyu}{Lu et~al.}{2013}]{lu_emigration_2013}
Lu, Yao, Zai Liang, and Miao~David Chunyu. 2013.
\newblock \enquote{Emigration from {China} in {Comparative} {Perspective}.}
\newblock {\em Social Forces\/} 92:631--658.

\bibitem[\protect\citeauthoryear{Lubbers, Verdery, and Molina}{Lubbers et~al.}{2020}]{lubbers_social_2020}
Lubbers, Miranda~Jessica, Ashton~M. Verdery, and José~Luis Molina. 2020.
\newblock \enquote{Social {Networks} and {Transnational} {Social} {Fields}: {A} {Review} of {Quantitative} and {Mixed}-{Methods} {Approaches}.}
\newblock {\em International Migration Review\/} 54:177--204.

\bibitem[\protect\citeauthoryear{Mabogunje}{Mabogunje}{1970}]{mabogunje_systems_1970}
Mabogunje, Akin~L. 1970.
\newblock \enquote{Systems {Approach} to a {Theory} of {Rural}-{Urban} {Migration}.}
\newblock {\em Geographical Analysis\/} 2:1--18.

\bibitem[\protect\citeauthoryear{Massey and Denton}{Massey and Denton}{1988}]{massey_dimensions_1988}
Massey, D.S. and N.A. Denton. 1988.
\newblock \enquote{The dimensions of residential segregation.}
\newblock {\em Social Forces\/} 67:281--315.

\bibitem[\protect\citeauthoryear{Massey and Denton}{Massey and Denton}{1993}]{massey_american_1993}
Massey, Douglas and Nancy~A. Denton. 1993.
\newblock {\em American {Apartheid}: {Segregation} and the {Making} of the {Underclass}\/}.
\newblock Harvard University Press.

\bibitem[\protect\citeauthoryear{Massey}{Massey}{1990}]{massey_social_1990}
Massey, Douglas~S. 1990.
\newblock \enquote{Social {Structure}, {Household} {Strategies}, and the {Cumulative} {Causation} of {Migration}.}
\newblock {\em Population Index\/} 56:3.

\bibitem[\protect\citeauthoryear{Massey, Arango, Hugo, Kouaouci, and Pellegrino}{Massey et~al.}{1999}]{massey_worlds_1999}
Massey, Douglas~S., Joaquin Arango, Graeme Hugo, Ali Kouaouci, and Adela Pellegrino. 1999.
\newblock {\em Worlds in {Motion}: {Understanding} {International} {Migration} at the {End} of the {Millennium}\/}.
\newblock Clarendon Press.

\bibitem[\protect\citeauthoryear{Massey, Arango, Hugo, Kouaouci, Pellegrino, and Taylor}{Massey et~al.}{1993}]{massey_theories_1993}
Massey, Douglas~S., Joaquin Arango, Graeme Hugo, Ali Kouaouci, Adela Pellegrino, and J.~Edward Taylor. 1993.
\newblock \enquote{Theories of {International} {Migration}: {A} {Review} and {Appraisal}.}
\newblock {\em Population and Development Review\/} 19:431--466.

\bibitem[\protect\citeauthoryear{Massey, Durand, and Pren}{Massey et~al.}{2016}]{massey_why_2016}
Massey, Douglas~S., Jorge Durand, and Karen~A. Pren. 2016.
\newblock \enquote{Why {Border} {Enforcement} {Backfired}.}
\newblock {\em American Journal of Sociology\/} 121:1557--1600.

\bibitem[\protect\citeauthoryear{Massey and Espinosa}{Massey and Espinosa}{1997}]{massey_whats_1997}
Massey, Douglas~S. and Kristin~E. Espinosa. 1997.
\newblock \enquote{What's {Driving} {Mexico}-{U}.{S}. {Migration}? {A} {Theoretical}, {Empirical}, and {Policy} {Analysis}.}
\newblock {\em American Journal of Sociology\/} 102:939--999.

\bibitem[\protect\citeauthoryear{Massey, Goldring, and Durand}{Massey et~al.}{1994a}]{massey_continuities_1994}
Massey, Douglas~S., Luin Goldring, and Jorge Durand. 1994a.
\newblock \enquote{Continuities in {Transnational} {Migration}: {An} {Analysis} of {Nineteen} {Mexican} {Communities}.}
\newblock {\em American Journal of Sociology\/} 99:1492--1533.

\bibitem[\protect\citeauthoryear{Massey, Gross, and Shibuya}{Massey et~al.}{1994b}]{massey_migration_1994}
Massey, Douglas~S., Andrew~B. Gross, and Kumiko Shibuya. 1994b.
\newblock \enquote{Migration, {Segregation}, and the {Geographic} {Concentration} of {Poverty}.}
\newblock {\em American Sociological Review\/} 59:425.

\bibitem[\protect\citeauthoryear{Massey and Tannen}{Massey and Tannen}{2018}]{massey_suburbanization_2018}
Massey, Douglas~S. and Jonathan Tannen. 2018.
\newblock \enquote{Suburbanization and segregation in the {United} {States}: 1970–2010.}
\newblock {\em Ethnic and Racial Studies\/} 41:1594--1611.

\bibitem[\protect\citeauthoryear{Mata-Codesal}{Mata-Codesal}{2015}]{mata-codesal_ways_2015}
Mata-Codesal, Diana. 2015.
\newblock \enquote{Ways of {Staying} {Put} in {Ecuador}: {Social} and {Embodied} {Experiences} of {Mobility}–{Immobility} {Interactions}.}
\newblock {\em Journal of Ethnic and Migration Studies\/} 41:2274--2290.

\bibitem[\protect\citeauthoryear{McFarland, Moody, Diehl, Smith, and Thomas}{McFarland et~al.}{2014}]{mcfarland_network_2014}
McFarland, Daniel~A., James Moody, David Diehl, Jeffrey~A. Smith, and Reuben~J. Thomas. 2014.
\newblock \enquote{Network {Ecology} and {Adolescent} {Social} {Structure}.}
\newblock {\em American Sociological Review\/} 79:1088--1121.

\bibitem[\protect\citeauthoryear{McMahan and McFarland}{McMahan and McFarland}{2021}]{mcmahan_creative_2021}
McMahan, Peter and Daniel~A. McFarland. 2021.
\newblock \enquote{Creative {Destruction}: {The} {Structural} {Consequences} of {Scientific} {Curation}.}
\newblock {\em American Sociological Review\/} 86:341--376.

\bibitem[\protect\citeauthoryear{McMillan}{McMillan}{2019}]{mcmillan_tied_2019}
McMillan, Cassie. 2019.
\newblock \enquote{Tied {Together}: {Adolescent} {Friendship} {Networks}, {Immigrant} {Status}, and {Health} {Outcomes}.}
\newblock {\em Demography\/} 56:1075--1103.

\bibitem[\protect\citeauthoryear{McPherson, Smith-Lovin, and Cook}{McPherson et~al.}{2001}]{mcpherson_birds_2001}
McPherson, Miller, Lynn Smith-Lovin, and James~M Cook. 2001.
\newblock \enquote{Birds of a {Feather}: {Homophily} in {Social} {Networks}.}
\newblock {\em Annual Review of Sociology\/} 27:415--444.

\bibitem[\protect\citeauthoryear{Miller and Page}{Miller and Page}{2009}]{miller_complex_2009}
Miller, John~H. and Scott Page. 2009.
\newblock {\em Complex {Adaptive} {Systems}: {An} {Introduction} to {Computational} {Models} of {Social} {Life}\/}.
\newblock Princeton University Press.
\newblock Publication Title: Complex Adaptive Systems.

\bibitem[\protect\citeauthoryear{{MIT Election Data {and} Science Lab}}{{MIT Election Data {and} Science Lab}}{2018}]{mit_election_data_and_science_lab_county_2018}
{MIT Election Data {and} Science Lab}. 2018.
\newblock \enquote{County {Presidential} {Election} {Returns} 2000-2016.}
\newblock Retrieved September 29, 2020 (https://doi.org/10.7910/DVN/VOQCHQ).

\bibitem[\protect\citeauthoryear{Molloy, Smith, and Wozniak}{Molloy et~al.}{2011}]{molloy_internal_2011}
Molloy, Raven, Christopher~L Smith, and Abigail Wozniak. 2011.
\newblock \enquote{Internal {Migration} in the {United} {States}.}
\newblock {\em Journal of Economic Perspectives\/} 25:173--196.

\bibitem[\protect\citeauthoryear{Molloy, Smith, and Wozniak}{Molloy et~al.}{2017}]{molloy_job_2017}
Molloy, Raven, Christopher~L. Smith, and Abigail Wozniak. 2017.
\newblock \enquote{Job {Changing} and the {Decline} in {Long}-{Distance} {Migration} in the {United} {States}.}
\newblock {\em Demography\/} 54:631--653.

\bibitem[\protect\citeauthoryear{Monras}{Monras}{2018}]{monras_economic_2018}
Monras, Joan. 2018.
\newblock \enquote{Economic {Shocks} and {Internal} {Migration}.}
\newblock {\em CEPR Discussion Paper No. DP12977\/} .

\bibitem[\protect\citeauthoryear{Moody}{Moody}{2001}]{moody_race_2001}
Moody, James. 2001.
\newblock \enquote{Race, {School} {Integration}, and {Friendship} {Segregation} in {America}.}
\newblock {\em American Journal of Sociology\/} 107:679--716.

\bibitem[\protect\citeauthoryear{Morris, Handcock, and Hunter}{Morris et~al.}{2008}]{morris_specification_2008}
Morris, Martina, Mark~S. Handcock, and David~R. Hunter. 2008.
\newblock \enquote{Specification of {Exponential}-{Family} {Random} {Graph} {Models}: {Terms} and {Computational} {Aspects}.}
\newblock {\em Journal of statistical software\/} 24:1548--7660.

\bibitem[\protect\citeauthoryear{Mouw, Chavez, Edelblute, and Verdery}{Mouw et~al.}{2014}]{mouw_binational_2014}
Mouw, Ted, Sergio Chavez, Heather Edelblute, and Ashton Verdery. 2014.
\newblock \enquote{Binational {Social} {Networks} and {Assimilation}: {A} {Test} of the {Importance} of {Transnationalism}.}
\newblock {\em Social Problems\/} 61:329--359.

\bibitem[\protect\citeauthoryear{Mueller and Gasteyer}{Mueller and Gasteyer}{2023}]{mueller_ethnically_2023}
Mueller, J.~Tom and Stephen Gasteyer. 2023.
\newblock \enquote{The ethnically and racially uneven role of water infrastructure spending in rural economic development.}
\newblock {\em Nature Water\/} 1:74--82.
\newblock Number: 1 Publisher: Nature Publishing Group.

\bibitem[\protect\citeauthoryear{Mummolo and Nall}{Mummolo and Nall}{2016}]{mummolo_why_2016}
Mummolo, Jonathan and Clayton Nall. 2016.
\newblock \enquote{Why {Partisans} {Do} {Not} {Sort}: {The} {Constraints} on {Political} {Segregation}.}
\newblock {\em The Journal of Politics\/} 79:45--59.
\newblock Publisher: The University of Chicago Press.

\bibitem[\protect\citeauthoryear{{National Bureau of Economic Research}}{{National Bureau of Economic Research}}{2016}]{national_bureau_of_economic_research_county_2016}
{National Bureau of Economic Research}. 2016.
\newblock \enquote{County {Distance} {Database}.}
\newblock Retrieved September 29, 2020 (https://data.nber.org/data/county-distance-database.html).

\bibitem[\protect\citeauthoryear{Nogle}{Nogle}{1994}]{nogle_systems_1994}
Nogle, June~Marie. 1994.
\newblock \enquote{The {Systems} {Approach} to {International} {Migration}: {An} {Application} of {Network} {Analysis} {Methods}.}
\newblock {\em International Migration\/} 32:329--342.

\bibitem[\protect\citeauthoryear{Palloni, Massey, Ceballos, Espinosa, and Spittel}{Palloni et~al.}{2001}]{palloni_social_2001}
Palloni, Alberto, Douglas S. Massey, Miguel Ceballos, Kristin Espinosa, and Michael Spittel. 2001.
\newblock \enquote{Social {Capital} and {International} {Migration}: {A} {Test} {Using} {Information} on {Family} {Networks}.}
\newblock {\em American Journal of Sociology\/} 106:1262--1298.

\bibitem[\protect\citeauthoryear{Papachristos, Hureau, and Braga}{Papachristos et~al.}{2013}]{papachristos_corner_2013}
Papachristos, Andrew~V., David~M. Hureau, and Anthony~A. Braga. 2013.
\newblock \enquote{The {Corner} and the {Crew}: {The} {Influence} of {Geography} and {Social} {Networks} on {Gang} {Violence}.}
\newblock {\em American Sociological Review\/} 78:417--447.

\bibitem[\protect\citeauthoryear{Partridge, Rickman, Olfert, and Ali}{Partridge et~al.}{2012}]{partridge_dwindling_2012}
Partridge, Mark~D., Dan~S. Rickman, M.~Rose Olfert, and Kamar Ali. 2012.
\newblock \enquote{Dwindling {U}.{S}. internal migration: {Evidence} of spatial equilibrium or structural shifts in local labor markets?}
\newblock {\em Regional Science and Urban Economics\/} 42:375--388.

\bibitem[\protect\citeauthoryear{Paul}{Paul}{2011}]{paul_stepwise_2011}
Paul, Anju~Mary. 2011.
\newblock \enquote{Stepwise {International} {Migration}: {A} {Multistage} {Migration} {Pattern} for the {Aspiring} {Migrant}.}
\newblock {\em American Journal of Sociology\/} 116:1842--86.

\bibitem[\protect\citeauthoryear{Paul}{Paul}{2017}]{paul_multinational_2017}
Paul, Anju~Mary. 2017.
\newblock {\em Multinational {Maids}: {Stepwise} {Migration} in a {Global} {Labor} {Market}\/}.
\newblock Cambridge University Press.

\bibitem[\protect\citeauthoryear{Piore}{Piore}{2018}]{piore_2018_dual}
Piore, Michael~J. 2018.
\newblock \enquote{The dual labor market: theory and implications.}
\newblock In {\em Social stratification\/}, pp. 629--640. Routledge.

\bibitem[\protect\citeauthoryear{Plantinga, Détang-Dessendre, Hunt, and Piguet}{Plantinga et~al.}{2013}]{plantinga_housing_2013}
Plantinga, Andrew~J., Cécile Détang-Dessendre, Gary~L. Hunt, and Virginie Piguet. 2013.
\newblock \enquote{Housing prices and inter-urban migration.}
\newblock {\em Regional Science and Urban Economics\/} 43:296--306.

\bibitem[\protect\citeauthoryear{Poot, Alimi, Cameron, and Maré}{Poot et~al.}{2016}]{poot_gravity_2016}
Poot, Jacques, Omoniyi Alimi, Michael~P. Cameron, and David~C. Maré. 2016.
\newblock \enquote{The {Gravity} {Model} of {Migration}: {The} {Successful} {Comeback} of an {Ageing} {Superstar} in {Regional} {Science}.}
\newblock {\em SSRN Electronic Journal\/} .

\bibitem[\protect\citeauthoryear{Preuhs}{Preuhs}{1999}]{preuhs_state_1999}
Preuhs, Robert~R. 1999.
\newblock \enquote{State {Policy} {Components} of {Interstate} {Migration} in the {United} {States}.}
\newblock {\em Political Research Quarterly\/} 52:527--549.

\bibitem[\protect\citeauthoryear{Preuhs}{Preuhs}{2020}]{preuhs_pack_2020}
Preuhs, Robert~R. 2020.
\newblock \enquote{Pack {Your} {Politics}! {Assessing} the {Vote} {Choice} of {Latino} {Interstate} {Migrants}.}
\newblock {\em The Journal of Race, Ethnicity, and Politics\/} 5:130--165.

\bibitem[\protect\citeauthoryear{Putnam}{Putnam}{2000}]{putnam_bowling_2000}
Putnam, Robert~D. 2000.
\newblock {\em Bowling {Alone}: {The} {Collapse} and {Revival} of {American} {Community}\/}.
\newblock Simon and Schuster.

\bibitem[\protect\citeauthoryear{Quillian}{Quillian}{2015}]{quillian_comparison_2015}
Quillian, Lincoln. 2015.
\newblock \enquote{A {Comparison} of {Traditional} and {Discrete}-{Choice} {Approaches} to the {Analysis} of {Residential} {Mobility} and {Locational} {Attainment}.}
\newblock {\em The ANNALS of the American Academy of Political and Social Science\/} 660:240--260.

\bibitem[\protect\citeauthoryear{Ravenstein}{Ravenstein}{1885}]{ravenstein_laws_1885}
Ravenstein, E.~G. 1885.
\newblock \enquote{The {Laws} of {Migration}.}
\newblock {\em Journal of the Royal Statistical Society\/} 48:167--235.

\bibitem[\protect\citeauthoryear{Raymer and Rogers}{Raymer and Rogers}{2007}]{raymer_applying_2007}
Raymer, James and Andrei Rogers. 2007.
\newblock \enquote{Applying {Model} {Migration} {Schedules} to {Represent} {Age}-{Specific} {Migration} {Flows}.}
\newblock In {\em International {Migration} in {Europe}\/}, edited by  James Raymer and Frans Willekens, pp. 175--192. Chichester, UK: John Wiley \& Sons, Ltd.

\bibitem[\protect\citeauthoryear{Riddell and Harvey}{Riddell and Harvey}{1972}]{riddell_urban_1972}
Riddell, J.~Barry and Milton~E. Harvey. 1972.
\newblock \enquote{The {Urban} {System} in the {Migration} {Process}: {An} {Evaluation} of {Step}-{Wise} {Migration} in {Sierra} {Leone}.}
\newblock {\em Economic Geography\/} 48:270.

\bibitem[\protect\citeauthoryear{Rogers and Castro}{Rogers and Castro}{1981}]{rogers_model_1981}
Rogers, Andrei and Luis~J Castro. 1981.
\newblock {\em Model {Migration} {Schedules}\/}.
\newblock Laxenburg: International Institute for Applied Systems Analysis.

\bibitem[\protect\citeauthoryear{Ryo}{Ryo}{2013}]{ryo_deciding_2013}
Ryo, Emily. 2013.
\newblock \enquote{Deciding to {Cross}: {Norms} and {Economics} of {Unauthorized} {Migration}.}
\newblock {\em American Sociological Review\/} 78:574--603.

\bibitem[\protect\citeauthoryear{Sakoda}{Sakoda}{1971}]{sakoda_checkerboard_1971}
Sakoda, James~M. 1971.
\newblock \enquote{The checkerboard model of social interaction.}
\newblock {\em The Journal of Mathematical Sociology\/} 1:119--132.

\bibitem[\protect\citeauthoryear{Schelling}{Schelling}{1969}]{schelling_models_1969}
Schelling, Thomas~C. 1969.
\newblock \enquote{Models of {Segregation}.}
\newblock {\em American Economic Review\/} 59:483--493.

\bibitem[\protect\citeauthoryear{Schelling}{Schelling}{1978}]{schelling_micromotives_1978}
Schelling, Thomas~C. 1978.
\newblock {\em Micromotives and {Macrobehavior}\/}.
\newblock W. W. Norton \& Company.

\bibitem[\protect\citeauthoryear{Schewel}{Schewel}{2020}]{schewel_understanding_2020}
Schewel, Kerilyn. 2020.
\newblock \enquote{Understanding {Immobility}: {Moving} {Beyond} the {Mobility} {Bias} in {Migration} {Studies}.}
\newblock {\em International Migration Review\/} 54:328--355.

\bibitem[\protect\citeauthoryear{Schildkraut, Jiménez, Dovidio, and Huo}{Schildkraut et~al.}{2019}]{schildkraut_tale_2019}
Schildkraut, Deborah~J, Tomás~R Jiménez, John~F Dovidio, and Yuen~J Huo. 2019.
\newblock \enquote{A {Tale} of {Two} {States}: {How} {State} {Immigration} {Climate} {Affects} {Belonging} to {State} and {Country} among {Latinos}.}
\newblock {\em Social Problems\/} 66:332--355.

\bibitem[\protect\citeauthoryear{Schiller, Basch, and Blanc}{Schiller et~al.}{1995}]{schiller_immigrant_1995}
Schiller, Nina~Glick, Linda Basch, and Cristina~Szanton Blanc. 1995.
\newblock \enquote{From {Immigrant} to {Transmigrant}: {Theorizing} {Transnational} {Migration}.}
\newblock {\em Anthropological Quarterly\/} 68:48--63.

\bibitem[\protect\citeauthoryear{Schmid and Desmarais}{Schmid and Desmarais}{2017}]{schmid_exponential_2017}
Schmid, Christian~S. and Bruce~A. Desmarais. 2017.
\newblock \enquote{Exponential random graph models with big networks: {Maximum} pseudolikelihood estimation and the parametric bootstrap.}
\newblock In {\em 2017 {IEEE} {International} {Conference} on {Big} {Data} ({Big} {Data})\/}, pp. 116--121.

\bibitem[\protect\citeauthoryear{Schroeder and Pacas}{Schroeder and Pacas}{2021}]{schroeder_across_2021}
Schroeder, Jonathan~P. and José~D. Pacas. 2021.
\newblock \enquote{Across the {Rural}–{Urban} {Universe}: {Two} {Continuous} {Indices} of {Urbanization} for {U}.{S}. {Census} {Microdata}.}
\newblock {\em Spatial Demography\/} 9:131--154.

\bibitem[\protect\citeauthoryear{Sharkey}{Sharkey}{2015}]{sharkey_geographic_2015}
Sharkey, Patrick. 2015.
\newblock \enquote{Geographic {Migration} of {Black} and {White} {Families} {Over} {Four} {Generations}.}
\newblock {\em Demography\/} 52:209--231.

\bibitem[\protect\citeauthoryear{Smith and Papachristos}{Smith and Papachristos}{2016}]{smith_trust_2016}
Smith, Chris~M. and Andrew~V. Papachristos. 2016.
\newblock \enquote{Trust {Thy} {Crooked} {Neighbor}: {Multiplexity} in {Chicago} {Organized} {Crime} {Networks}.}
\newblock {\em American Sociological Review\/} 81:644--667.

\bibitem[\protect\citeauthoryear{Smith, McPherson, and Smith-Lovin}{Smith et~al.}{2014}]{smith_social_2014}
Smith, Jeffrey~A., Miller McPherson, and Lynn Smith-Lovin. 2014.
\newblock \enquote{Social {Distance} in the {United} {States}: {Sex}, {Race}, {Religion}, {Age}, and {Education} {Homophily} among {Confidants}, 1985 to 2004.}
\newblock {\em American Sociological Review\/} 79:432--456.

\bibitem[\protect\citeauthoryear{Snijders}{Snijders}{2002}]{snijders_markov_2002}
Snijders, Tom A~B. 2002.
\newblock \enquote{Markov {Chain} {Monte} {Carlo} {Estimation} of {Exponential} {Random} {Graph} {Models}.}
\newblock {\em Journal of Social Structure\/} 3:1--40.

\bibitem[\protect\citeauthoryear{Sparrowe and Liden}{Sparrowe and Liden}{1997}]{sparrowe_process_1997}
Sparrowe, Raymond~T. and Robert~C. Liden. 1997.
\newblock \enquote{Process and {Structure} in {Leader}-{Member} {Exchange}.}
\newblock {\em Academy of Management Review\/} 22:522--552.

\bibitem[\protect\citeauthoryear{Spring, Mulder, Thomas, and Cooke}{Spring et~al.}{2021}]{spring_migration_2021}
Spring, Amy, Clara~H. Mulder, Michael~J. Thomas, and Thomas~J. Cooke. 2021.
\newblock \enquote{Migration after union dissolution in the {United} {States}: {The} role of non-resident family.}
\newblock {\em Social Science Research\/} 96:102539.

\bibitem[\protect\citeauthoryear{Srivastava and Banaji}{Srivastava and Banaji}{2011}]{srivastava_culture_2011}
Srivastava, Sameer~B. and Mahzarin~R. Banaji. 2011.
\newblock \enquote{Culture, {Cognition}, and {Collaborative} {Networks} in {Organizations}.}
\newblock {\em American Sociological Review\/} 76:207--233.

\bibitem[\protect\citeauthoryear{Steinbeck}{Steinbeck}{1939}]{steinbeck_grapes_1939}
Steinbeck, John. 1939.
\newblock {\em The {Grapes} of {Wrath}\/}.
\newblock Penguin.

\bibitem[\protect\citeauthoryear{Stockdale and Haartsen}{Stockdale and Haartsen}{2018}]{stockdale_editorial_2018}
Stockdale, Aileen and Tialda Haartsen. 2018.
\newblock \enquote{Editorial introduction: {Putting} rural stayers in the spotlight.}
\newblock {\em Population, Space and Place\/} 24:e2124.

\bibitem[\protect\citeauthoryear{Strauss and Ikeda}{Strauss and Ikeda}{1990}]{strauss_pseudolikelihood_1990}
Strauss, David and Michael Ikeda. 1990.
\newblock \enquote{Pseudolikelihood {Estimation} for {Social} {Networks}.}
\newblock {\em Journal of the American Statistical Association\/} 85:204--212.

\bibitem[\protect\citeauthoryear{Tam~Cho, Gimpel, and Hui}{Tam~Cho et~al.}{2013}]{tam_cho_voter_2013}
Tam~Cho, Wendy~K., James~G. Gimpel, and Iris~S. Hui. 2013.
\newblock \enquote{Voter {Migration} and the {Geographic} {Sorting} of the {American} {Electorate}.}
\newblock {\em Annals of the Association of American Geographers\/} 103:856--870.

\bibitem[\protect\citeauthoryear{Thomas, Huang, Yin, Xu, Almquist, Hipp, and Butts}{Thomas et~al.}{2022}]{thomas_geographical_2022}
Thomas, Loring~J., Peng Huang, Fan Yin, Junlan Xu, Zack~W. Almquist, John~R. Hipp, and Carter~T. Butts. 2022.
\newblock \enquote{Geographical patterns of social cohesion drive disparities in early {COVID} infection hazard.}
\newblock {\em Proceedings of the National Academy of Sciences\/} 119:e2121675119.

\bibitem[\protect\citeauthoryear{Tiebout}{Tiebout}{1956}]{tiebout_pure_1956}
Tiebout, Charles~M. 1956.
\newblock \enquote{A {Pure} {Theory} of {Local} {Expenditures}.}
\newblock {\em Journal of Political Economy\/} 64:416--424.

\bibitem[\protect\citeauthoryear{Tocqueville}{Tocqueville}{1834}]{tocqueville_democracy_1834}
Tocqueville, Alexis. 1834.
\newblock {\em Democracy in {America}: {And} {Two} {Essays} on {America}\/}.
\newblock Penguin UK.

\bibitem[\protect\citeauthoryear{Todaro}{Todaro}{1976}]{todaro_internal_1976}
Todaro, Michael~P. 1976.
\newblock {\em Internal migration in developing countries\/}.
\newblock Genève, Switzerland: International Labour Office.

\bibitem[\protect\citeauthoryear{Tolnay}{Tolnay}{2003}]{tolnay_african_2003}
Tolnay, Stewart~E. 2003.
\newblock \enquote{The {African} {American} “{Great} {Migration}” and {Beyond}.}
\newblock {\em Annual Review of Sociology\/} 29:209--232.

\bibitem[\protect\citeauthoryear{Treyz, Rickman, Hunt, and Greenwood}{Treyz et~al.}{1993}]{treyz_dynamics_1993}
Treyz, George~I., Dan~S. Rickman, Gary~L. Hunt, and Michael~J. Greenwood. 1993.
\newblock \enquote{The {Dynamics} of {U}.{S}. {Internal} {Migration}.}
\newblock {\em The Review of Economics and Statistics\/} 75:209--214.

\bibitem[\protect\citeauthoryear{{U.S. Census Bureau}}{{U.S. Census Bureau}}{2013}]{us_census_bureau_census_2013}
{U.S. Census Bureau}. 2013.
\newblock \enquote{Census {Regions} and {Divisions} of the {United} {States}.}
\newblock Retrieved September 29, 2020 (https://www2.census.gov/geo/pdfs/maps-data/maps/reference/us\_regdiv.pdf).

\bibitem[\protect\citeauthoryear{van Duijn, Gile, and Handcock}{van Duijn et~al.}{2009}]{van_duijn_framework_2009}
van Duijn, Marijtje A.~J., Krista~J. Gile, and Mark~S. Handcock. 2009.
\newblock \enquote{A framework for the comparison of maximum pseudo-likelihood and maximum likelihood estimation of exponential family random graph models.}
\newblock {\em Social Networks\/} 31:52--62.

\bibitem[\protect\citeauthoryear{Verdery, Mouw, Edelblute, and Chavez}{Verdery et~al.}{2018}]{verdery_communication_2018}
Verdery, Ashton~M., Ted Mouw, Heather Edelblute, and Sergio Chavez. 2018.
\newblock \enquote{Communication flows and the durability of a transnational social field.}
\newblock {\em Social Networks\/} 53:57--71.

\bibitem[\protect\citeauthoryear{Vogel}{Vogel}{2007}]{vogel_knockout_2007}
Vogel, Gretchen. 2007.
\newblock \enquote{A {Knockout} {Award} in {Medicine}.}
\newblock {\em Science\/} 318:178--179.

\bibitem[\protect\citeauthoryear{von Reichert, Cromartie, and Arthun}{von Reichert et~al.}{2014a}]{von_reichert_impacts_2014}
von Reichert, Christiane, John~B. Cromartie, and Ryan~O. Arthun. 2014a.
\newblock \enquote{Impacts of {Return} {Migration} on {Rural} {U}.{S}. {Communities}.}
\newblock {\em Rural Sociology\/} 79:200--226.

\bibitem[\protect\citeauthoryear{von Reichert, Cromartie, and Arthun}{von Reichert et~al.}{2014b}]{von_reichert_reasons_2014}
von Reichert, Christiane, John~B. Cromartie, and Ryan~O. Arthun. 2014b.
\newblock \enquote{Reasons for {Returning} and {Not} {Returning} to {Rural} {U}.{S}. {Communities}.}
\newblock {\em The Professional Geographer\/} 66:58--72.

\bibitem[\protect\citeauthoryear{Vögtle and Windzio}{Vögtle and Windzio}{2022}]{vogtle_global_2022}
Vögtle, EvaMaria and Michael Windzio. 2022.
\newblock \enquote{The ‘{Global} {South}’ in the transnational student mobility network. {Effects} of institutional instability, reputation of the higher education systems, post-colonial ties, and culture.}
\newblock {\em Globalisation, Societies and Education\/} pp. 1--19.

\bibitem[\protect\citeauthoryear{Waldinger}{Waldinger}{2013}]{waldinger_immigrant_2013}
Waldinger, Roger. 2013.
\newblock \enquote{Immigrant transnationalism.}
\newblock {\em Current Sociology\/} 61:756--777.

\bibitem[\protect\citeauthoryear{Waldinger and Fitzgerald}{Waldinger and Fitzgerald}{2004}]{waldinger_transnationalism_2004}
Waldinger, Roger and David Fitzgerald. 2004.
\newblock \enquote{Transnationalism in {Question}.}
\newblock {\em American Journal of Sociology\/} 109:1177--1195.
\newblock Publisher: The University of Chicago Press.

\bibitem[\protect\citeauthoryear{Wallace and Karra}{Wallace and Karra}{2020}]{wallace_true_2020}
Wallace, Tim and Krishna Karra. 2020.
\newblock \enquote{The {True} {Colors} of {America}’s {Political} {Spectrum} {Are} {Gray} and {Green}.}
\newblock {\em The New York Times\/} .

\bibitem[\protect\citeauthoryear{Wallerstein}{Wallerstein}{2011}]{wallerstein_modern_2011}
Wallerstein, Immanuel. 2011.
\newblock {\em The {Modern} {World}-{System} {I}: {Capitalist} {Agriculture} and the {Origins} of the {European} {World}-{Economy} in the {Sixteenth} {Century}\/}.
\newblock University of California Press.
\newblock Google-Books-ID: JZqhKZ9ucc0C.

\bibitem[\protect\citeauthoryear{Wasserman and Pattison}{Wasserman and Pattison}{1996}]{wasserman_logit_1996}
Wasserman, Stanley and Philippa Pattison. 1996.
\newblock \enquote{Logit models and logistic regressions for social networks: {I}. {An} introduction to {Markov} graphs and p*.}
\newblock {\em Psychometrika\/} 61:401--425.

\bibitem[\protect\citeauthoryear{Weber}{Weber}{1922}]{weber_Wirtschaft_1922}
Weber, Max. 1922.
\newblock {\em Wirtschaft und {Gesellschaft}: {Grundriß} der {Verstehenden} {Soziologie}\/}.
\newblock Duncker \& Humblot.

\bibitem[\protect\citeauthoryear{White and Liang}{White and Liang}{1998}]{white_effect_1998}
White, Michael~J. and Zai Liang. 1998.
\newblock \enquote{The effect of immigration on the internal migration of the native-born population, 1981–1990.}
\newblock {\em Population Research and Policy Review\/} 17:141--166.

\bibitem[\protect\citeauthoryear{Wilson, O'Brien, and Sesma}{Wilson et~al.}{2009}]{wilson_human_2009}
Wilson, David~Sloan, Daniel~Tumminelli O'Brien, and Artura Sesma. 2009.
\newblock \enquote{Human prosociality from an evolutionary perspective: variation and correlations at a city-wide scale.}
\newblock {\em Evolution and Human Behavior\/} 30:190--200.

\bibitem[\protect\citeauthoryear{Wimmer and Lewis}{Wimmer and Lewis}{2010}]{wimmer_beyond_2010}
Wimmer, Andreas and Kevin Lewis. 2010.
\newblock \enquote{Beyond and {Below} {Racial} {Homophily}: {ERG} {Models} of a {Friendship} {Network} {Documented} on {Facebook}.}
\newblock {\em American Journal of Sociology\/} 116:583--642.

\bibitem[\protect\citeauthoryear{Windzio}{Windzio}{2018}]{windzio_network_2018}
Windzio, Michael. 2018.
\newblock \enquote{The network of global migration 1990–2013.}
\newblock {\em Social Networks\/} 53:20--29.

\bibitem[\protect\citeauthoryear{Windzio, Teney, and Lenkewitz}{Windzio et~al.}{2019}]{windzio_network_2019}
Windzio, Michael, Céline Teney, and Sven Lenkewitz. 2019.
\newblock \enquote{A network analysis of intra-{EU} migration flows: how regulatory policies, economic inequalities and the network-topology shape the intra-{EU} migration space.}
\newblock {\em Journal of Ethnic and Migration Studies\/} 47:951--969.

\bibitem[\protect\citeauthoryear{Wright, Ellis, and Reibel}{Wright et~al.}{1997}]{wright_linkage_1997}
Wright, Richard~A., Mark Ellis, and Michael Reibel. 1997.
\newblock \enquote{The {Linkage} between {Immigration} and {Internal} {Migration} in {Large} {Metropolitan} {Areas} in the {United} {States}.}
\newblock {\em Economic Geography\/} 73:234--254.

\bibitem[\protect\citeauthoryear{Xie and Zhang}{Xie and Zhang}{2019}]{xie_long-term_2019}
Xie, Yu and Chunni Zhang. 2019.
\newblock \enquote{The long-term impact of the {Communist} {Revolution} on social stratification in contemporary {China}.}
\newblock {\em Proceedings of the National Academy of Sciences\/} 116:19392--19397.

\bibitem[\protect\citeauthoryear{Zhou}{Zhou}{1992}]{zhou_chinatown_1992}
Zhou, Min. 1992.
\newblock {\em Chinatown: {The} {Socioeconomic} {Potential} of an {Urban} {Enclave}\/}.
\newblock Temple University Press.

\bibitem[\protect\citeauthoryear{Zipf}{Zipf}{1946}]{zipf_p1_1946}
Zipf, George~Kingsley. 1946.
\newblock \enquote{The {P1} {P2}/{D} {Hypothesis}: {On} the {Intercity} {Movement} of {Persons}.}
\newblock {\em American Sociological Review\/} 11:677--686.

\bibitem[\protect\citeauthoryear{Zipf}{Zipf}{1949}]{zipf_human_1949}
Zipf, George~Kingsley. 1949.
\newblock {\em Human behavior and the principle of least effort\/}.
\newblock Oxford, England: Addison-Wesley Press.

\end{thebibliography}
\end{document}